\documentclass[pra,aps,showpacs,floatfix]{revtex4}
\usepackage{amsfonts}
\usepackage{graphicx}
\usepackage[ansinew]{inputenc}
\usepackage{array}
\usepackage{color}
\usepackage{amsmath}
\usepackage{amsxtra}
\usepackage{amstext}
\usepackage{amssymb}
\usepackage{latexsym}
\usepackage{dsfont}
\usepackage{verbatim}

\begin{document}

\title{The classical and quantum synchronization between two scattering modes in
Bose-Einstein condensates generated by the standing-wave laser beams}
\author{Lin Zhang$^1$}
\email{zhanglincn@snnu.edu.cn}
\author{Xiaoting Xu$^1$}
\author{Xing Liu$^1$}
\author{Weiping Zhang$^{2,3,4}$}
\email{wpzhang@phy.ecnu.edu.cn}
\affiliation{$^1$School of physics and Information Technology, Shaanxi Normal University,
Xi'an 710119, China}
\affiliation{$^{2}$Department of Physics and Astronomy, Shanghai Jiao Tong University,
Shanghai 200240, P. R. China}
\affiliation{$^{3}$Quantum Institute for Light and Atoms, Department of Physics,
East China Normal University, No.500,  Shanghai 200241, P. R. China}
\affiliation{$^{4}$Collaborative Innovation Center of Extreme Optics, Shanxi University,
Taiyuan, Shanxi 030006, P. R. China}

\begin{abstract}
Both classical and quantum dynamics of the synchronization between two
nonlinear mechanical modes scattered from Bose-Einstein condensates (BECs)
by the standing-wave laser beam are comparatively investigated. As the
ultra-low dissipations of the momentum modes in the atomic BECs, the
synchronized dynamics are studied in a framework of closed-system theory
in order to track down both the classical and the quantum synchronizations from an
angle of quantum control. The classical synchronization and the relevant
dynamics of measure synchronization, the quantum synchronization and two
different types of measures proposed by Mari and estimated by mutual information
based on $Q$-function are studied respectively in order to reveal both the macroscopic
and the microscopic signatures of synchronized behaviors in a closed quantum system.
The results demonstrate that the ``revival and collapse" of the quantum fluctuations
beyond the classical mean-value dynamics due to long-lasting mode coherence
discriminates the quantum synchronization from the classical one, which not only
excludes the possibilities of an exact synchronization and a perfect density overlap
in phase space, but also leads to upper limitations to Mari measure and large
unceasing fluctuations to mutual information between two scattering modes. We reveal
a close dynamic connection between Mari measure and the mutual information of two nonlinear
momentum modes in closed BEC systems by demonstrating an opposite mean-value
behavior but a similar fluctuation variation with respect to their respective
evolutionary scales.
\end{abstract}

\maketitle

\section{Introduction}

In recent years, many attentions are paid to the quantum control problems for
exploring classically driven devices which are actually working in a quantum
framework \cite{Brif,Wiseman}. For the quantum control on a collection of nearly
identical microscopic quantum units from a macroscopic level down to atomic/molecular
scale, only the synchronized behavior emerged from the microscopic scale can be
identified by the classical devices as a macroscopic signature, and thus the synchronization
behaviors at the macroscopic scale, that is, the classical synchronization (CS), play a
critical role to trace down (classically detect or control) the coherent quantum dynamics
among the microscopic quantum units. However, the synchronized dynamics between different
quantum units, which induce dynamical transitions of macroscopic temporal-spacial orders,
is intrinsically determined by the microscopic synchronization within a decoherence time for
a quantum evolution. Therefore, the quantum synchronization (QS) as another important signature
for the quantum coherent dynamics have been extensively studied with the nanomechanical
resonators \cite{Bagheri,Shim,Ying}, the spins and the atomic qubits \cite{Orth,Hush}.
The dynamics of the coupled oscillators leading to QS have also been extensively considered
in the optomechanical systems \cite{Holmes,Shlomi,Heinrich,Zhang}.

In this paper, we want to push this topic by exploring both classical and quantum natures
of the synchronization between two nonlinear collective modes in a macroscopic quantum
matter: Bose-Einstein condensate (BEC). In BEC, the synchronized behaviors between a large number
of atoms are complicated because the dimension of BEC is huge, but the main features of synchronization
can also be revealed by only analyzing the dynamics between several collective modes
coherently excited from it, especially by the simplest case of a two-mode excitation. In a two-mode
quantum system, one mode can be treated as a passive oscillator which is easily initialized to a
ground state and the other is the active one which produces quantum control actions via their mutual
couplings. In this driven-response closed loops, the dynamical measures of two quantum oscillators
to be synchronized within a decoherence time are the useful signatures to guide efficient controls
on the quantum evolution of the passive oscillator starting from a specific initial state.
Recently, many measures are presented for the judgement of QS \cite{Mari,Ameri} and most of
the works considered two coupled nonlinear quantum oscillators exposed to different dissipative
baths \cite{Goy,Zhi,Agrawal,Lee,Math,Walter,Gie}. These studies are mainly concerned about QS
phenomena in the open quantum systems, and the results demonstrated that QS of an open quantum
system is eventually determined by the statistical properties of the external baths \cite{Breuer}.
Further studies indicate that two oscillators coupled to a common reservoir are more likely to be
synchronized than that they separately couple to two different reservoirs \cite{Giorgi}.
These dissipative synchronizations are all related to the stable steady states or the dynamical
attractors to which the coupled systems can eventually settle down due to the dissipation and
the decoherence induced by the baths \cite{Xue}. However, for a control problem, a full real-time
evolution of the correlated states between coupled units within a decoherence time is more important
than the longtime asymptotic behavior. Therefore the damping and the couplings with the baths can
be neglected in a short time interval which enables an assumption of a closed quantum system.
As the conservation of the phase volume for a closed quantum system, the synchronization without
dissipations will be very different from that of an open quantum system. In this paper, we will
consider the dynamical synchronization between two autonomous nonlinear modes generated in the
atomic BECs \cite{Weiping} by investigating the full synchronized behaviors in both the classical
and the quantum regimes.

As discussed above, a complete synchronization for a closed quantum system must
include both the macroscopic and the microscopic signatures of the synchronized dynamics, that is,
both the CS and QS. In the classical regime, the synchronization phenomena for a dissipative system
have been studied for nearly half a centaury from the coupled Huygens clock (in 1665)
to chemical reactions, biological and social behaviors \cite{Janson,Pikovsky}.
Many CS measures have been proposed in differen ways, such as complete synchronization,
generalized synchronization, phase synchronization, lag synchronization,
projective synchronization and so on \cite{Pikovsky,Zhou}. However, from a microscopic point of view, how to estimate
QS of quantum systems is still a problem. A
natural way to define a measure for QS is to generalize
the classical counterparts directly to the quantum regime. However, the
generalization of the classical measures to quantum case is not always
straightforward because not all the classical counterparts in
the quantum regime can be uniquely found, for example, the phase synchronization is related to the
Hermite phase operator which, obviously, is difficult to be defined in the
quantum regime \cite{Barnett}. While for the coupled conservative
Hamiltonian systems, an interesting CS called measure synchronization (MS)
was found \cite{Zanette} and, recently, its QS case was considered in
the BEC system \cite{Qiu}.
Up to now, two types of measure for QS are proposed in the literatures, one is based on the
distance between two quantum states in the Hilbert space and the other relies on
the quantum state correlations. However the distance of the quantum states in
Hilbert space is not unique and different norms or inner products
can be adopted for different QS definitions along this line. Moreover the
QS measures defined by using state correlations are even more
complicated because there exist many quantum correlations for two quantum states, such as quantum discord,
quantum fidelity, quantum coherence, quantum entropy, and even different quantum
entanglements \cite{Manzano}. Therefore, the measure to judge QS
between two coupled quantum systems might not be theoretically
unique and which one we should pick is only determined by what microscopic property we want to resolve.
In this paper, we will investigate the dynamics of QS between
two atomic scattering modes in BEC by using $Q$-function as a tool.
As $Q$-function is positive and bounded, we can easily combine
two types of QS measure together (the distance and the correlation) to
explore the dynamical features of QS in a closed quantum system by
investigating either the ``distance" (overlap) from a classical aspect or
the ``correlation" based on the mutual information theory.

\section{The Theoretical model}

Theoretically, we consider only two low-energy scattering modes $A$ and $B$
stimulated in BEC by a standing-wave laser beam. The coherent atomic BEC can
be prepared either in a number state or in a coherent state (see Ref.\cite{Weiping}
for the state preparation of BEC). The Hamiltonian describing two scattering
modes within a BEC beam in a number state $|N\rangle$ can be described by \cite{Weiping}%
\begin{equation}
\hat{H}_N=\hat{H}_{A}+\hat{H}_{B}+\hat{H}_{AB},  \label{H1}
\end{equation}%
\begin{equation*}
\hat{H}_{A}=\hbar \omega _{A}\hat{a}_{A}^{\dag }\hat{a}_{A}+\hbar \chi \hat{a%
}_{A}^{\dag 2}\hat{a}_{A}^{2},
\end{equation*}%
\begin{equation*}
\hat{H}_{B}=\hbar \omega _{B}\hat{a}_{B}^{\dag }\hat{a}_{B}+\hbar \chi \hat{a%
}_{B}^{\dag 2}\hat{a}_{B}^{2},
\end{equation*}%
\begin{equation*}
\hat{H}_{AB}=\hbar \left[ g+\chi (\hat{N}-1) \right] \left( \hat{%
a}_{A}^{\dag }\hat{a}_{B}+\hat{a}_{B}^{\dag }\hat{a}_{A}\right) +4\hbar \chi
\hat{a}_{A}^{\dag }\hat{a}_{A}\hat{a}_{B}^{\dag }\hat{a}_{B},
\end{equation*}%
where $\omega_A$ and $\omega_B$ are the mechanical frequencies of two
momentum modes induced by the collective atomic recoil \cite{Lin}, $g=|\Omega|^2/4\Delta$
is the linear coupling rate with $\Omega$ being the Rabi frequency of the
atom interacting with the standing-wave field and $\Delta$ being the detuning of
the laser field from the atomic transition frequency, and $\chi$ is the nonlinear coupling rate
which is due to the interatomic dipole-dipole interaction (photon exchange)
\cite{Weiping,Smerzi}. For the scattering modes generated in a BEC of number state,
the total atomic number $\hat{N}=\hat{N}_{A}+\hat{N}_{B}=%
\hat{a}_{A}^{\dag }\hat{a} _{A}+\hat{a}_{B}^{\dag }\hat{a}_{B}$
is conserved in a sense of $[\hat{N},\hat{H}_N]=0$.

System (\ref{H1}) can be
treated as two coupled nonlinear oscillators in Hilbert space of $\mathcal{H}_N$, where $\hat{H}_{A,B}$ denotes
two free Kerr-like nonlinear oscillators with free energy of $\hbar \omega
_{A,B}+\hbar \chi (N_{A,B}-1)$ in a number-state representation of $\hat{N}%
_{A,B}\left\vert N_{A,B}\right\rangle =N_{A,B}\left\vert
N_{A,B}\right\rangle $. The first term in $\hat{H}_{AB}$ describes the
mutual coupling of two nonlinear Boson oscillators which induces mode
transitions and the second term of $\hat{H}_{AB}$ describes intensity
interaction analogous to two light modes in Kerr medium [20]. Surely, the above
model can also be obtained in the BEC system with a double-well potential with
two-mode approximation (left-well and right-well modes) \cite{Smerzi}.
Usually, for a two-mode system, an interaction picture with respect to the
middle energy of
$\hat{H}_{0}=\frac{1}{2}\hbar \left( \omega _{A}+\omega _{B}\right) (
\hat{a}_{A}^{\dag }\hat{a}_{A}+\hat{a}_{B}^{\dag }\hat{a}_{B})$
can be introduced as%
\begin{equation}
\hat{H}^{\prime }_{N}=\hat{H}_N-\hat{H}_{0}=\hbar \delta \left( \hat{a}_{A}^{\dag }%
\hat{a}_{A}-\hat{a}_{B}^{\dag }\hat{a}_{B}\right) +\hbar \chi \hat{a}%
_{A}^{\dag 2}\hat{a}_{A}^{2}+\hbar \chi \hat{a}_{B}^{\dag 2}\hat{a}_{B}^{2}+%
\hat{H}_{AB},  \label{H2}
\end{equation}%
where the free half-level gap between the two modes is defined by $\delta =\frac{1}{2}\left( \omega
_{A} -\omega _{B}\right)$.

Surely, we can also investigate the scattering modes generated from BEC
in a coherent state that is prepared by coherently mixing of several BEC beams with different atomic
number of $N$ \cite{Weiping}. In this sense, the above quantum system (\ref{H2}) can be generalized
into a two-mode system in a
Hilbert space of $\mathcal{H}=\bigoplus_{N}\mathcal{H}_{N}$.
In the following sections we will consider both cases in the
interaction picture by omitting the prime of $\hat{H}^{\prime}_{N}$ for
simplicity.

\section{Nonlinear dynamics of CS: measure synchronization}

Firstly, we consider the macroscopic synchronized behavior of two scattering modes in a classical point of view and
the Hamiltonian system (\ref{H2}) admits a well-known CS: measure synchronization (MS) \cite%
{Zanette}. The operator equations for the two momentum modes
in the BEC beam with a total atomic number of $N$ will be determined by%
\begin{eqnarray}
i\frac{d\hat{a}_{A}}{dt} &=&\delta \hat{a}_{A}+g\hat{a}_{B}+2\chi %
\left(\hat{N}+\hat{N}_{B}\right) \hat{a}_{A}+\chi\left(\hat{N}+\hat{N}%
_{A}\right) \hat{a}_{B}+\chi \hat{a}_{B}^{\dag }\hat{a}_{A}^{2},  \label{a} \\
i\frac{d\hat{a}_{B}}{dt} &=&-\delta \hat{a}_{B}+g\hat{a}_{A}+2\chi%
\left(\hat{N}+\hat{N}_{A}\right) \hat{a}_{B}+\chi \left( \hat{N}+\hat{N}%
_{B}\right) \hat{a}_{A}+\chi \hat{a}_{A}^{\dag }\hat{a}_{B}^{2},  \label{b}
\end{eqnarray}%
where $\hat{N}=\hat{N}_{A}+\hat{N}_{B}$.
Equivalently, we can write it by two parts, the linear and the
nonlinear parts, as%
\begin{equation*}
i\frac{d}{dt}\left(
\begin{array}{c}
\hat{a}_{A} \\
\hat{a}_{B}%
\end{array}%
\right) =\left(
\begin{array}{cc}
\delta & g \\
g & -\delta%
\end{array}%
\right) \left(
\begin{array}{c}
\hat{a}_{A} \\
\hat{a}_{B}%
\end{array}%
\right) +\chi \left[
\begin{array}{cc}
2\left( \hat{N}+\hat{N}_{B}\right) +\hat{a}_{B}^{\dag }\hat{a}_{A} & \hat{N}+%
\hat{N}_{A} \\
\hat{N}+\hat{N}_{B} & 2\left( \hat{N}+\hat{N}_{A}\right) +\hat{a}_{A}^{\dag }%
\hat{a}_{B}%
\end{array}%
\right] \left(
\begin{array}{c}
\hat{a}_{A} \\
\hat{a}_{B}%
\end{array}%
\right) .
\end{equation*}%
If the nonlinear interatomic scattering coefficient (due to dipole-dipole
interaction) $\chi =0$, the Hamiltonian will reduce to a typical JC model
with an effective Rabi frequency of $\Omega _{R}=\sqrt{\delta ^{2}+g^{2}}$. The
nonlinear dipole-dipole coefficient $\chi $ will modify $\delta $ and $g$ by
introducing a number-dependent Kerr interaction between two
momentum modes, leading to an approximate Rabi
frequency of $ \Omega _{R}\approx \sqrt{( \delta +2\chi \hat{N}) ^{2}
+(g+\chi \hat{N}) ^{2}}$.
%%Fig1%%
\begin{figure}[t]
\begin{center}
\includegraphics[width=440pt]{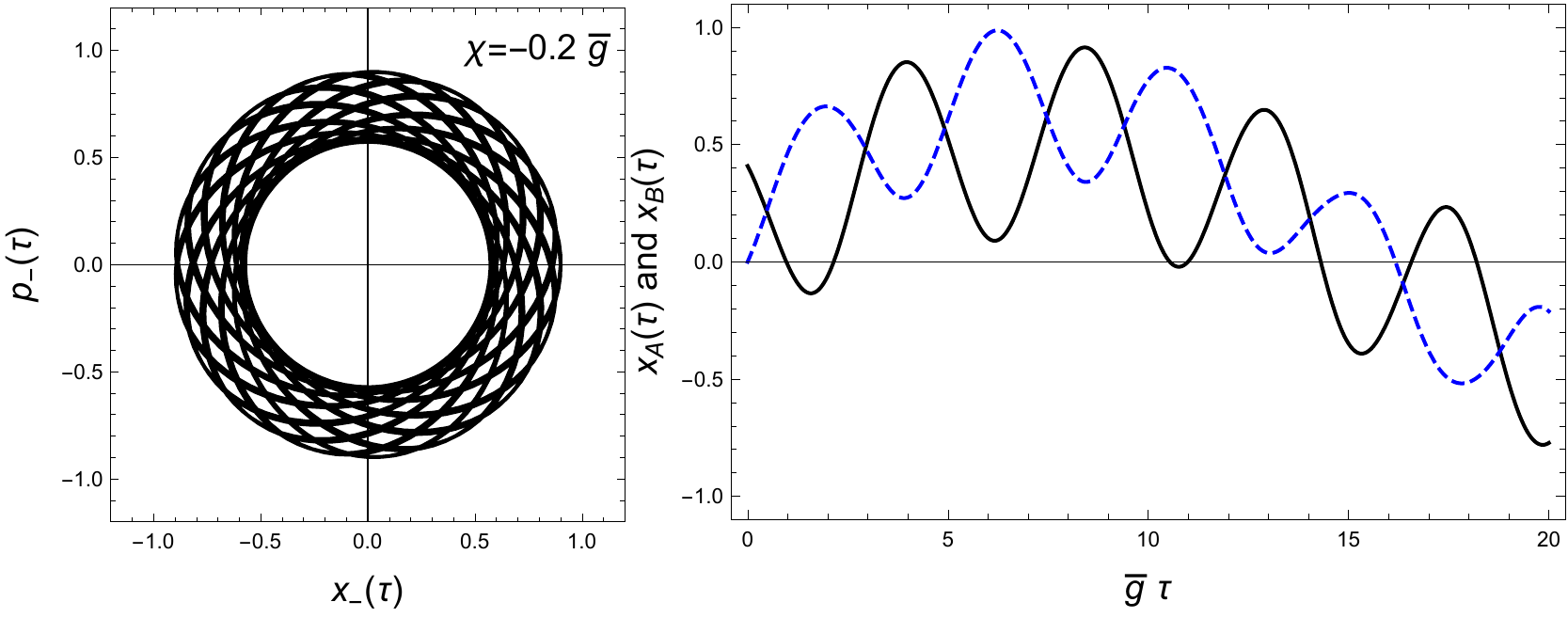}\\[0pt]
\includegraphics[width=440pt]{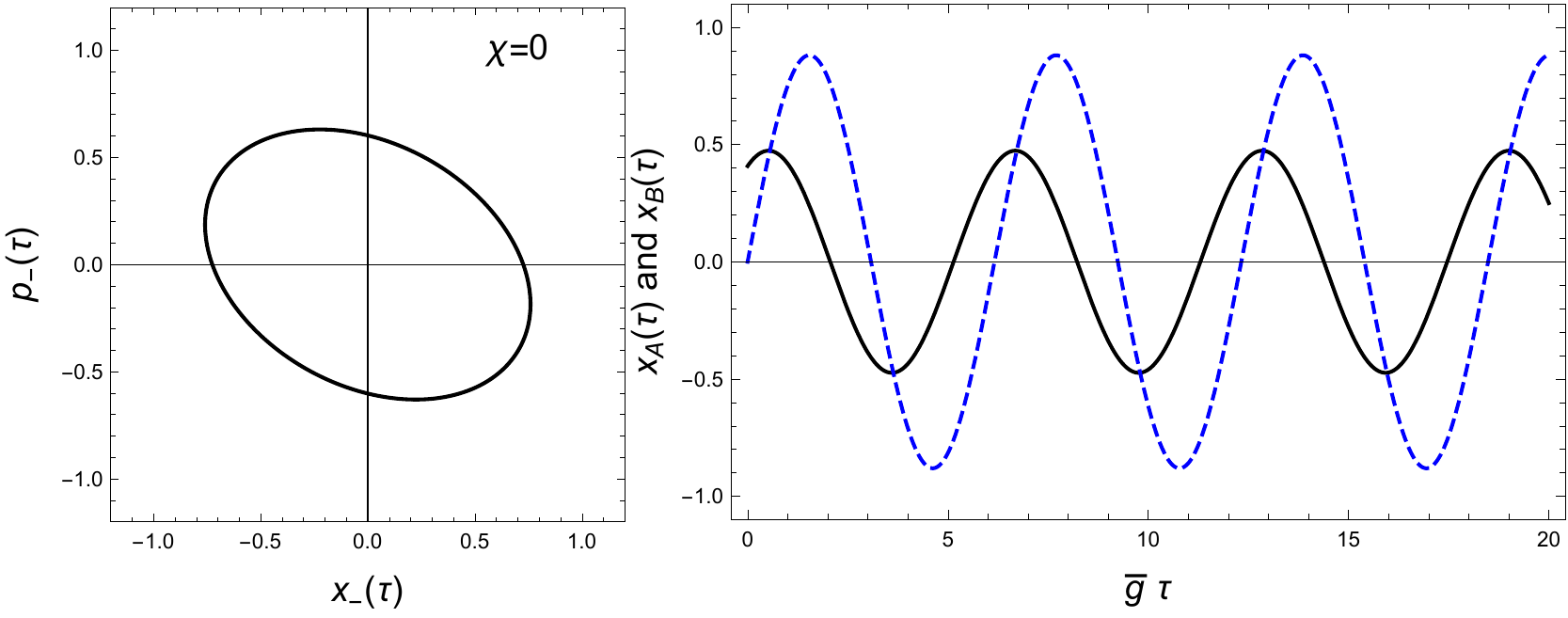}\\[0pt]
\includegraphics[width=440pt]{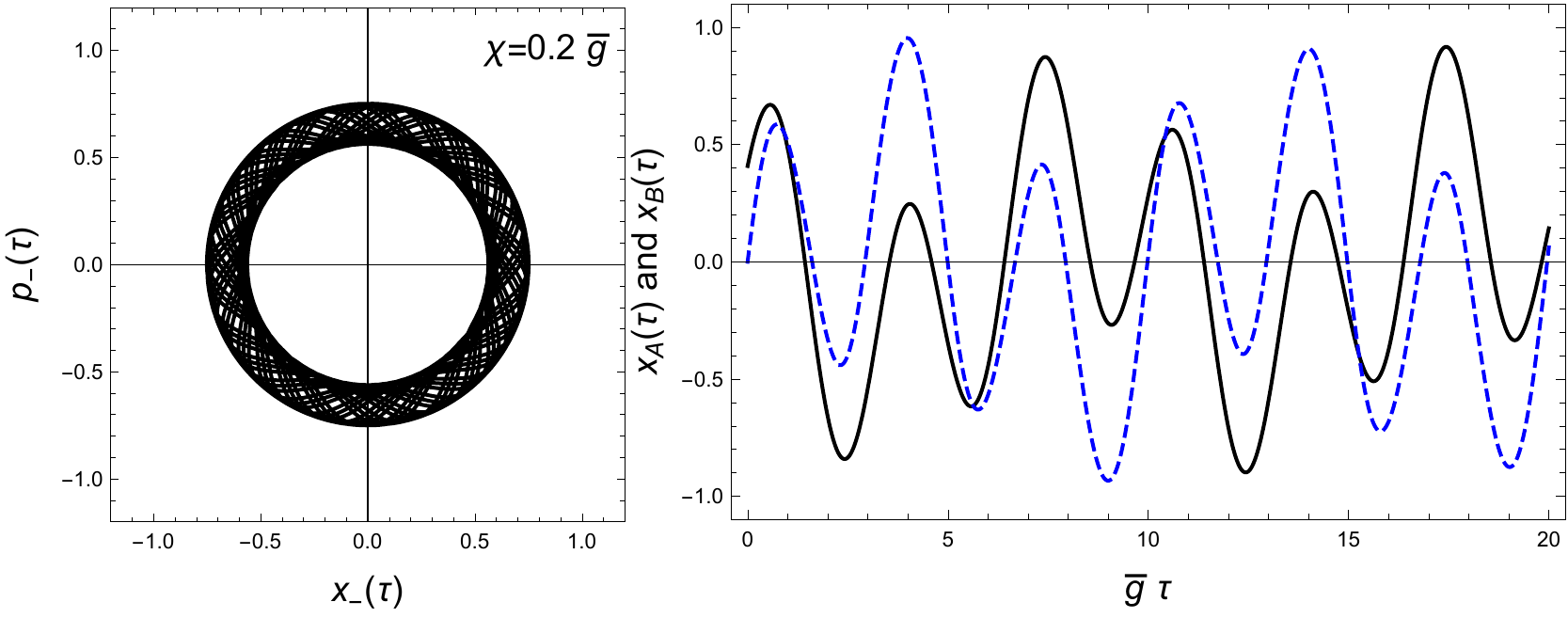}
\end{center}
\caption{The dynamics of the quadratures $x_A$ (solid lines), $x_B$ (dashed
lines) and their corresponding errors ($x_{-}$ and $p_{-}$) in the phase
space under different nonlinear coupling rates $\protect\chi=-0.2\bar{g}$
(upper panel), $\protect\chi=0$ (middle panel) and $\protect\chi%
=0.2\bar{g}$ (lower panel). The mode detuning is $\protect\bar{\delta}%
=-0.2\bar{g}$ and the initial conditions are $\protect\alpha(0)=(1+2i)/%
\protect\sqrt{6}$ and $\protect\beta(0)=i/\protect\sqrt{6}$.}
\label{figure1}
\end{figure}

\begin{figure}[b]
\begin{center}
\includegraphics[width=162pt]{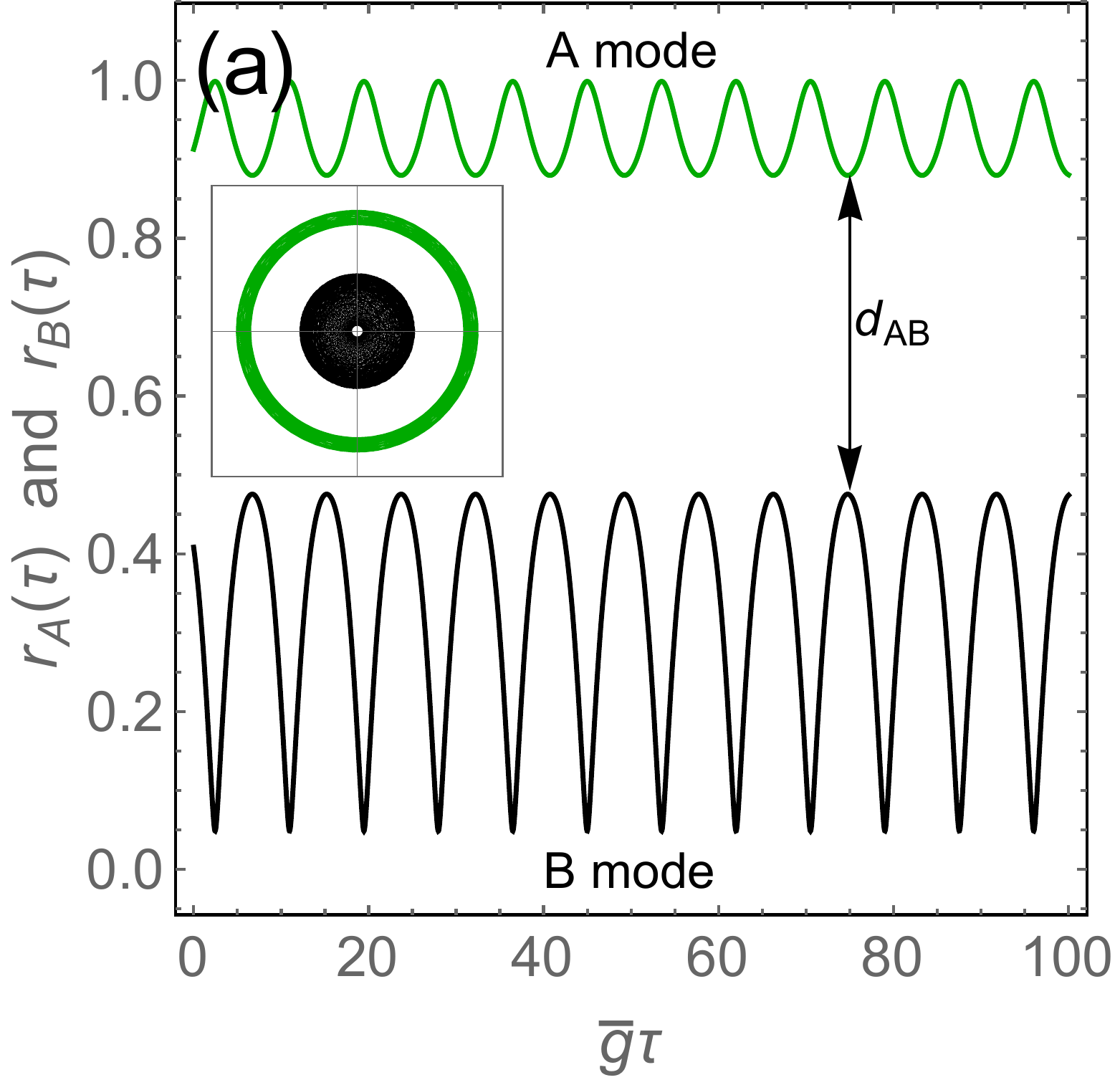} %
\includegraphics[width=162pt]{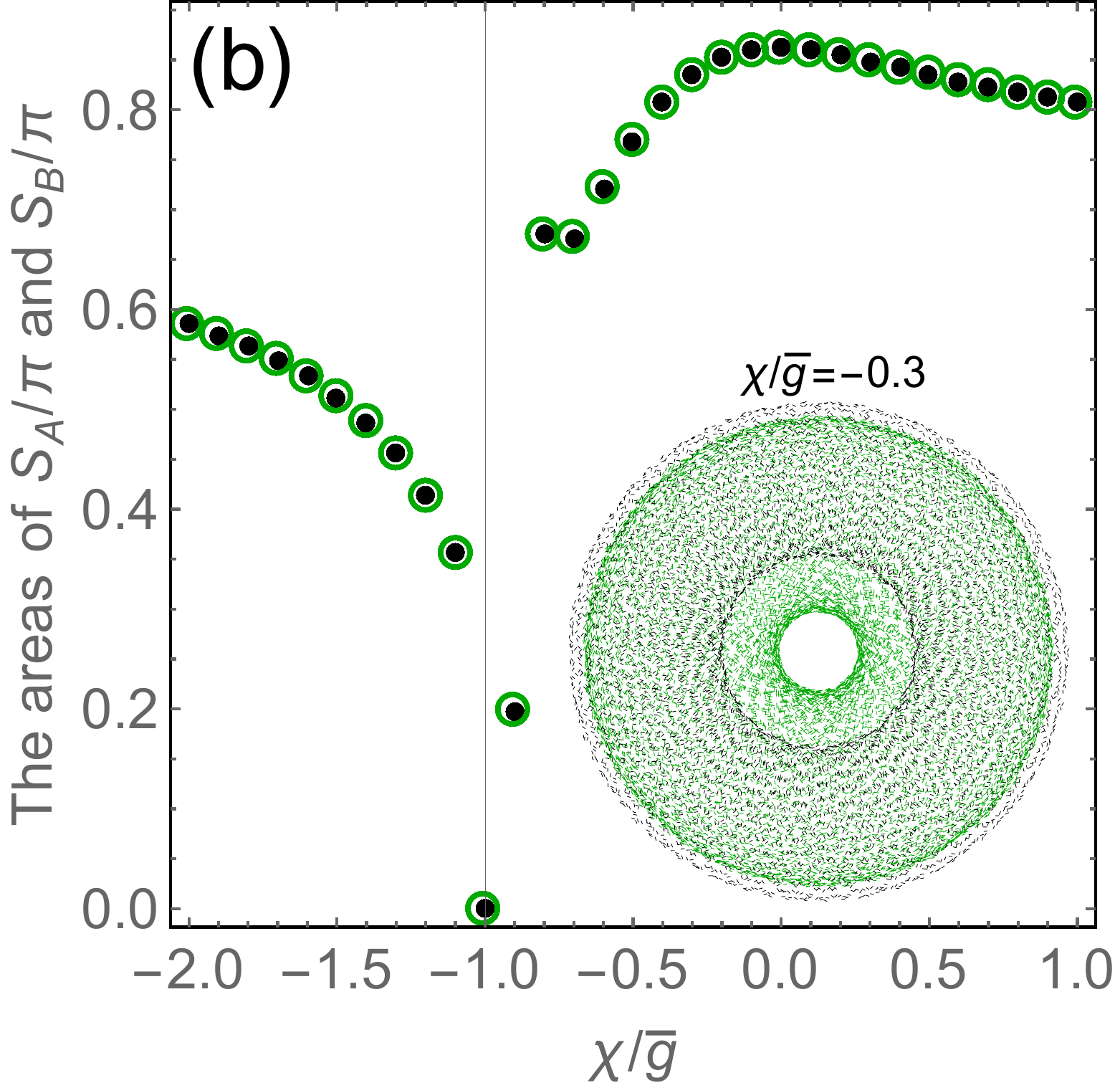} %
\includegraphics[width=164pt]{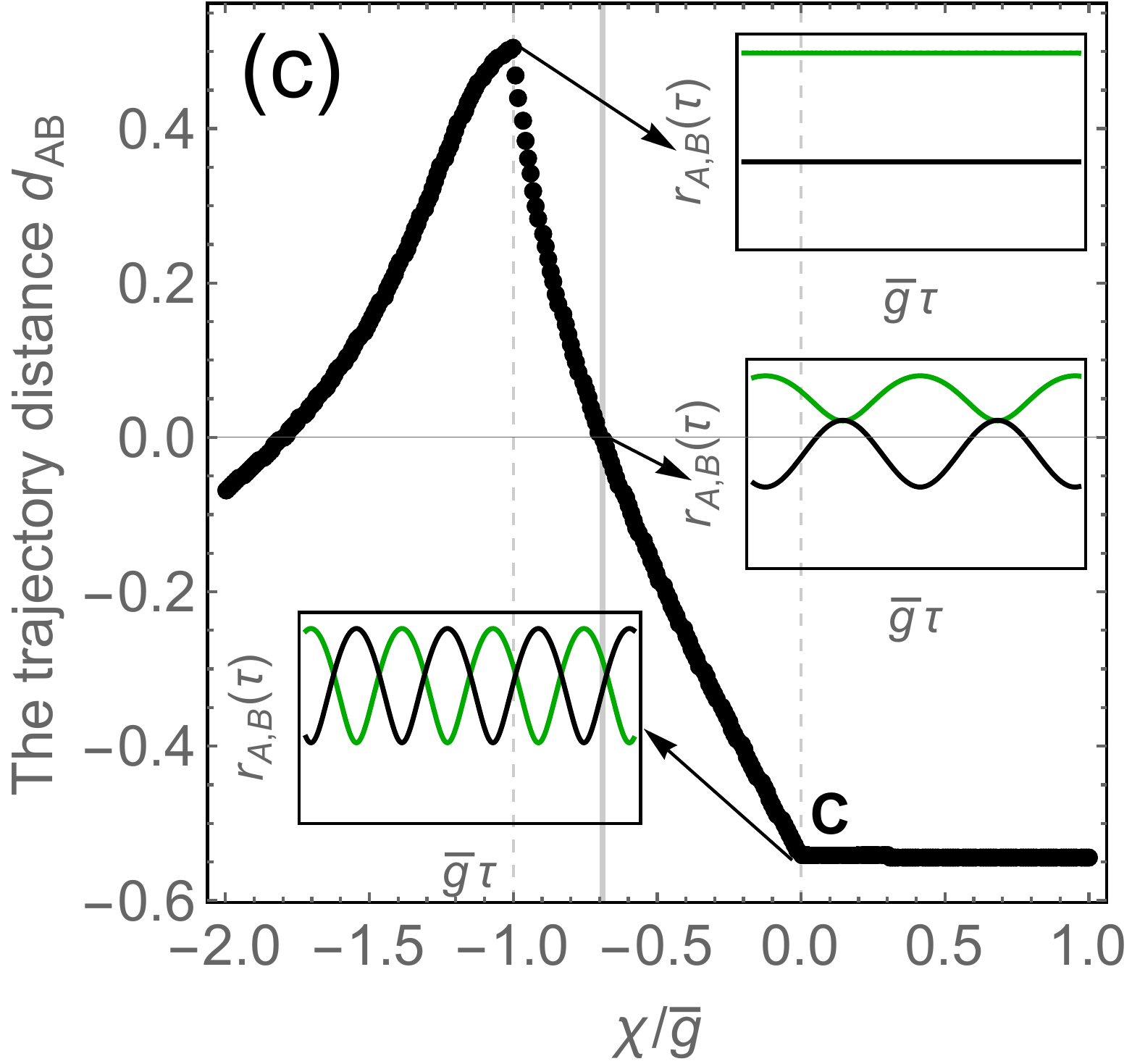}\\[0pt]
\includegraphics[width=122pt]{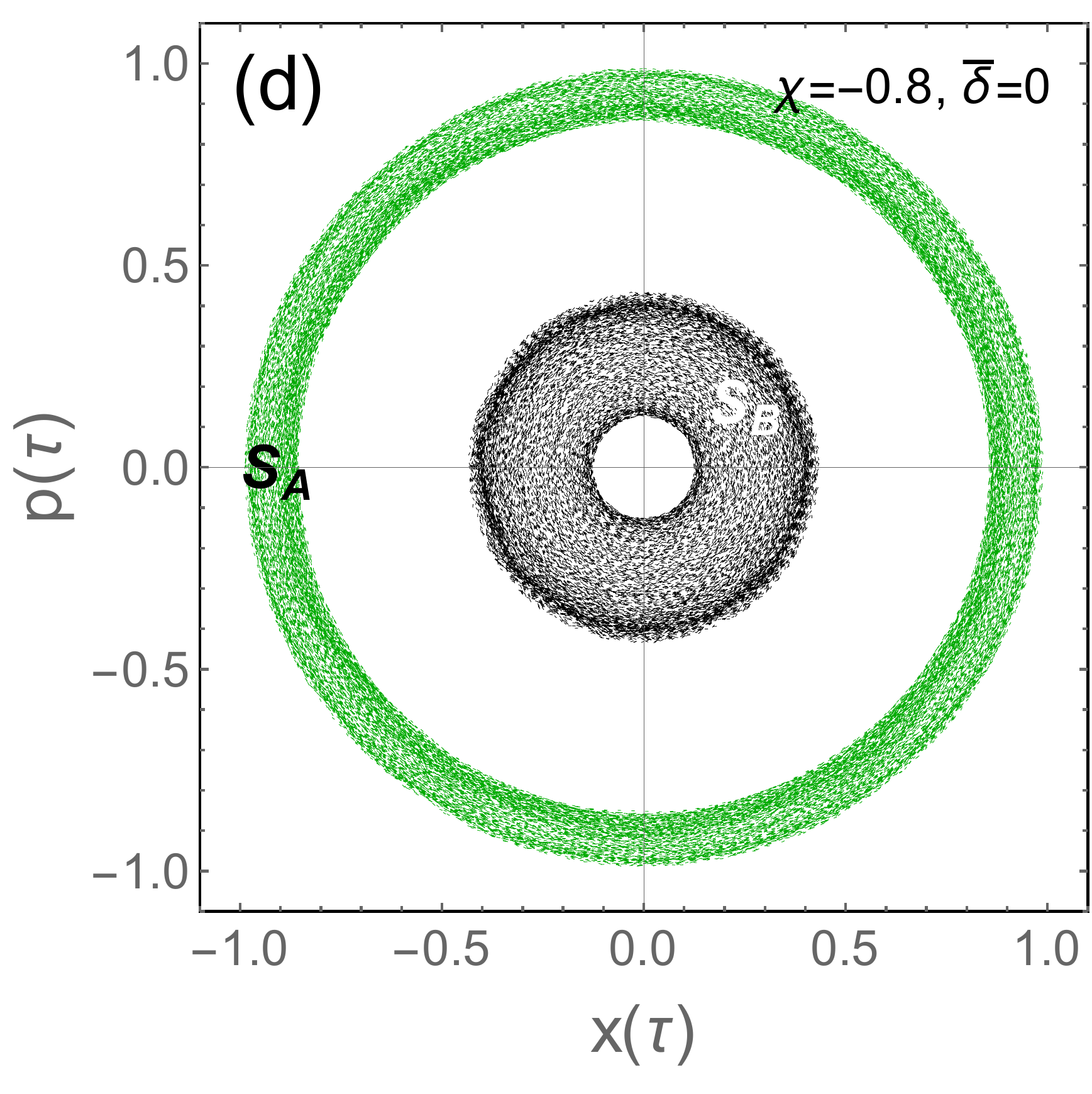}%
\includegraphics[width=122pt]{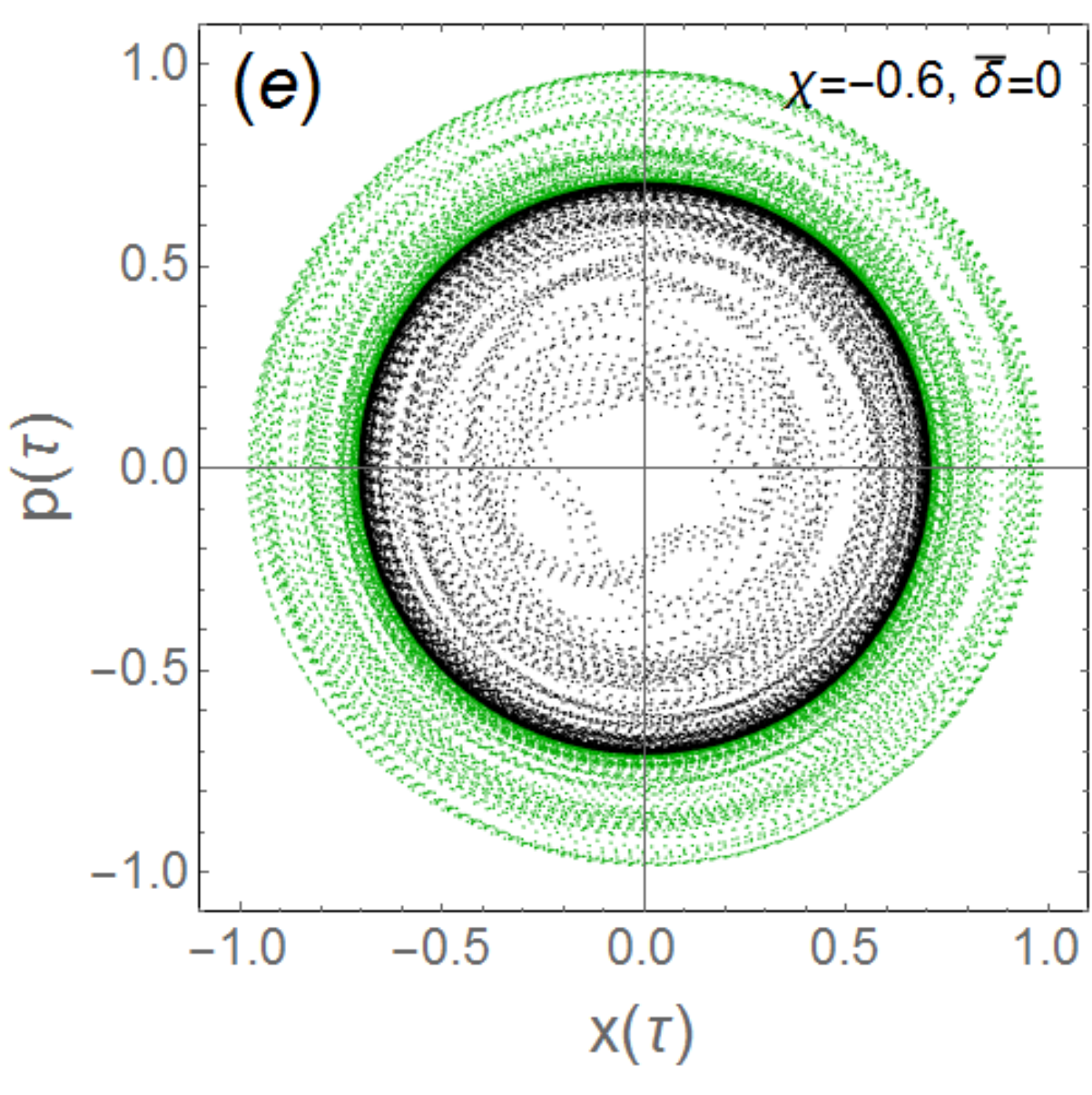}%
\includegraphics[width=122pt]{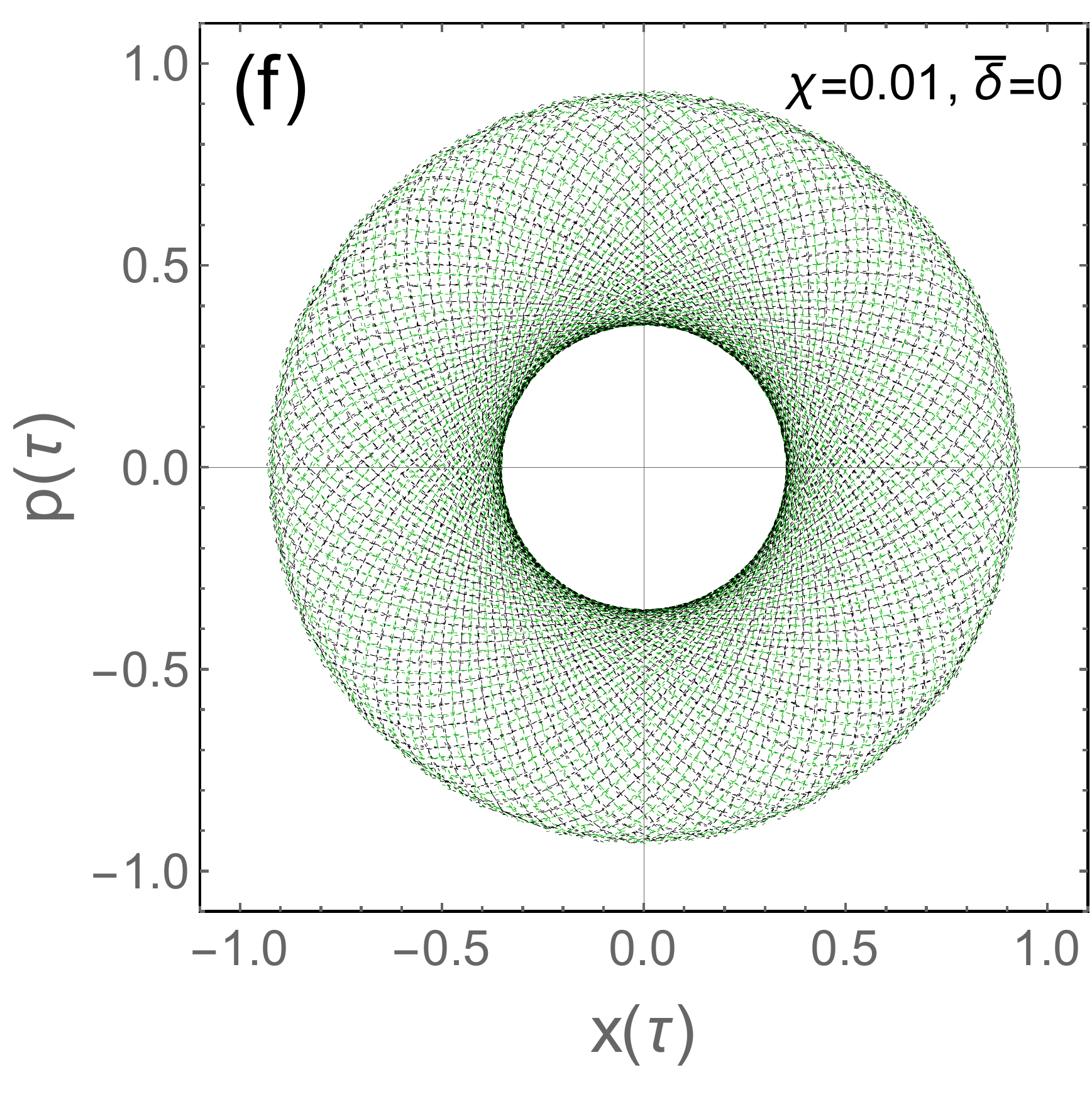}%
\includegraphics[width=122pt]{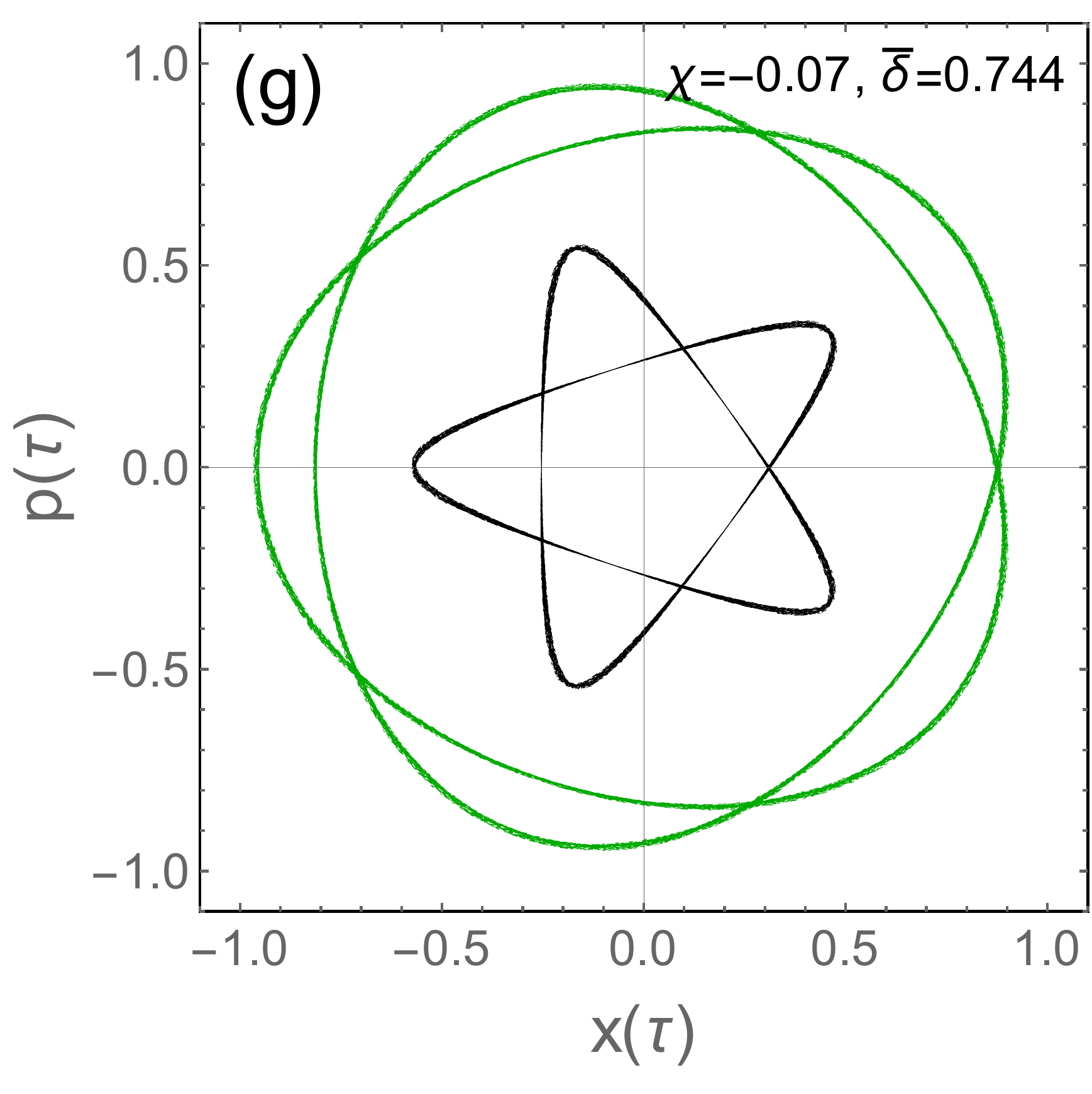}
\end{center}
\caption{(a) The temporal dynamics of $r_A(\protect\tau)$ and $r_B(\protect\tau%
)$ and the corresponding trajectories in the phase space (inset) with the
parameters of $\bar{\delta}=-0.2, \protect\chi=-0.8$;
(b) The covering areas of modes $A$ (black dots) and $B$ (green circles) in phase space
versus the coupling rate of $\protect\chi$ under mode detuning of $\protect%
\bar{\delta}=-0.4$. Inset: A sample phase picture of the trajectories in a limited time interval
with $\chi=-0.3$; (c) The trajectory distance between modes $A$ and $B$ modified by the coupling
rate of $\protect\chi$ under mode detuning of $\bar{\delta}=-1$;
(d)-(f) The long-time orbital overlap in the phase
space modified by the nonlinear couplings of (d) $\chi=-0.8$, (e) $%
\chi=-0.6$ and (f) $\chi=0.01$ with $\bar{\delta}=0$; (g)
An accidental periodic orbit of two modes with very specific
parameters of $\chi=-0.07, \protect\bar{\delta}=0.744$. All the parameters are
scaled by $\bar{g}$ and the initial conditions
are the same as in Fig.\ref{figure1}.}%
\label{figure3}
\end{figure}

The classical dynamics of the two-coupled modes
can be investigated by merely putting them into a
coherent state of $|\alpha_{A,B}\rangle$ and just replacing $\hat{a}_{A,B}\rightarrow \alpha
_{A,B}$ with $\hat{N}\rightarrow N=\left\vert \alpha _{A}\right\vert
^{2}+\left\vert \alpha _{B}\right\vert ^{2}$ in Eq.(\ref{a}) and Eq.(\ref{b}%
), and then we have%
\begin{eqnarray}
i\dot{\alpha}(\tau) &=&\bar{\delta} \alpha+\bar{g}\beta+2\chi \left( 1+\left\vert
\beta\right\vert ^{2}\right) \alpha+\chi \left( 1+\left\vert
\alpha\right\vert^{2}\right) \beta+\chi \beta^{\ast }\alpha^{2},  \label{dya}
\\
i\dot{\beta}(\tau) &=&-\bar{\delta} \beta+\bar{g}\alpha+2\chi \left( 1+\left\vert
\alpha\right\vert ^{2}\right) \beta+\chi \left( 1+\left\vert
\beta\right\vert^{2}\right) \alpha+\chi \alpha^{\ast }\beta^{2},  \label{dyb}
\end{eqnarray}%
where the normalized variables $\alpha=\alpha _{A}/\sqrt{N}, \beta=\alpha _{B}/\sqrt{N}$
for $|\alpha|^2+|\beta|^2=1$ and the parameters $\tau =Nt, \bar{\delta}=\delta /N, \bar{g}=g/N$
are rescaled by the total atomic number of $N$. For a classical dynamics, the
MS between two scattering modes beyond dissipation can be found in Fig.\ref{figure1} when the coupling
rate $\chi$ enters into a certain parametric region.
The right-hand column of Fig.\ref{figure1} shows the quadrature dynamics of
$\alpha(t)=(x_A +i p_A)/\sqrt{2}$, $\beta(t)=(x_B +i p_B)/\sqrt{2}$ and the
corresponding quadrature errors of
\begin{equation}
x_{-}=\frac{1}{\sqrt{2}}\left(x_A-x_B\right),\ p_{-} =\frac{1}{\sqrt{2}}%
\left(p_A-p_B\right),
\end{equation}
are shown in the left column in phase space. Fig.\ref{figure1} illustrates that the
quadrature errors between modes $A$ and $B$ generally conduct a quasi-periodic motion and never
can reach to a zero-measure orbit (a periodic circle or a constant point) in the phase space
when the two modes start from different initial
states, which indicates only a partial dynamical synchronization between two coupled modes
without damping. As Eq.(\ref{dya}) and Eq.(\ref{dyb}) are invariant for exchange
of $A$ and $B$, we can see a complete classical synchronization (when $x_{-}\rightarrow 0$
and $p_{-}\rightarrow 0$) only happens for $\bar{\delta}=0$ starting from exactly the
same initial states of $\alpha(0)$ and $\beta(0)$ (symmetric case). The dynamics of the
two quadratures of modes $A$ and $B$ perform a partial synchronization from local anti-phase
motion to in-phase motion when the nonlinear coupling rate $\chi$ changes from $-0.2\bar{g}$
to $0.2\bar{g}$ as shown in the right column of Fig.\ref{figure1}. For some specific parameters
and initial states, the error dynamics will follow strict periodic orbits in the phase
space with a fixed relative phase between two mechanical modes as shown in
the middle panel of Fig.\ref{figure1} or in Fig.\ref{figure3}(g). We notice
that the local anti-phase motion of two modes covers a larger ring area of
error orbit than that for a in-phase motion in the phase space, which indicates
a weaker energy exchanging between two modes of synchronized motions than that of
the anti-synchronized motions. As the conservation of phase volume, the separated trajectories
of two modes can not run into a same orbit due to the energy exchange between them through
the mode coupling. Therefore, when two modes synchronized, two interwinding trajectories
are formed and they will cover the same area in the phase space since the energy exchange meets the detailed equilibrium.%

The impossibility of a complete synchronization in a conservative Hamiltonian
system was first investigated by Hampton and Zanette \cite{Zanette} and they found
the interesting MS of two coupled Hamiltonian systems. Since then MS in many
non-dissipative systems have been extensively studied by many authors \cite%
{Wangwy,Tianjing,Vincent}. As the conservation of an overall phase volume,
MS exhibits a typical trajectory overlap of two coupled oscillators in the
phase space.
%%Fig1%%
By using the polar form of a complex variables as
\begin{equation}
\alpha (\tau )=r_{A}(\tau )e^{i\theta _{A}(\tau )},\,\beta (\tau
)=r_{B}(\tau )e^{i\theta _{B}(\tau )},
\label{polar}
\end{equation}%
we can easily check the trajectory overlap of two scattering modes by modifying
the coupling rate $\chi $ as well as the mode detuning $\bar{\delta} $ as shown in
Fig.\ref{figure3}. A typical dynamics of $r_{A}(\tau )$, $r_{B}(\tau )$ and their
longtime trajectories (inset) are given in Fig.\ref{figure3}(a). The covering areas
of the quasi-periodic trajectories of modes $A$ and $B$ in the phase space can be
estimated by%
\begin{equation*}
S_{A,B} =\pi \lim_{T\rightarrow \infty }\left( \max \left[
r_{A,B}^{2}(\tau )\right] _{T}-\min \left[ r_{A,B}^{2}(\tau )\right] _{T}\right),
\end{equation*}%
where $\max \left[ r(\tau) \right] _{T}$ ($\min \left[ %
r(\tau) \right] _{T}$) means the maximum (minimum) value
of $r\left( \tau \right) $ during a time interval of $T$.
As shown in Fig.\ref{figure3}(b), the two modes keep nearly a same phase area ($S_{A}\approx S_{B}$ with
differences below $10^{-3}$) in a large parametric region of $\chi $, displaying an invariant
measure of two modes. We can also define a trajectory distance between two modes by calculating
$d_{AB}=\lim_{T\rightarrow \infty }\left( \min\left[ r_{A}(\tau )\right] _{T}-\max \left[ r_{B}%
(\tau )\right] _{T}\right)$ to see the trajectory overlap in the phase space as shown in Fig.\ref{figure3}%
(c). If $d_{AB}\geq 0$, the ring radius of trajectory $A$ is larger than that of mode $B$ and there
is no overlap between them. When $d_{AB}<0$, the overlap begins and it reaches its maximum value at
a critical point C shown in Fig.\ref{figure3}(c). Fig.\ref{figure3}(c) reveals that the overlap of the
quasi-periodic trajectories in the phase space happens only in a very specific parametric window of $\chi$
and $\bar{\delta} $, indicating the resonant exchange of energy between two scattering modes.
Fig.\ref{figure3}(d)-(f) demonstrate the trajectory overlap with the increment of coupling
rate $\chi$ between two symmetric modes of $\bar{\delta}=0$. We can find a clear evidence of orbital
attraction in MS as shown in Fig.\ref{figure3}(c) for that the orbital merging is more quickly
than orbital separation with respect to the coupling rate of $\chi$ (the slope of $d_{AB}$ dramatically
decreases at the point of C in Fig.\ref{figure3}(c)). For some specific initial states and parameters,
the trajectories of modes $A$ and $B$ will run into the interesting synchronized periodic orbits (the correlated
Lissajous curves with zero measures) such as shown by Fig.\ref{figure3}(g). We can numerically verify that there are many
correlated Lissajous orbits in the phase space determined by the initial conditions and the
parameters of $\chi $ and $\bar{\delta} $. Therefore, according to the classical dynamics of Eq.(\ref{dya})
and Eq.(\ref{dyb}), the two nonlinear scattering modes can obtain a partial synchronized dynamics in certain
parametric windows in a sense of covering the same phase space of their trajectories with invariant
measures or conducting correlated Lissajous orbits with zero measures.
\begin{figure}[h]
\begin{center}
\includegraphics[width=165pt]{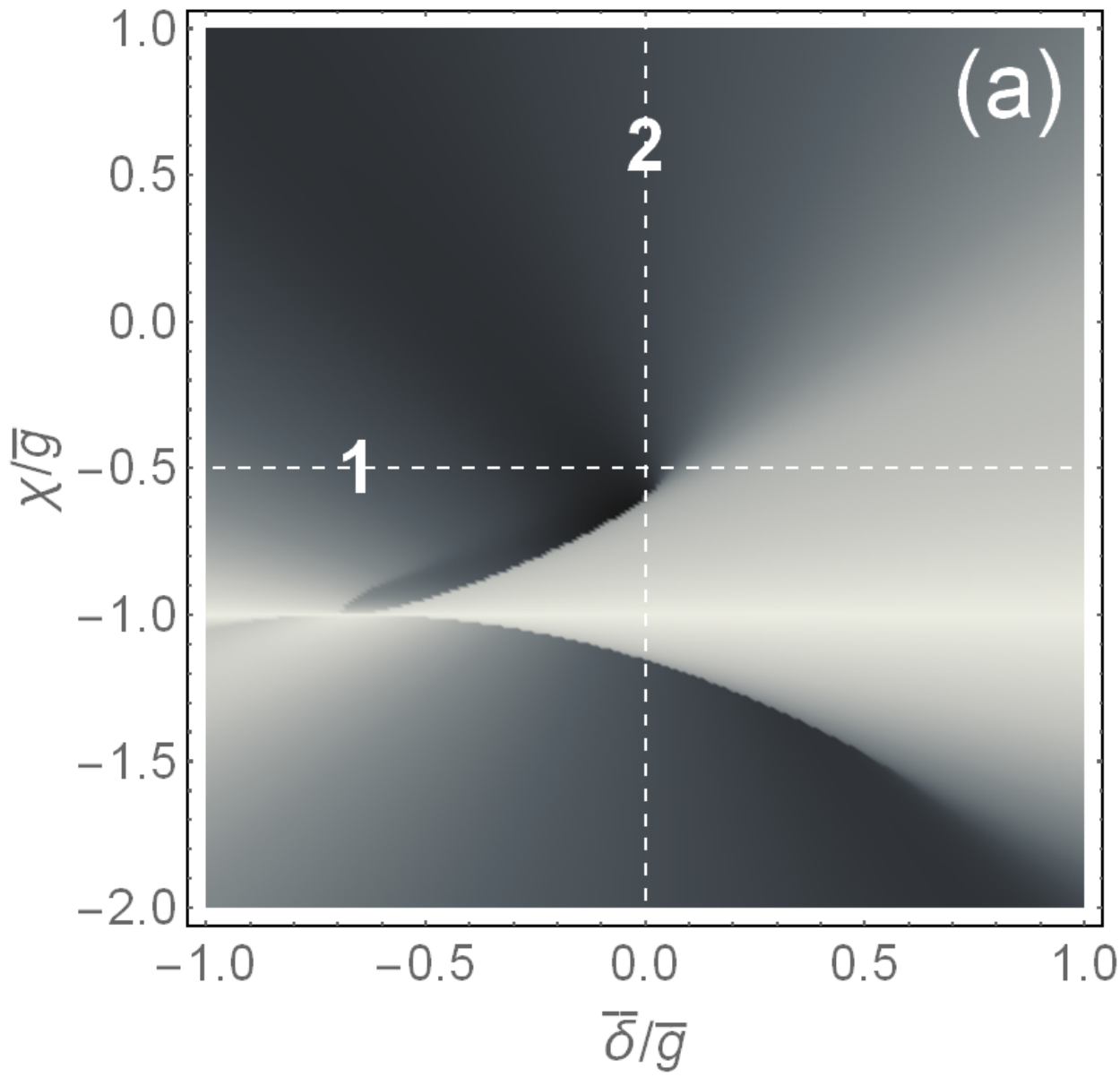}%
\includegraphics[width=160pt]{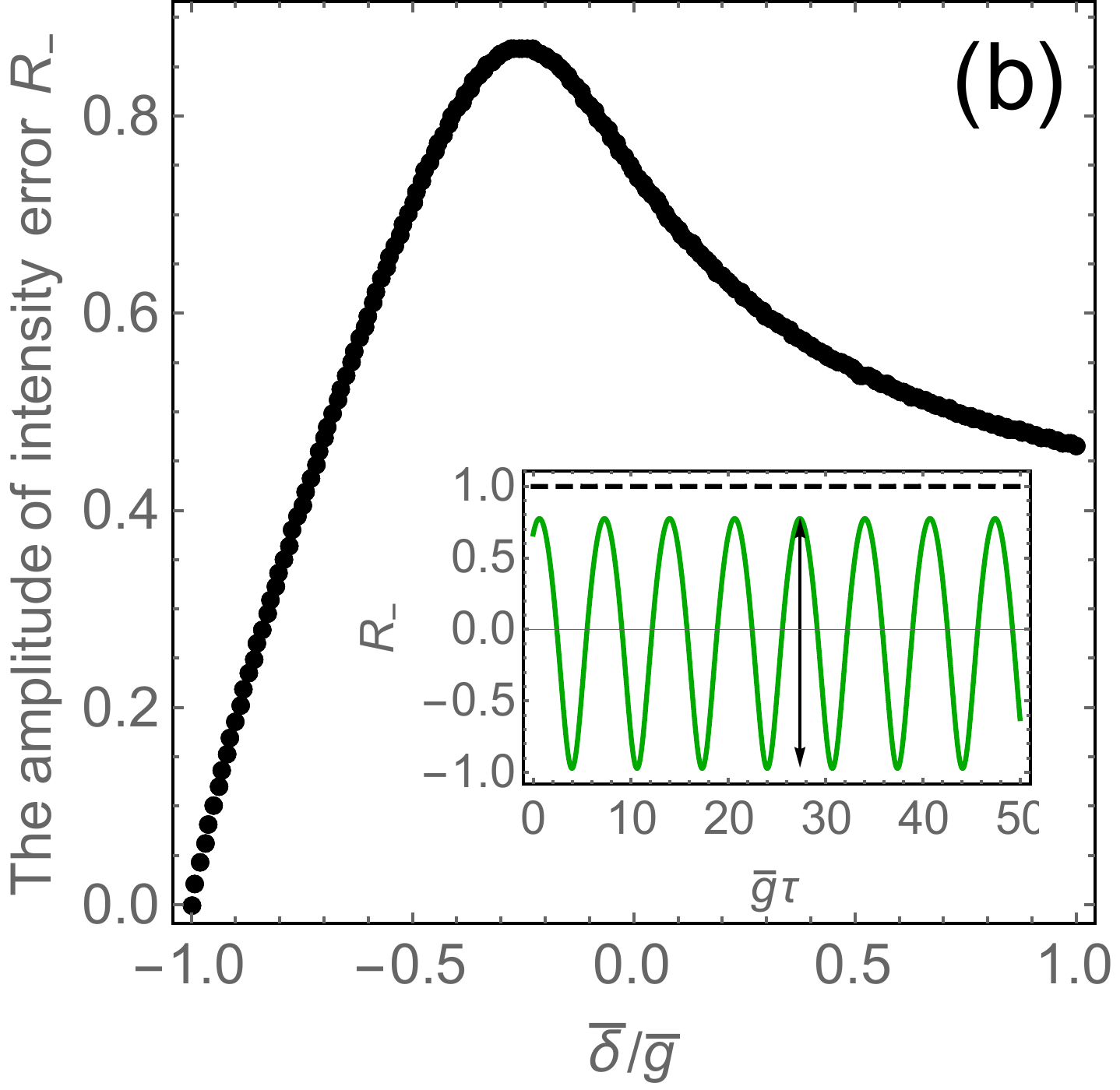}%
\includegraphics[width=160pt]{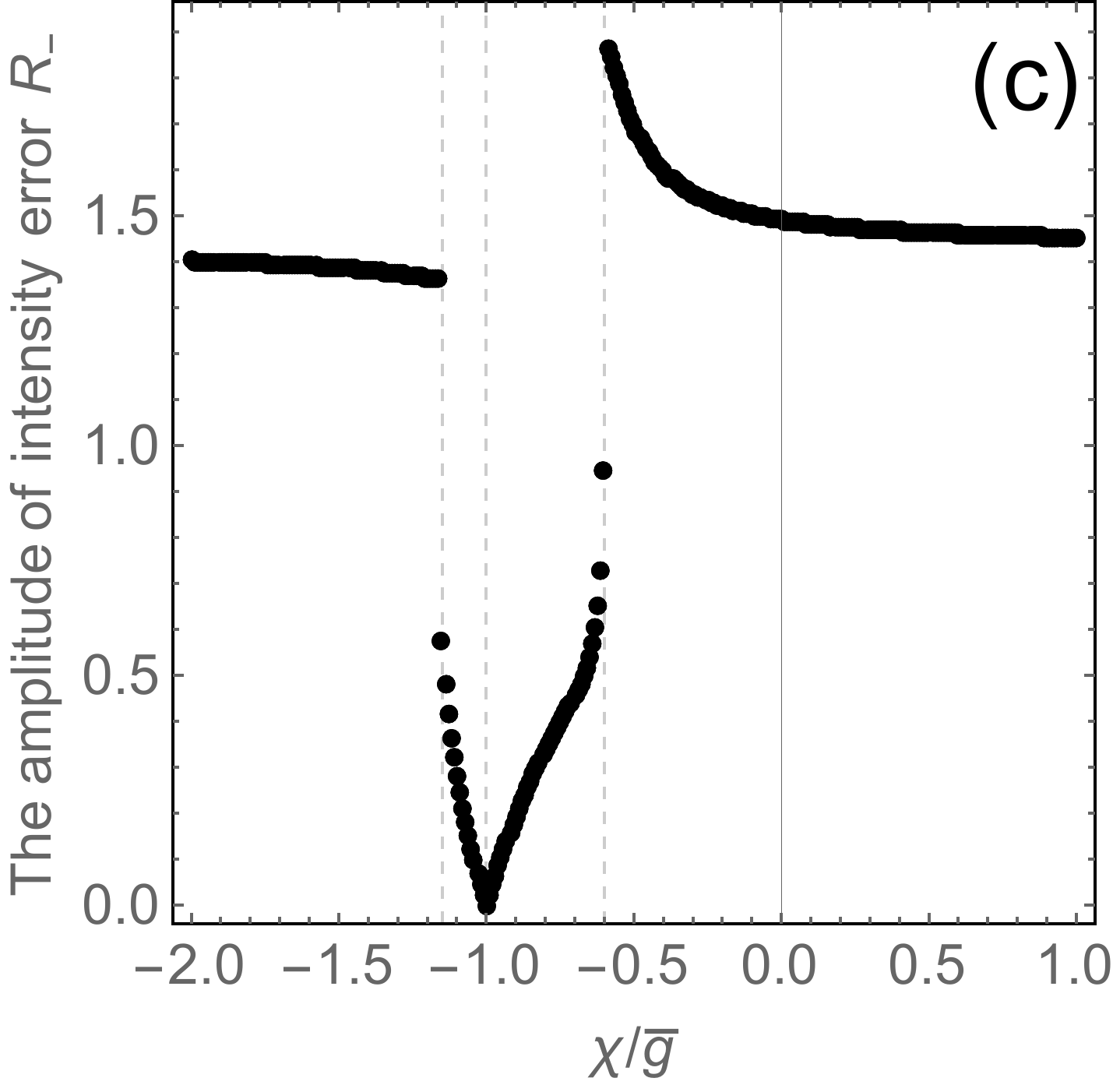}\\%
\includegraphics[width=170pt]{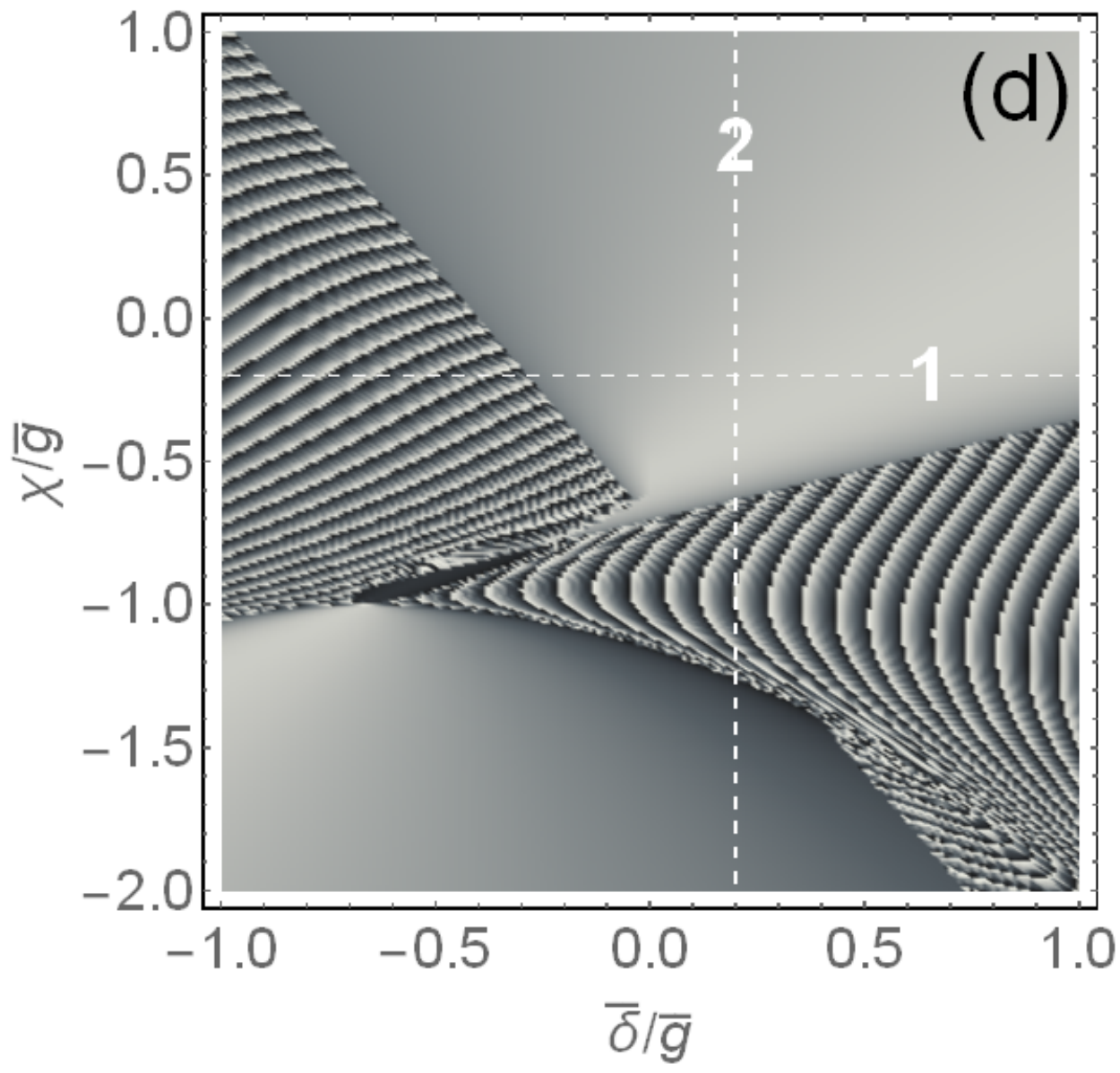} %
\includegraphics[width=160pt]{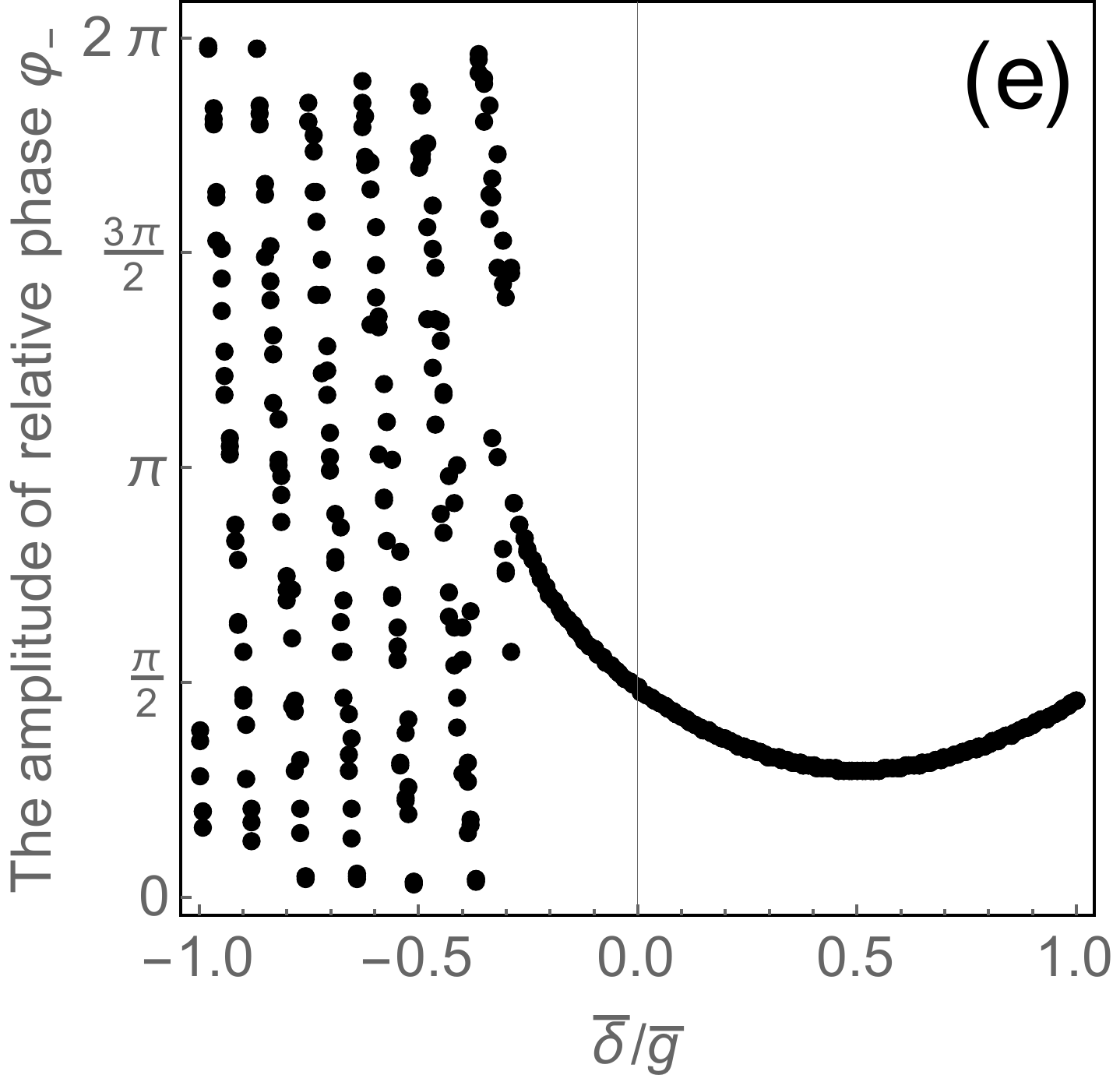} %
\includegraphics[width=160pt]{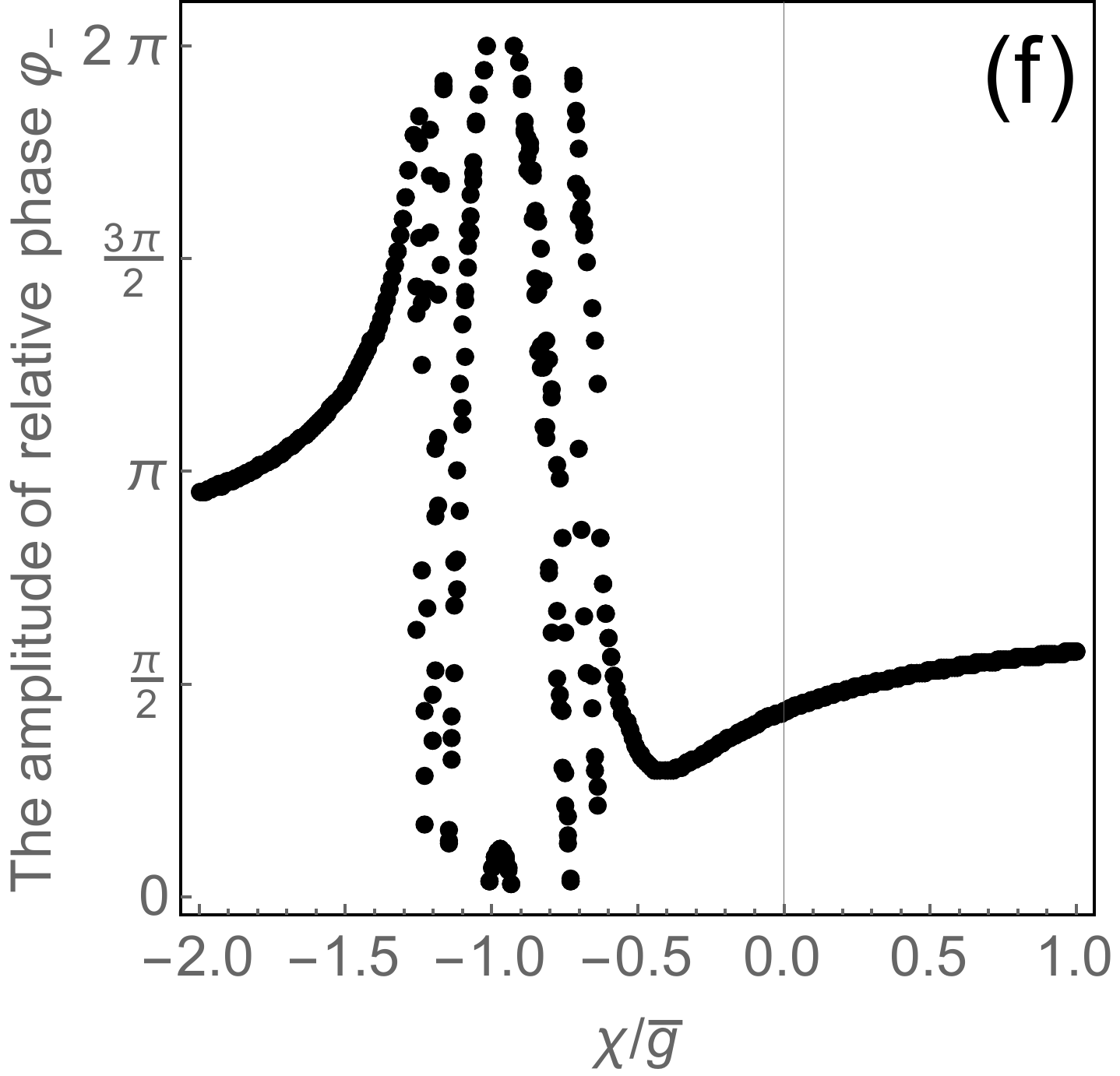}%
\end{center}
\caption{The amplitudes of $R_{-}$ and $\varphi_{-}$ of the two mechanical modes
modified by $\bar{\delta}$ and $\chi$. (a) The density plot of the amplitude of
$R_{-}(\tau)$ in the plane of $\bar{\delta}$ and $\chi$; The amplitude of
$R_{-}(\tau)$ changes with (b) $\bar{\delta}$ or with (c) $\chi$ under the specific
coupling $\chi=-0.5$ or detuning $\bar{\delta}=0$ indicated by the dashed line 1 and
2 in (a); (d) The density plot of amplitude of phase difference $\varphi_{-}$
in the plane of $\bar{\delta}$ and $\chi$; (e)(f) The amplitude of $\varphi_{-}$ versus
$\bar{\delta}$ or $\chi$ for fixed $\chi =-0.2$ or $\bar{\delta}=0.2$. All the frequencies
are scalded by $\bar{g}$ and the initial conditions are set to be the same as those in the
previous figures.}
\label{figure2}
\end{figure}

As for a non-dissipative Hamiltonian system, the trajectory should be very sensitive to
the initial conditions because the total phase volume of two coupled systems is determined
by the initial states. The synchronized behavior of MS can exhibit a different characteristic
if we consider the dynamics via intensity imbalance and relative phase between two
modes starting from different initial states. Then the trajectory overlap means that two
scattering modes can possess a correlated intensity relation or lock to a constant relative
phase as implied by Fig.\ref{figure1} and Fig.\ref{figure3}. The instaneous intensity imbalance
and phase difference between two mechanical modes can be defined by
\begin{equation}
R_{-}=|\alpha|^2-|\beta|^2=r^{2}_{A}-r^{2}_{B}, \quad \varphi_{-}%
=\arg(\alpha)-\arg(\beta)=\theta_A-\theta_B.
\end{equation}
As a general quasi-periodic behavior of $R_{-}(\tau)$ (see the inset of
Fig.\ref{figure2}(b)), the oscillating amplitude of $R_{-}(\bar{\delta},\chi,\tau)$
changing with the detuning $\bar{\delta}$ and $\chi$ is calculated in
Fig.\ref{figure2}(a). Fig.\ref{figure2}(b) and Fig.\ref{figure2}(c) are
the oscillating amplitudes of $R_{-}$ with respect to
$\bar{\delta}$ and $\chi$ for the specific $\chi=-0.5$ and $\bar{\delta}=0$, respectively,
indicated by the dashed lines of 1 and 2
in Fig.\ref{figure2}(a). Fig.\ref{figure2}(a) reveals that the
amplitude of the intensity imbalance can reach a
constant value (AC behavior) and it will disappear for
$\chi=-\bar{g}$ or at a large value of $\bar{\delta}$ (DC behavior).
Similarly, the amplitudes of the relative phase $\varphi_{-}(\tau)$
are also shown in the lower frames of Fig.\ref{figure2}.
Fig.\ref{figure2}(d) presents the density plot of oscillating amplitude of $\varphi_{-}(\tau)$ in the
parametric plane of $\bar{\delta}$ and $\chi$. Fig.\ref{figure2}(e) and (f) display
the amplitudes of $\varphi_{-}(\tau)$ changing with respect to parameters
$\bar{\delta}$ and $\chi$ indicated by two dashed lines of 1 and 2 in
Fig.\ref{figure2}(d), respectively. We can see that, in some parametric windows of $\bar{\delta}$ and $\chi$,
the oscillating amplitude of $\varphi_{-}(\tau)$ is randomly dependent on the parameters. Fig.\ref{figure2}(d)-(f) demonstrate
that only in specific parametric windows of $\bar{\delta}$ and $\chi$ can the phase amplitude
obtain a stable value for a partial phase synchronization.
\begin{figure}[b]
\begin{center}
\includegraphics[width=160pt]{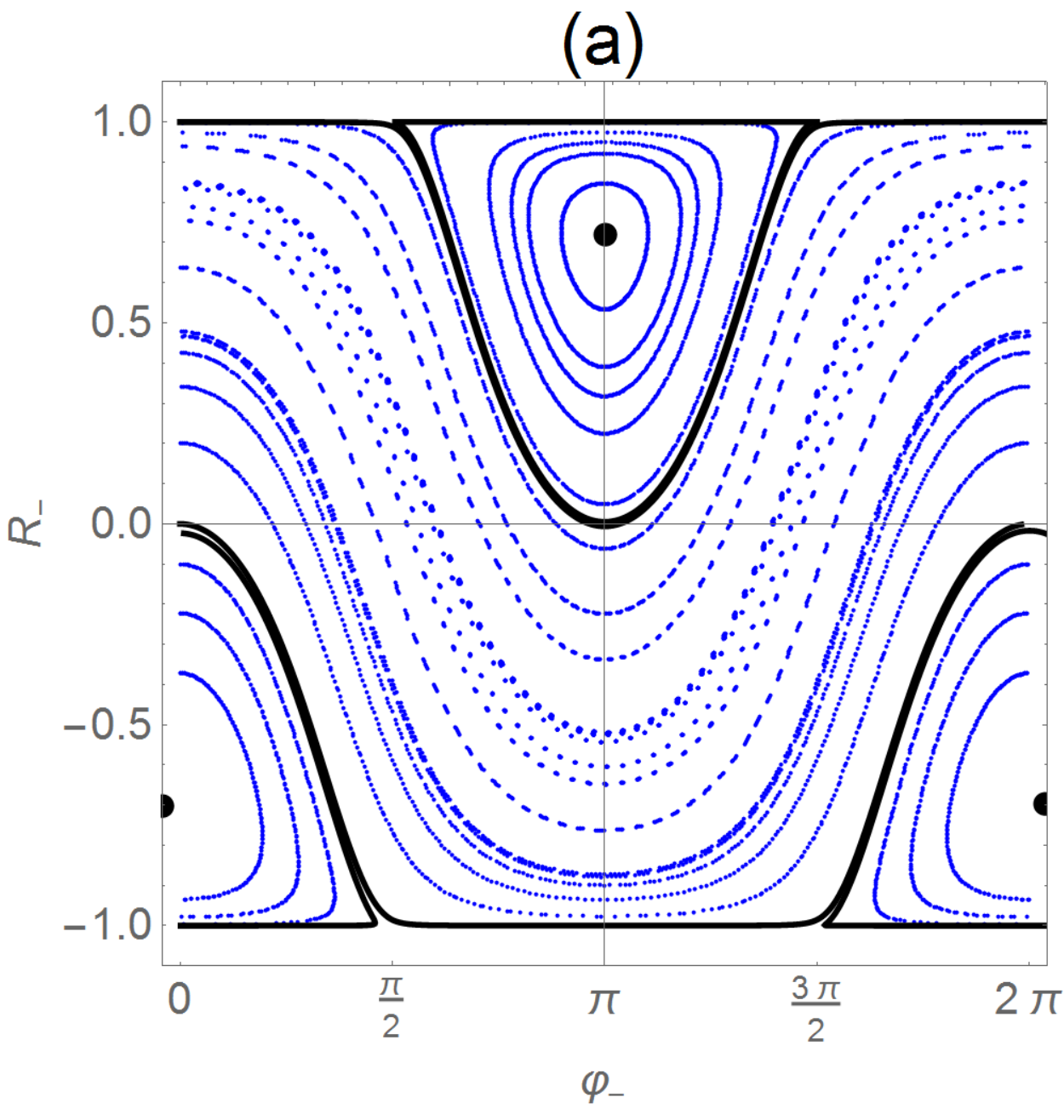} %
\includegraphics[width=160pt]{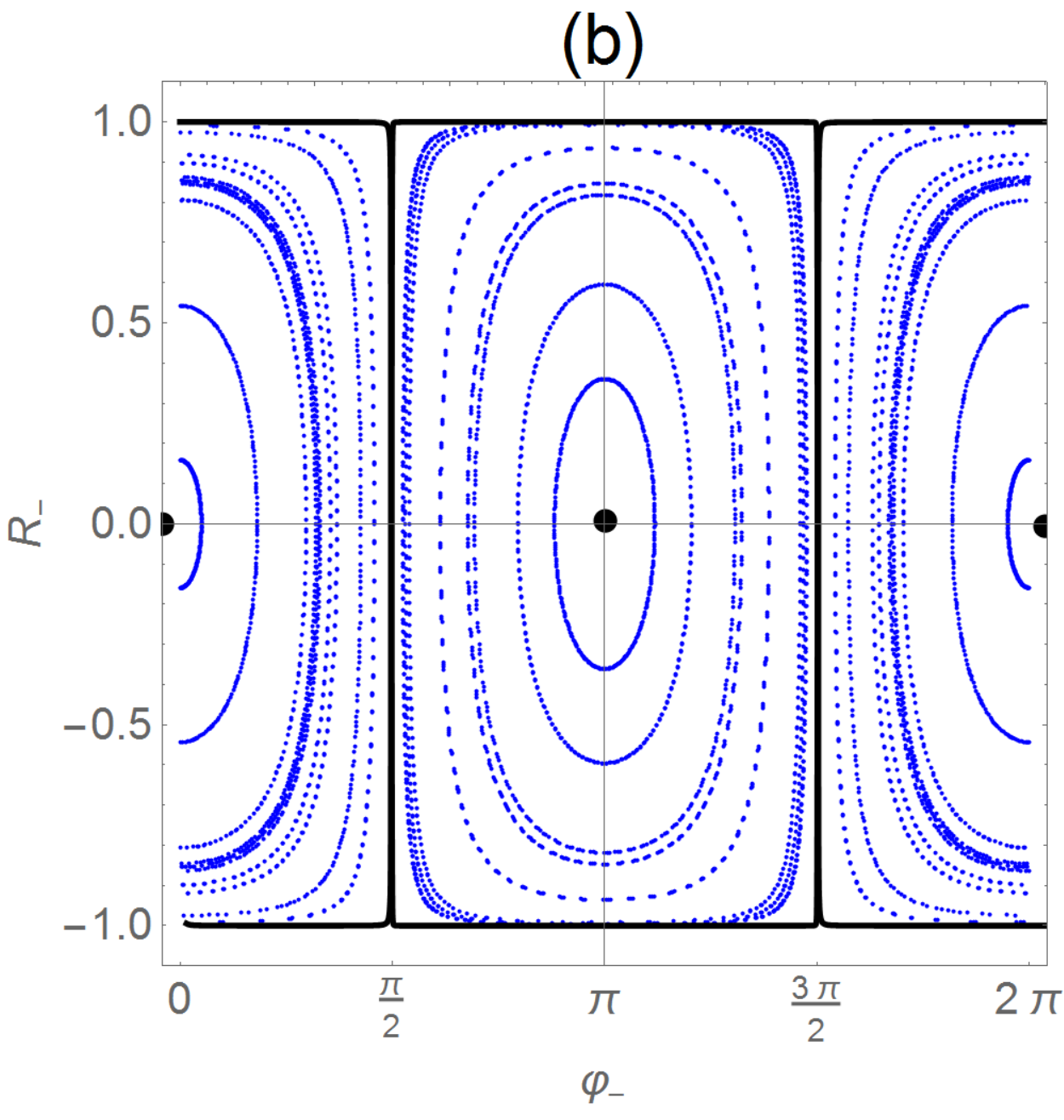} %
\includegraphics[width=160pt]{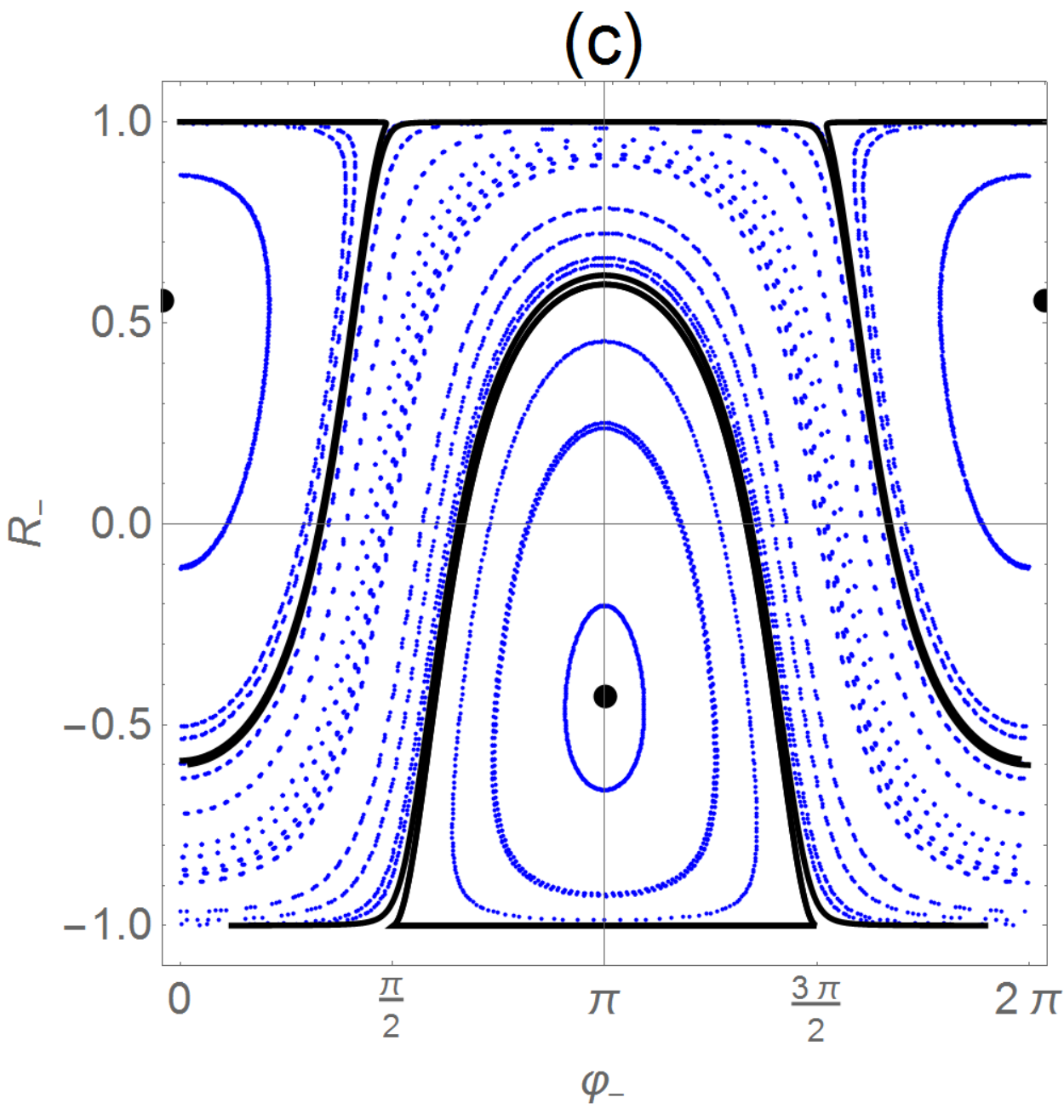}%
\end{center}
\caption{The evolutions of the intensity imbalance $R_{-}$
and phase difference $\varphi_{-}$ in the phase space with
(a) $\bar{\delta}=-1$, (b) $\bar{\delta}=0$ and (c) $\bar{\delta}=0.5$,
starting from the random initial conditions.
The mode coupling rate is $\chi =-0.01$ and all the frequencies are scalded by
$\bar{g}$. The thick black lines are the separatrices of the orbits and the black dots
are the stable steady states.}
\label{phase}
\end{figure}

Alternatively, the intensity imbalance
or the phase error being an indicator of the
MS \cite{Zanette} can be revealed by the equations of
\begin{eqnarray}
\dot{r} &=&-\left( \bar{g}+\chi \right) \left( r^{2}+1\right) \sin
\varphi _{-}, \label{eqr}\\
\dot{\varphi}_{-} &=&-2\bar{\delta} +2\chi \frac{r^{2}-1}{r^{2}+1}+\left(
\bar{g}+\chi \right) \left( r-\frac{1}{r}\right) \cos \varphi _{-},
\label{eph}
\end{eqnarray}%
where we introduce the amplitude ratio of $r=r_{A}/r_{B}$ as a convenient variable to
describe the amplitude balance between two mechanical modes with a number conservation of $r_{A}^2+r_{B}^2=1$.
We can easily verify that $R_{-}=(r^2-1)/(r^2+1)$ and a fixed amplitude synchronization for
$r_{A}(\tau)\propto r_{B}(\tau)$ only happens when $\chi=-\bar{g}$ or $\varphi=n\pi, n\in\mathbb{Z}$.
In the case of $\chi=-\bar{g}$, the solution for the relative phase will be
$\varphi_{-}(t)=\varphi_0+2(\chi R_{0}-\bar{\delta})t$, $R_{-}(0)\equiv R_{0}=(r_{0}^2-1)/(r_{0}^2+1)$,
where the initial vales $r_0=r(0)$ and $\varphi_0=\varphi_{-}(0)$. Specifically, $R_0=\bar{\delta}/\chi$
is a steady state of Eq.(\ref{eqr}) and (\ref{eph}) under $\chi=-\bar{g}$ for $\varphi_{-}=\varphi_0$.
For $\varphi=n\pi$, a quartic equation of $r$ should be solved to get the steady states in this case
(the complete synchronization is related to the stable steady states corresponding the fixed points shown in Fig.\ref{phase}).
Fig.\ref{phase} displays the orbits in the phase space of $(\varphi_{-},R_{-})$
with random initial $r_0$ and $\varphi_0$ when the frequency detuning $\bar{\delta}$ changes from negative to positive.
The orbital flow in the phase space indicates that the amplitude synchronization or the phase
synchronization only happens under specific initial conditions, such as $R_{-}(0)=\pm 1$
or $\varphi_{-}(0)=\pi/2, 3\pi/2$ as shown in Fig.\ref{phase} by the separatrices (thick black lines).
The phase transitions from self-trapping case on mode $A$ in Fig.\ref{phase}(a) to the tunneling case in Fig.\ref{phase}(b)
and to the self-trapping case on mode $B$ in Fig.\ref{phase}(c) are all connected by the asymptotic synchronization
lines passing through the same dynamical frustrated points of $\varphi_{-}=\pi/2$ and $\varphi_{-}=3\pi/2$ ($R_{-}=\pm 1$).
The trajectory flow shown in Fig.\ref{phase} reveals that only a partial phase synchronization can easily
be obtained for $\bar{\delta}=0$ (Fig.\ref{phase}(b)) and the exact synchronized motion between two conservative
modes is indeed an accidental case corresponding only to the stable steady states which can be shared by both modes.

As described in Ref.\cite{Zanette}, a complete CS that two different orbits of the system collapse
into the same one is impossible in the coupled Hamiltonian systems because the phase volume must
be preserved (no damping). However, a weak or partial synchronization can be established between
two coupled conservative modes with a signature of two attractive interwinding orbits formed in phase space
due to mutual exchange of energy. According to the above calculations on mean-value dynamics,
we see that the partial two-mode synchronized motion can safely be acquired through the mode coupling
under specific parametric conditions in a MS point of view. If the initial conditions
are specially chosen, a full synchronization to reach the stable steady states can still be realized
in this system classically. However, in order to consider the synchronization from a quantum
aspect, we should give up the trajectory picture to investigate the probability density
distribution in the phase space instead. Recently, a work studied the MS
in a quantum regime to reveal some quantum properties of two coupled modes in
BEC system \cite{Qiu}. As the fluctuations are always involved in the quantum dynamics,
the CS measure by just investigating the overlap of the mean-value
trajectories in the phase space is obviously too rough for a quantum dynamics.

\section{The coherent dynamics of quantum synchronization}
\subsection{QS of scattering modes from BEC in a number state}

\begin{figure}[htp]
  \centering
  \includegraphics[width=125pt]{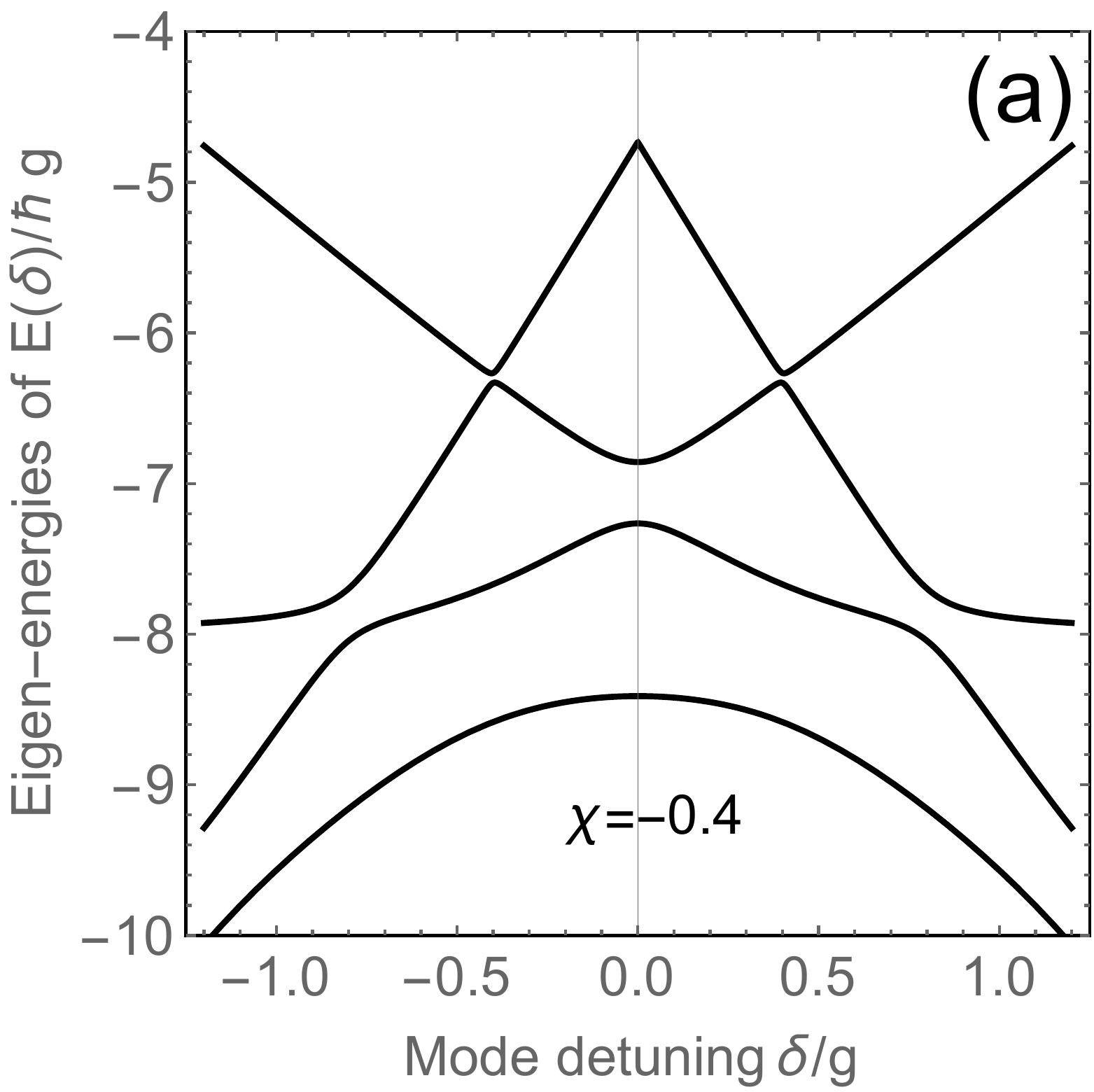}%
  \includegraphics[width=125pt]{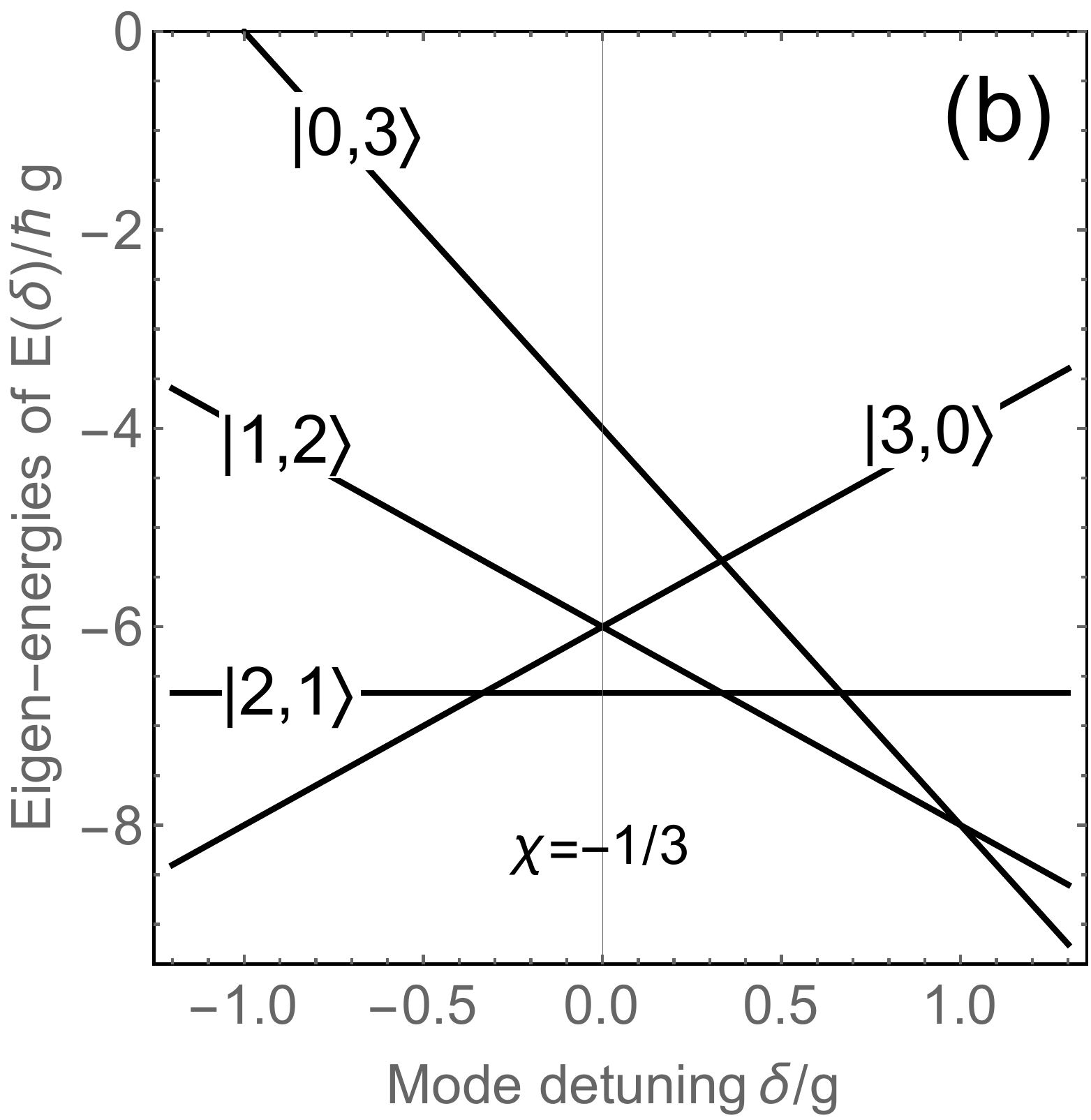}%
  \includegraphics[width=125pt]{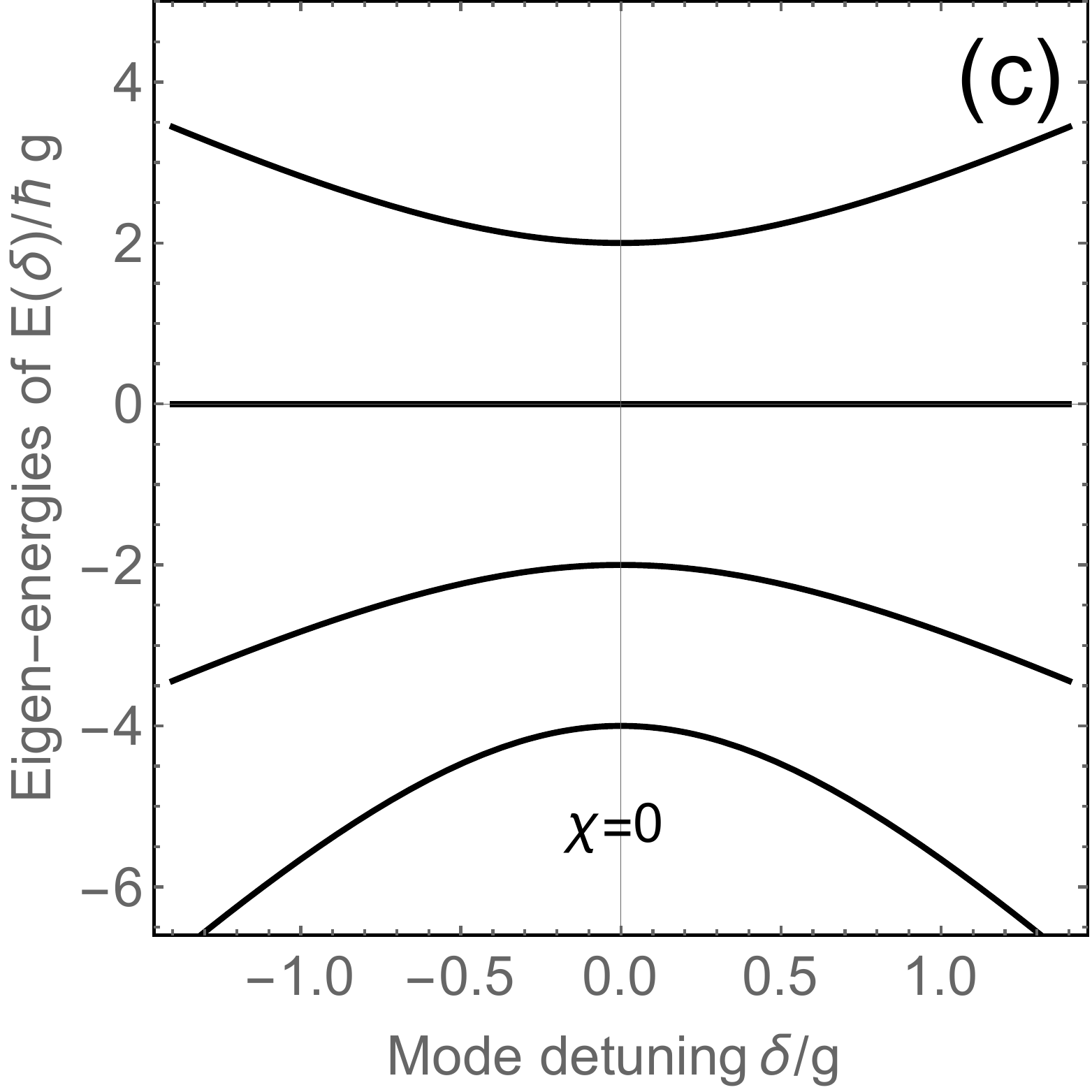}%
  \includegraphics[width=125pt]{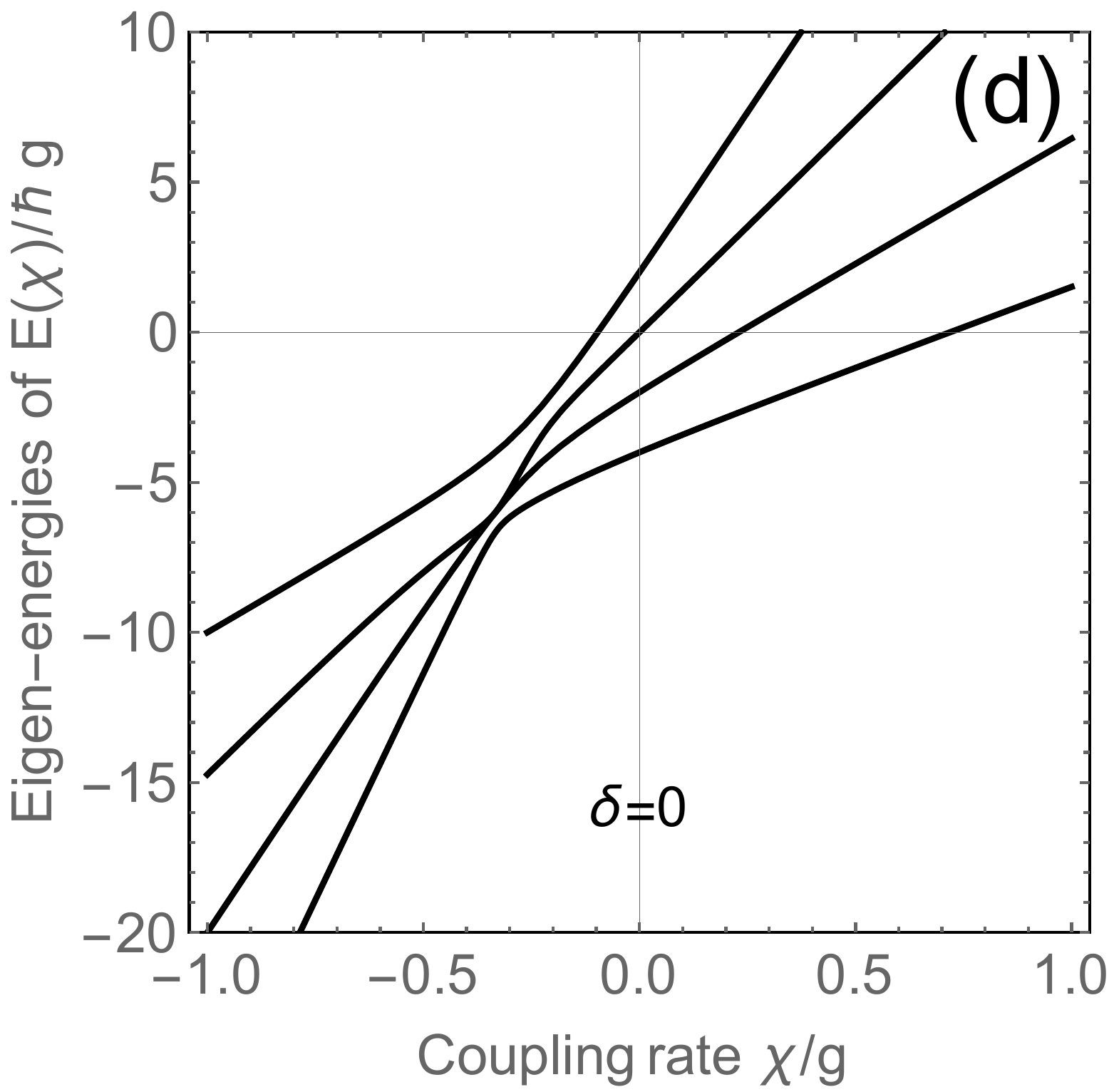}%
  \caption{The energy levels of the BEC modes of $3$ atoms changing with different nonlinear
  couplings of (a) $\chi=-0.4$, (b) $\chi=-1/3$, (c) $\chi=0$, and with (d) $\delta=0$.}\label{level}
\end{figure}

Because of the intrinsic fluctuations in a quantum system, the QS
is essentially different from the classical one and a
complete synchronization of two coupled quantum modes is definitely
avoided due to the uncertainty principle. In order to show the dynamics of
QS in a closed quantum system, the total state
of two excited modes in BEC should follow the Schr\"{o}dinger
equation of%
\begin{equation}
i\hbar \frac{d}{dt}\left\vert \psi \left( t\right) \right\rangle =\hat{H}%
_{N}\left\vert \psi \left( t\right) \right\rangle ,  \label{Hn}
\end{equation}%
where the wave function $|\psi(t)\rangle$ in
the Hilbert space of $\mathcal{H}_{N}$ can be expanded by%
\begin{equation}
\left\vert \psi \left( t\right) \right\rangle
=\sum_{j=0}^{N}C_{j}(t)\left\vert j,N-j\right\rangle,  \label{psi}
\end{equation}%
whose form is due to the conservation of $N$ ($N$ denotes the total atomic number of BEC
in number state $|N\rangle$ and thus the dimension of $\mathcal{H}_{N}$ is $D=N+1$) and
the unitary evolution keeps $\sum_{j=0}^{N}\left\vert C_{j}(t)\right\vert ^{2}=1 $.
Substituting Eq.(\ref{psi}) into Eq.(\ref{Hn}), we have the equation of motion for
the coefficients $C_{j}(t)$ as%
\begin{eqnarray}
i\dot{C}_{j}(t) &=&\sqrt{j\left( N-j+1\right) }\left[ g+\chi \left(
N-1\right) \right] C_{j-1}(t)
+\left[ \chi N^{2}-\left( \chi +\delta \right) N+2j\left( \delta +\chi
N\right) -2\chi j^{2}\right] C_{j} (t)  \notag \\
&&+\sqrt{\left( j+1\right) (N-j)}\left[ g+\chi \left( N-1\right) \right]
C_{j+1}(t).  \label{Cjs}
\end{eqnarray}
Clearly, the above coupled equations produce a symmetric tridiagonal matrix of
Hamiltonian $\hat{H}_N$ in the Fock representation and the algorithm to calculate
its eigenvalues are well-known (e.g., bisection algorithm \cite{bisect}).
Fig.\ref{level} gives a simple case of energy levels for BEC with only $3$ atoms ($D=4$) modified by
the nonlinear couplings of $\chi$. There are four energy levels in this case whose eigenstates
are the superposition of four bare number states of $|0,3\rangle$, $|1,2\rangle$, $|2,1\rangle$
and $|3,0\rangle$.
Clearly, the population distributions of modes $A$ and $B$
are dynamically determined by the avoided level crossings of the states which can easily
be modified by $\delta$ and $\chi$. A clear degenerate case of $g+\chi (N-1)=0$ shown in Fig.\ref{level}(b)
indicates a transitionless evolution between four bare states without energy avoided crossings for
\begin{equation}
C_j (t)=C_j (0)e^{-i\left[ \chi N^{2}-\left( \chi +\delta \right) N+2j\left( \delta +\chi
N\right) -2\chi j^{2}\right]t}.
\label{cj}
\end{equation}
This corresponds to the classical case of $\chi=-\bar{g}\equiv -g/N$ for
a synchronized intensity balance (see Eq.(\ref{eqr}) for $\dot{r}=0$) that only the
relative phase is linearly increasing just like that in Eq.(\ref{cj}).

Actually, an overlap of the quantum dynamics of modes $A$ and $B$ in the phase-space should be investigated
by using the probability distributions of the total quantum sates. The $Q$-function based on the coherent
state representation possesses a good behavior to provide positive and bounded probability distributions
in the phase space. The coherent properties of a quantum state $|\psi
(t)\rangle $ can be revealed by a continuous function which is a projection
of the quantum state onto a reference coherent state $|\alpha\rangle
$ by $\psi(\alpha ,t)=\langle \alpha |\psi (t)\rangle$,
and its probability density is called $Q$-function as
\begin{equation*}
Q(\alpha ,t)=\left\vert \psi(\alpha ,t)\right\vert ^{2},
\end{equation*}
If the $Q$-function goes to 1 in the phase space, then the state $%
|\psi (t)\rangle $ will be completely described by a coherent state $%
|\alpha\rangle $ with an overall amplitude and phase. Therefore,
the quantum MS of two quantum states can then be
traced by the overlap of probability distributions in the phase space revealed by $Q$-function.
The $Q$-function for the two-mode state of BEC in a number state $|N\rangle$ is
\begin{equation}
Q_{N}(\alpha _{A},\alpha _{B},t)=\langle \alpha _{A},\alpha
_{B}|\hat{\rho} _{N}|\alpha _{A},\alpha _{B}\rangle =\left\vert \langle \alpha
_{A},\alpha _{B}|\psi (t)\rangle \right\vert ^{2},
\end{equation}%
where the density operator $\hat{\rho} _{N}=|\psi (t)\rangle\langle\psi (t)|$.
In terms of $C_{j}(t)$, the above $Q$-function can be written as
\begin{equation}
Q_{N}(\alpha _{A},\alpha _{B},t)=\left\vert \sum_{j=0}^{N}C_{j}\left(
t\right) \frac{\left( \alpha _{A}^{\ast }\right) ^{j}\left( \alpha
_{B}^{\ast }\right) ^{N-j}}{\sqrt{j!\left( N-j\right) !}}\right\vert
^{2}e^{-\left( \left\vert \alpha _{A}\right\vert ^{2}+\left\vert \alpha
_{B}\right\vert ^{2}\right) }\geq 0.  \label{Q}
\end{equation}
By using Cauchy's inequality for a certain $N$, the above $Q$-function has an
upper limit function of
\begin{equation}
Q_{N}(\alpha _{A},\alpha _{B},t)\leq \frac{1}{N!}\left( \left\vert \alpha
_{A}\right\vert ^{2}+\left\vert \alpha _{B}\right\vert ^{2}\right)
^{N}e^{-\left( \left\vert \alpha _{A}\right\vert ^{2}+\left\vert \alpha
_{B}\right\vert ^{2}\right) }\equiv \bar{Q}_{N},  \label{Q1}
\end{equation}%
where the normalization condition in the Hilbert space
$\mathcal{H}_N$ is used. Clearly, the above bounded $Q$-function $\bar{Q}_{N}$ gives
a central symmetric distribution for modes $\alpha _{A}$ and $%
\alpha _{B}$ and no phase information is included.

%figure%
\begin{figure}[t]
\begin{center}
\includegraphics[width=120pt]{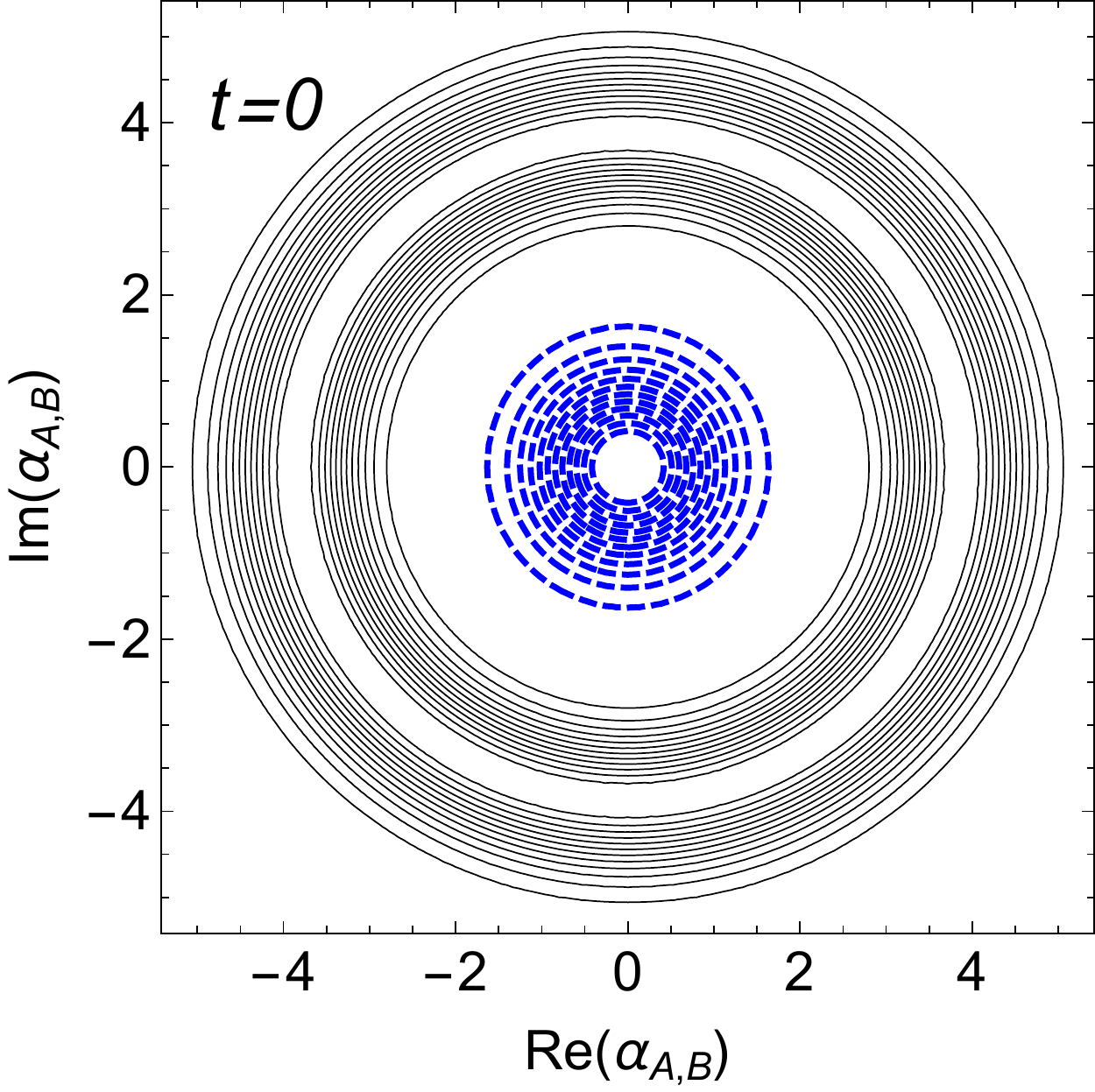}%
\includegraphics[width=120pt]{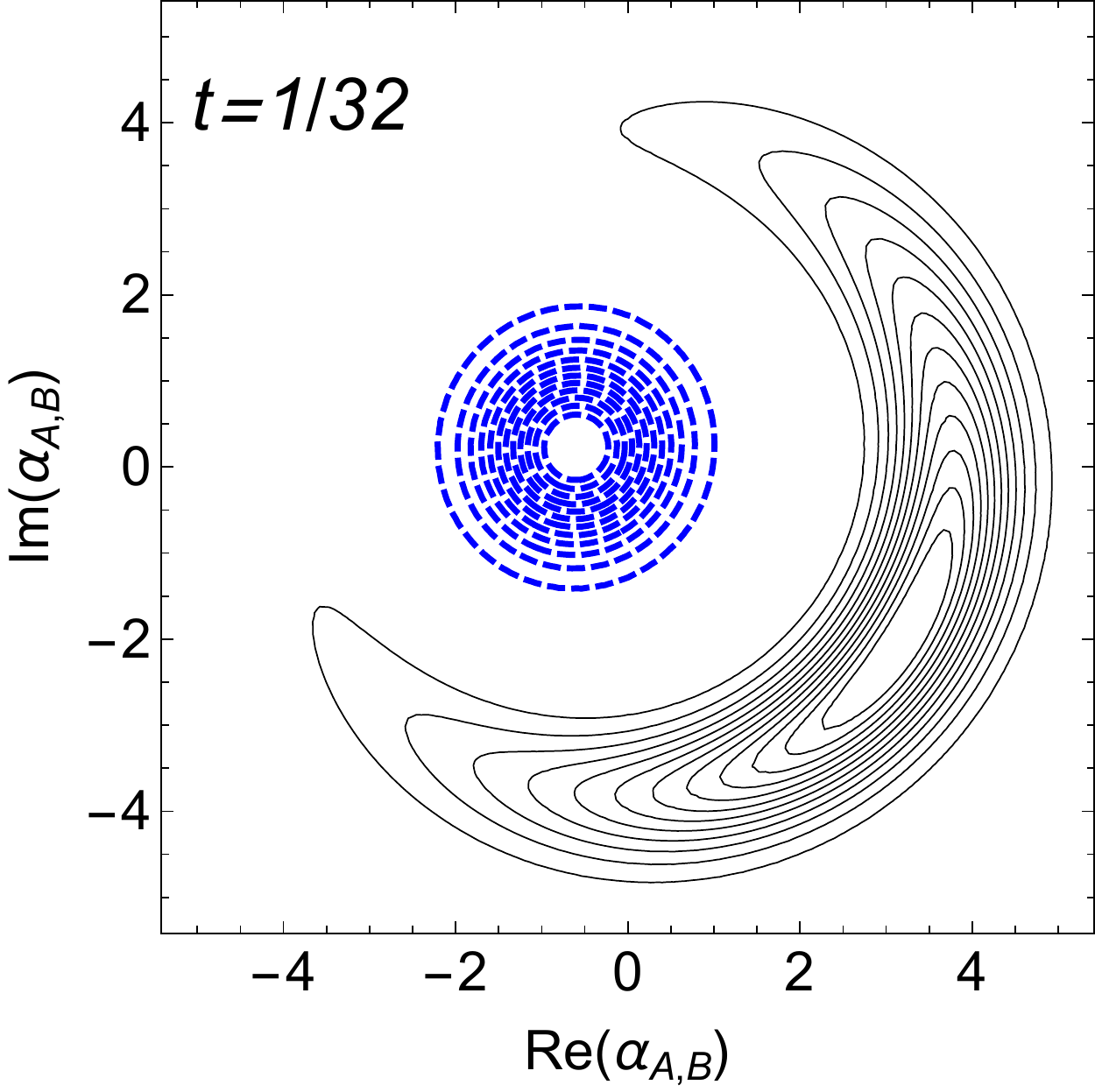}%
\includegraphics[width=120pt]{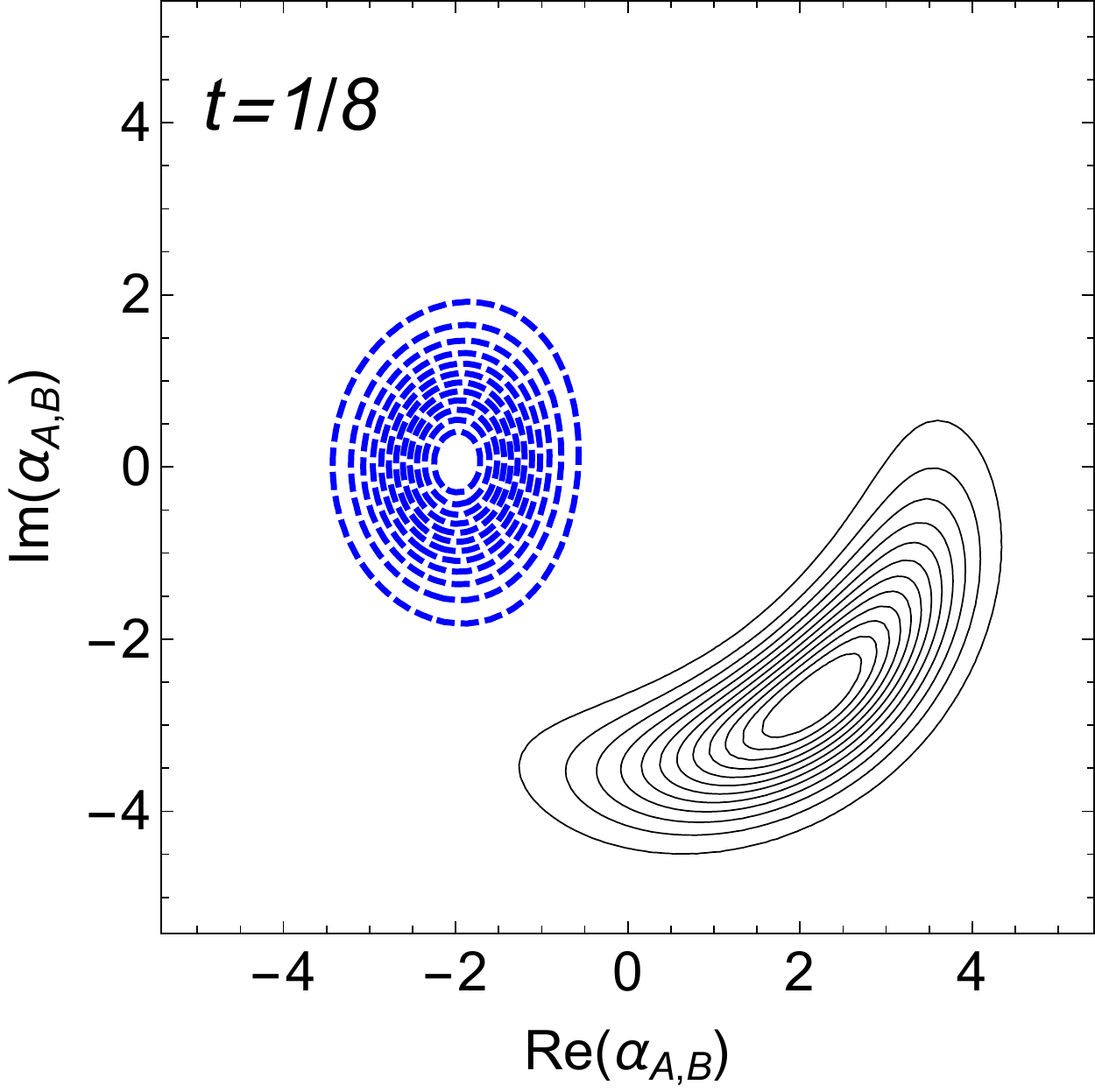}%
\includegraphics[width=120pt]{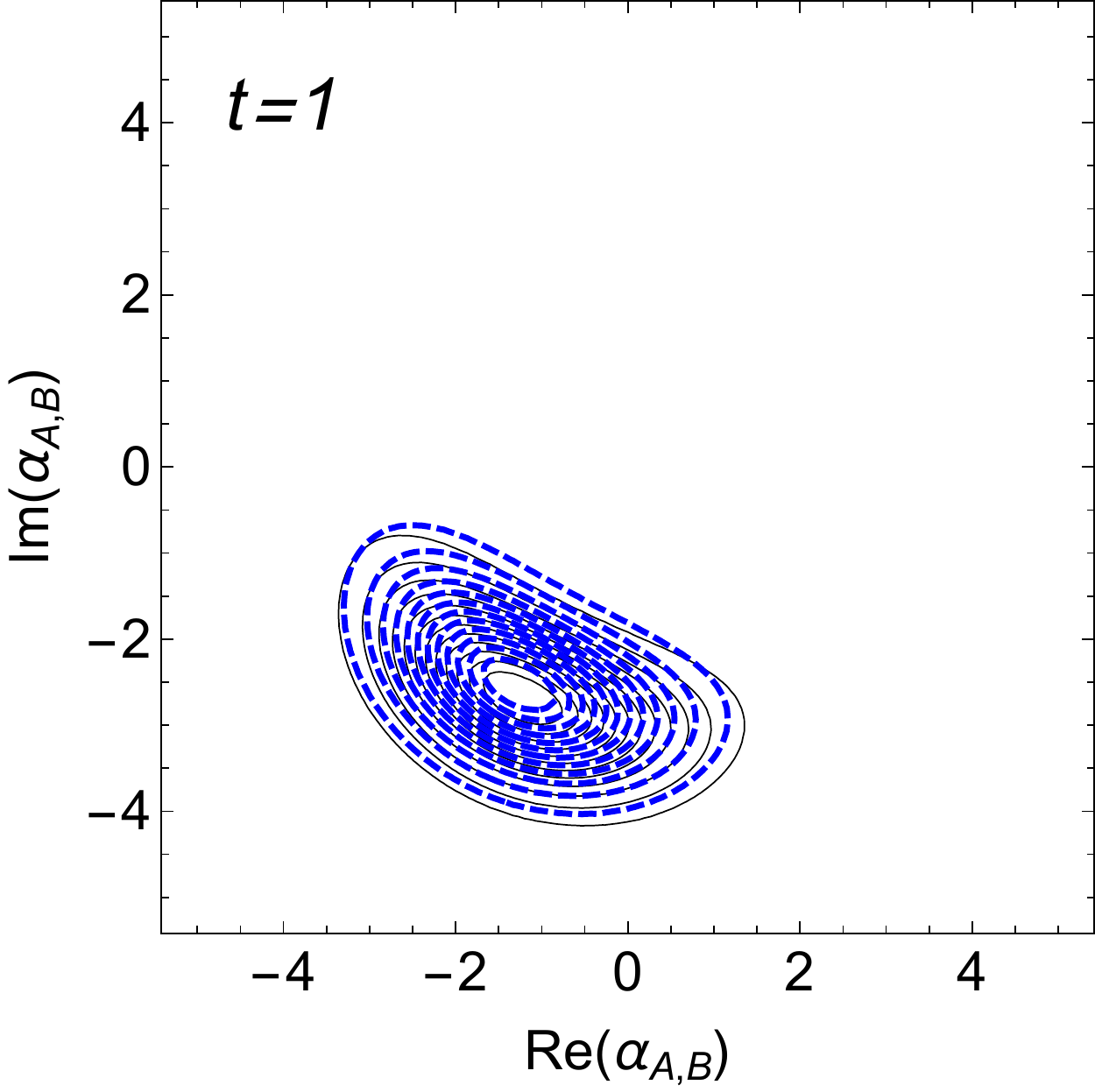}\\%
\includegraphics[width=120pt]{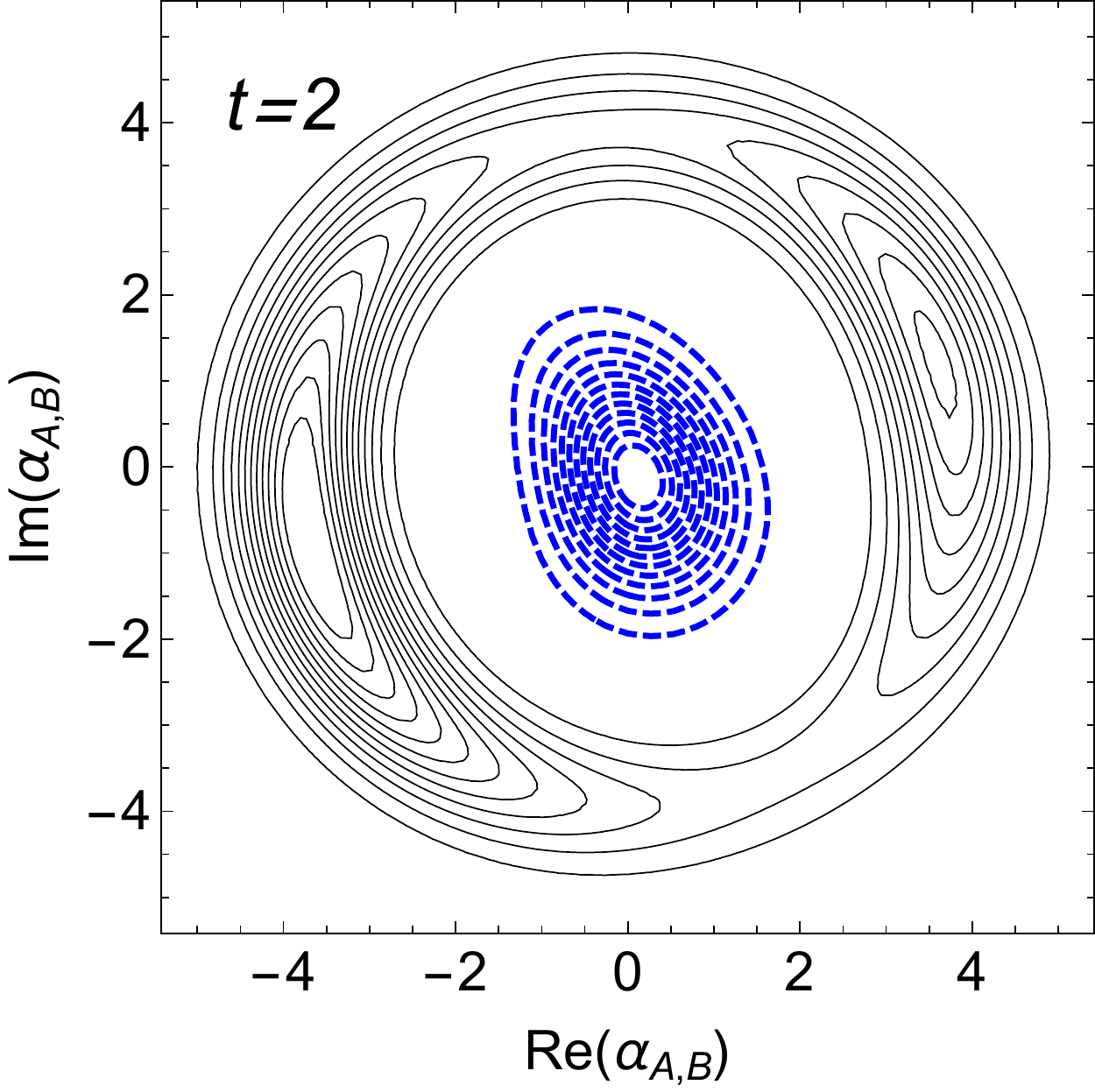}%
\includegraphics[width=120pt]{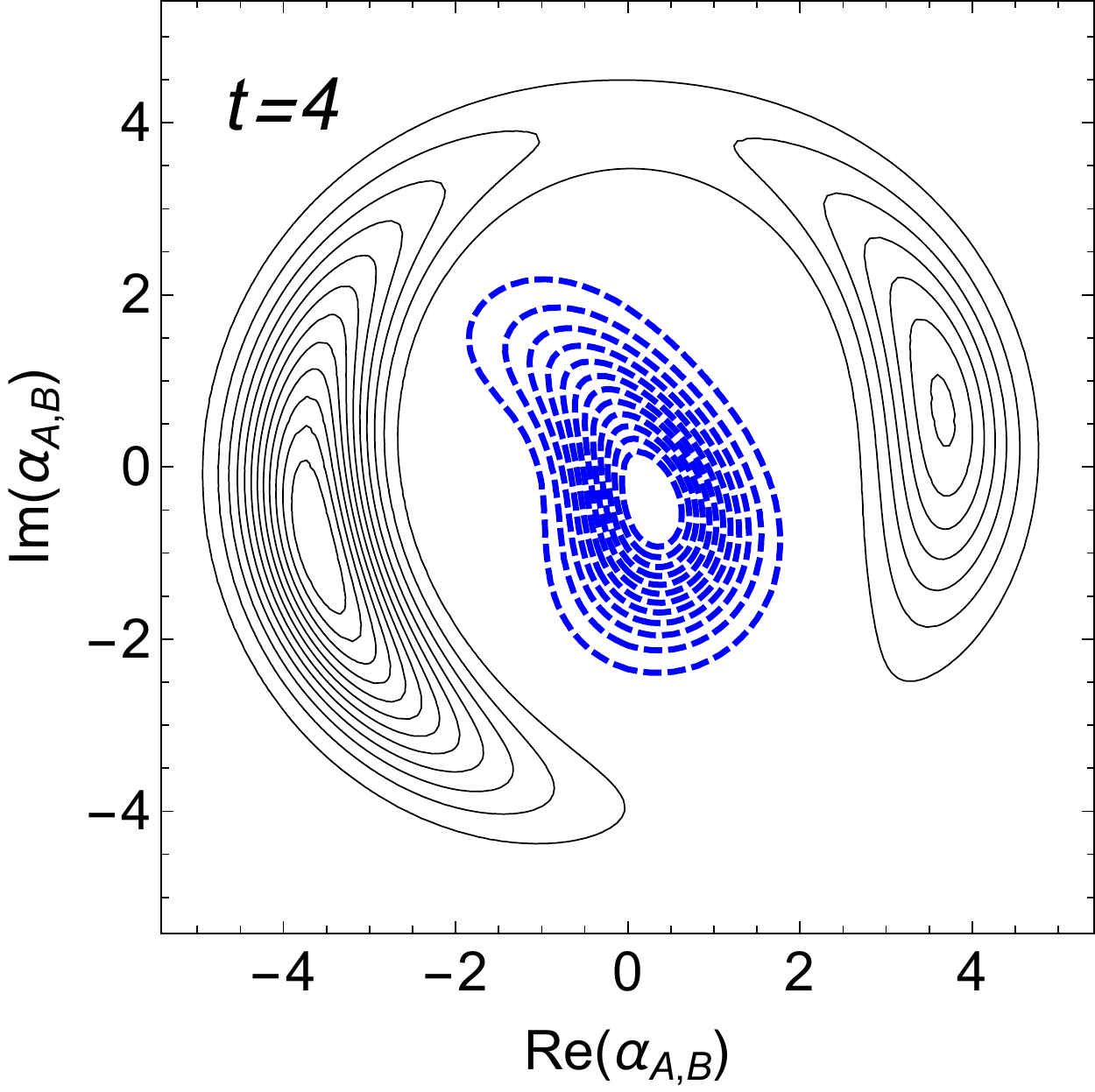}%
\includegraphics[width=120pt]{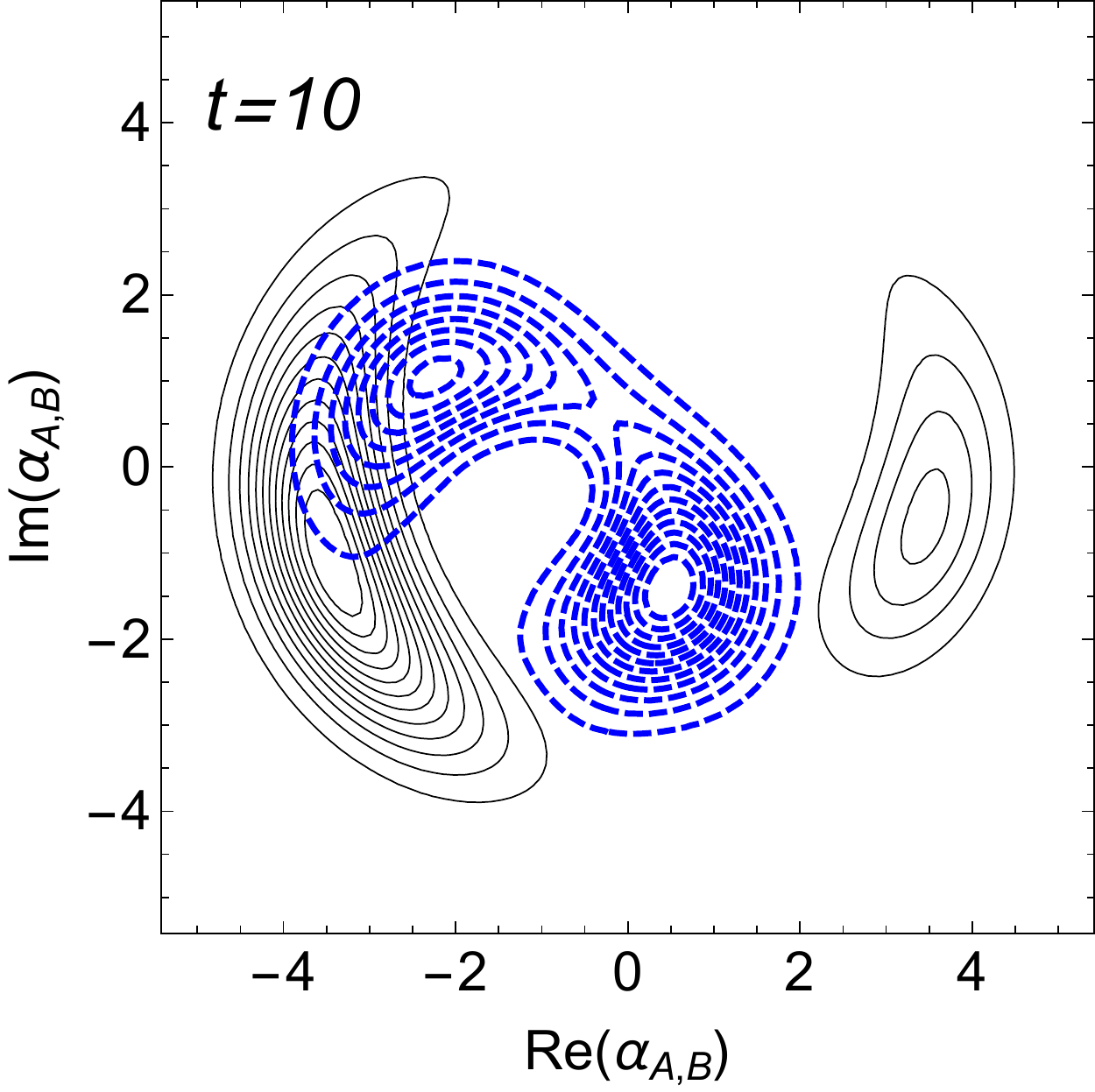}%
\includegraphics[width=120pt]{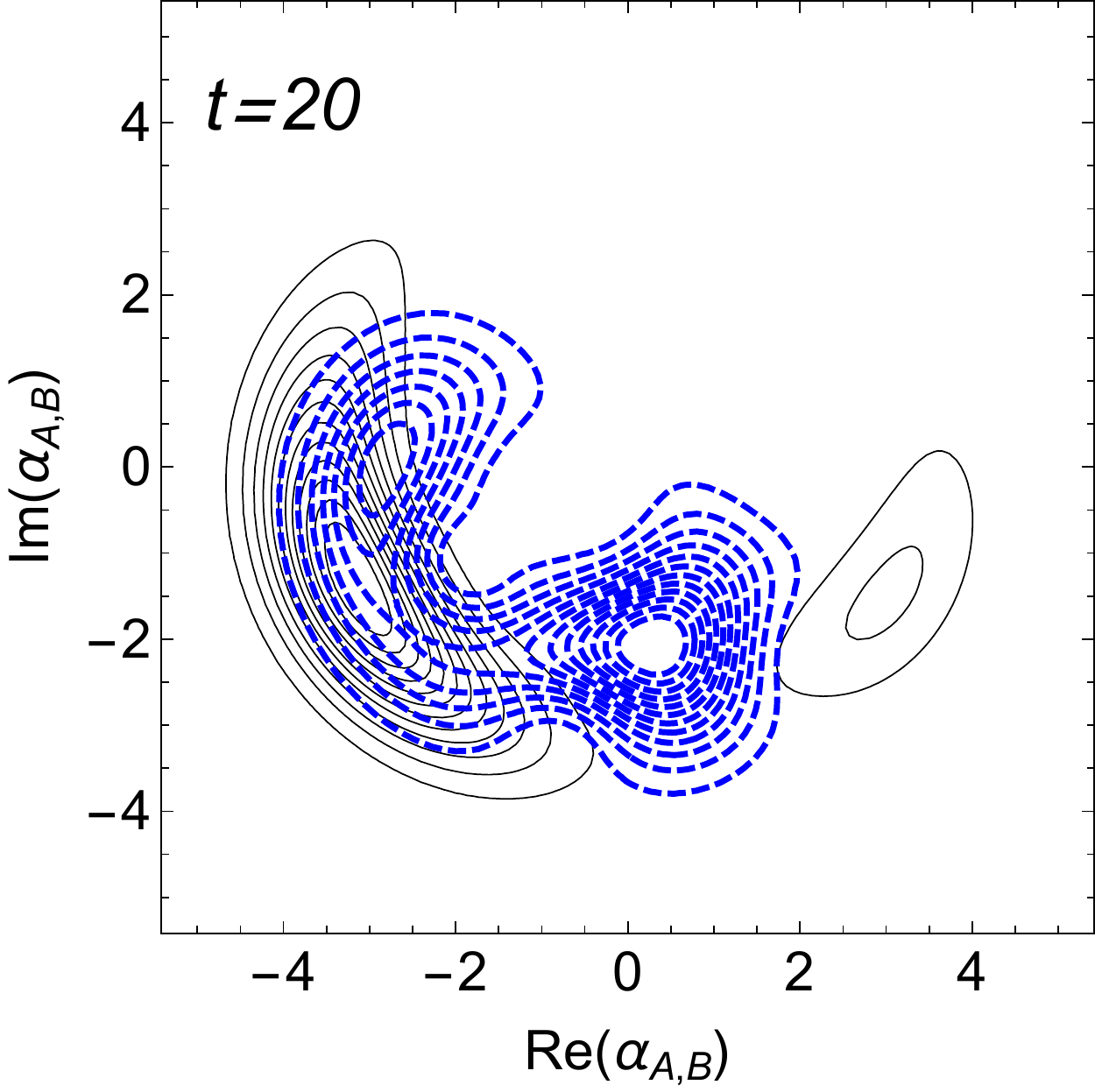}%
\end{center}
\caption{The temporal density distributions of the $Q$-functions for two
modes $Q(\protect\alpha _{A},\protect\alpha _{B},t)$ with specific $\protect%
\alpha _{B}=1+2i$ (solid lines) or with specific $\protect\alpha _{A}=1+2i$ (dashed lines)
at different evolution times $t$. The parameters are $\chi/g=-0.01$ and $%
\protect\delta/g=-1$. The time unit of the evolution is $\protect%
\pi/[1+(N-1)\protect\chi/g]$.}
\label{figure5}
\end{figure}

\begin{figure}[thp]
\begin{center}
\includegraphics[width=500pt]{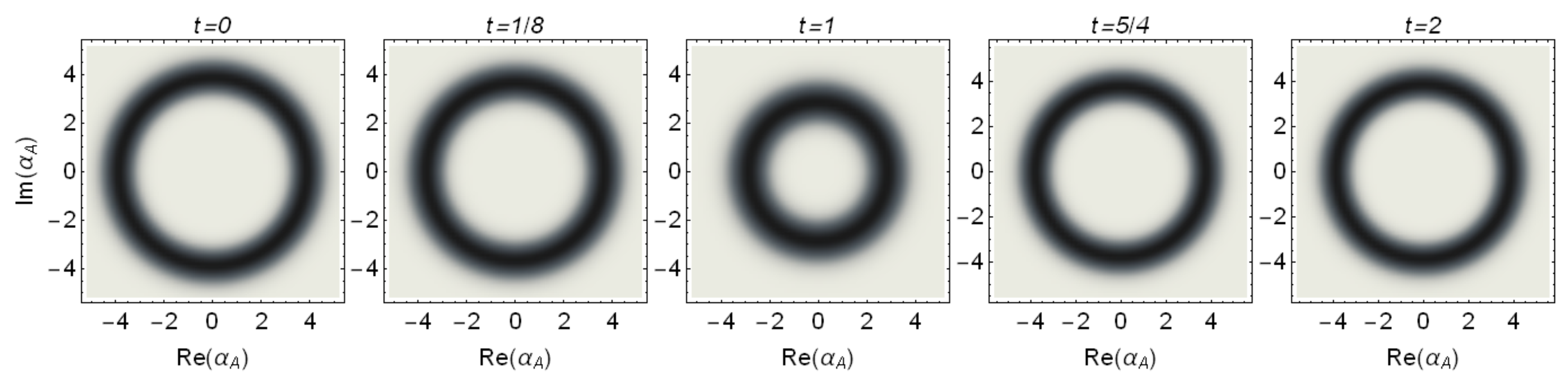}\\[0pt]
\includegraphics[width=500pt]{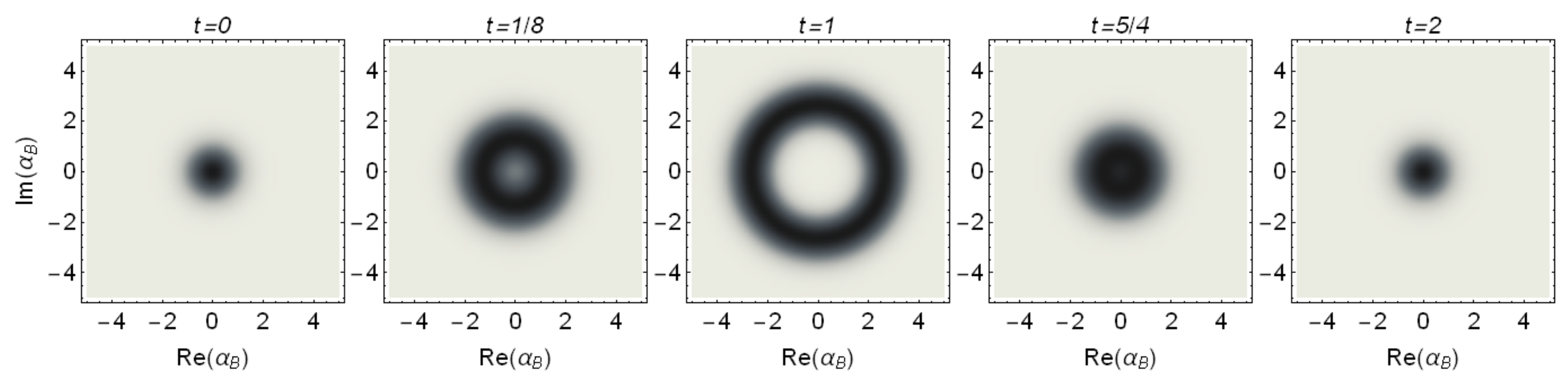}
\end{center}
\caption{The reversed dynamical phase distributions of two modes revealed by marginal $Q$-functions
$Q(\alpha _{A},t)$ (upper panel) and $Q(\alpha _{B},t)$ (lower
panel) with the same parameters, time scale and initial conditions as that in
Fig.\ref{figure5}.}
\label{figure6}
\end{figure}

The detailed evolution of the $Q$-function can be tracked down by exactly
solving coefficients of $C_{j}$s of Eq.(\ref{Cjs}) under
certain initial conditions. As the total atomic number $N$ is conserved
in the space of $\mathcal{H}_N$, the initial condition for $A$ mode is assumed to
be in the ground number state and $B$ mode is in a vacuum state, that
is, $\left\vert \psi \left( 0\right) \right\rangle =\left\vert
N\right\rangle_A \otimes \left\vert 0\right\rangle_B =\left\vert
N,0\right\rangle $ for $ C_{j}(0)=0, j=0,\cdots N-1$ and $C_{N}\left( 0\right) =1$.
Then the initial two-mode $Q$-function is%
\begin{equation}
\label{Nini}
Q_{N}(\alpha _{A},\alpha _{B},0)=\frac{\left( \left\vert \alpha
_{A}\right\vert ^{2}\right) ^{N}}{N!}e^{-\left( \left\vert \alpha
_{A}\right\vert ^{2}+\left\vert \alpha _{B}\right\vert ^{2}\right) },
\end{equation}%
which gives two well-separated probability distributions of two modes with no
overlap in the phase space as shown in Fig.\ref{figure5} ($t=0$). The
evolution of the probability distributions in the phase space with $N=15$ for
modes $A$ and $B$ are subsequently shown in Fig.\ref{figure5}. As the
high dimension of $Q(\alpha _{A},\alpha _{B},t)$, only the joint distributions of
one-mode with a specific value of the other can be depicted in
Fig.\ref{figure5}. The main property shown in Fig.\ref{figure5} is that the
local overlaps of the phase distributions can be obtained during the state
evolution, and a transient complete overlap can reach at a certain time
(e.g. $t=1$ in Fig.\ref{figure5}). The gradually merging of distributions of modes $A$ and $B$
in the phase space indicates that only a partial QS can be maintained between
two coupling modes in a Hilbert space of $%
\mathcal{H}_{N}$ after a long time evolution. Surely, the unperfect overlaps of the phase
distributions can be enhanced by properly adjusting the coupling rate of $\chi $
and the frequency detuning $\delta $ as well as the initial states of two modes.
As Fig.\ref{figure5} only shows temporal joint distributions for one mode when
the other one is in a specific value, so a marginal single-mode
quasi-probability function of $A$ or $B$ can be achieved by
integrating all the values of the other
\begin{eqnarray}
Q(\alpha _{A},t) &=&\int Q(\alpha _{A},\alpha _{B},t)\frac{d^{2}\alpha _{B}}{%
\pi }=e^{-\left\vert \alpha _{A}\right\vert ^{2}}\sum_{j=0}^{N}\Gamma \left(
N-j+\frac{1}{2}\right) \left\vert C_{j}(t)\right\vert ^{2}\frac{\left(
\left\vert \alpha _{A}\right\vert ^{2}\right) ^{j}}{j!\left( N-j\right) !},
\label{Qa} \\
Q(\alpha _{B},t) &=&\int Q(\alpha _{A},\alpha _{B},t)\frac{d^{2}\alpha _{A}}{%
\pi }=e^{-\left\vert \alpha _{B}\right\vert ^{2}}\sum_{j=0}^{N}\Gamma \left(
j+\frac{1}{2}\right) \left\vert C_{j}(t)\right\vert ^{2}\frac{\left(
\left\vert \alpha _{B}\right\vert ^{2}\right) ^{N-j}}{j!\left( N-j\right) !}.
\label{Qb}
\end{eqnarray}%
The above results clearly show that the marginal $Q$-functions for $A$ and $B$ modes
produce symmetric circular distributions which conduct breathing
motions in the phase space as shown in Fig.\ref{figure6}.
The breathing modes of the ringlike distribution mask all the irregular details of distribution patterns induced by the
nonlinear dynamics of two coupling modes. Although no stable distribution can be reached
beyond dissipations, a correlated out-of-phase breathing motion of the ring distributions of $A$ and $B$
is revealed and a transient perfect overlap of the marginal $Q$-functions at
certain time ($t=1$) is also found. The dynamics of the probability distributions demonstrated by
the marginal $Q$-functions in Fig.\ref{figure6} also verifies the partial QS behavior between
two scattering momentum modes in BEC. %
%figure%

In order to reveal the avoidance of a perfect QS between two
scattering modes generated in the BEC, we can
investigate the error (or deviation) operators between $A$ and $B$ modes defined by%
\begin{eqnarray*}
\hat{x}_{-} &=&\frac{1}{\sqrt{2}}\left( \hat{x}_{A}-\hat{x}_{B}\right)
\equiv \frac{1}{\sqrt{2}}\left[ \left( \hat{a}_{A}^{\dag }+\hat{a}%
_{A}\right) -\left( \hat{a}_{B}^{\dag }+\hat{a}_{B}\right) \right] , \\
\hat{p}_{-} &=&\frac{1}{\sqrt{2}}\left( \hat{p}_{A}-\hat{p}_{B}\right)
\equiv \frac{i}{\sqrt{2}}\left[ \left( \hat{a}_{A}^{\dag }-\hat{a}%
_{A}\right) -\left( \hat{a}_{B}^{\dag }-\hat{a}_{B}\right) \right] ,
\end{eqnarray*}%
where the zero-point-of-fluctuation units of
$x^{A,B}_{\mathrm{zpf}}=\sqrt{\frac{\hbar }{2m_{A,B}\omega _{A,B}}}$ and $p^{A,B}_{\mathrm{%
zpf}}=\sqrt{\frac{\hbar m_{A,B}\omega _{A,B}}{2}}$
for quadratures of modes $A,B$ ($\hat{x}_{A,B}$ and $\hat{p}_{A,B}$)
are used respectively. Then the error operators obey $\left[ \hat{x}_{-},\hat{p}_{-}%
\right] =2i$ and which leads to a fluctuation uncertainty for any quantum
state in the Hilbert space as%
\begin{equation}
\sigma ^{2}\left( x_{-}\right) \sigma ^{2}\left( p_{-}\right) \geq \left(
\frac{1}{2i}\left\langle \left[ \Delta \hat{x}_{-},\Delta \hat{p}_{-}\right]
\right\rangle \right) ^{2}=\left( \frac{1}{2i}\left\langle \left[ \hat{x}%
_{-},\hat{p}_{-}\right] \right\rangle \right) ^{2}=1,  \label{xp}
\end{equation}%
where the deviation of an arbitrary operator $\hat{O}$ is defined by $\Delta \hat{O}%
\equiv \hat{O}-\langle \hat{O}\rangle $, its variance is $\sigma
^{2}(O)\equiv \langle (\Delta \hat{O})^{2}\rangle $ and the fluctuation
above mean value is characterized by $\sigma (O)\equiv \sqrt{\langle (\Delta
\hat{O})^{2}\rangle }$.
The inequality of Eq.(\ref{xp}) indicates that if
one quadrature ($\hat{x}_{A,B}$) of the two modes is synchronized ($\hat{x}%
_{-}\rightarrow 0$), then the other ($\hat{p}_{A,B}$) will diverge, which
means that a complete QS between two coupled modes
is impossible in quantum regime. As the quantum states in a closed Hilbert
space of $\mathcal{H}_{N}$ are always on a pure states following Schr%
ödinger equation, the uncertainty of fluctuation will always
plays the role to destroy the dynamical synchronization between
two modes in quantum regime.

In order to reveal why two nonlinear classical
oscillators can synchronize completely via their mutual couplings but, in
the quantum regime, they never do, we
now consider the dynamics of the error operators with both the mean values and
their variances, simultaneously. The classical dynamics of the two modes are
always characterized by the mean values of their positions
$x_{A} \equiv \left\langle \hat{x}_{A}\right\rangle,
x_{B} \equiv \left\langle \hat{x}_{B}\right\rangle$
and momenta $p_{A} \equiv \left\langle \hat{p}_{A}\right\rangle,
p_{B} \equiv \left\langle \hat{p}_{B}\right\rangle $, and, often,
the CS is estimated only by
\begin{equation}
x_{-}\left( t\right) =\frac{1}{\sqrt{2}}\left[ x_{A}\left( t\right)
-x_{B}\left( t\right) \right] , \quad
p_{-}\left( t\right) =\frac{1}{\sqrt{2}}\left[ p_{A}\left( t\right)
-p_{B}\left( t\right) \right] ,
\end{equation}%
because the higher-order correlations decays very quickly with respect
to the classical time scale and the variances are always much smaller than
the mean values in a real system. Therefore, a complete CS is only estimated
by a vanishing of the mean errors after a longtime evolution. A partial synchronization,
such as the phase or amplitude synchronization, is achieved when the differences
between amplitudes of $R_{A,B}=x_{A,B}^{2}\left( t\right) +p_{A,B}^{2}\left( t\right)$
or the phases of $\varphi
_{A,B}\left( t\right) =\tan^{-1} \left[ p_{A,B}(t)/x_{A,B}(t)\right]$
are finally locked to a constant value, i.e., the deviation%
\begin{equation*}
R_{-}=R_{A}(t)-R_{B}(t) \quad \text{or} \quad \varphi _{-}\left( t\right) =\varphi
_{A}\left( t\right) -\varphi _{B}\left( t\right)
\end{equation*}%
asymptotically converges to a constant of $R_{0}\geq 0$ or $\varphi _{0}\in %
\left[ 0,2\pi \right] $ (see Fig.\ref{phase}).
\begin{figure}[t]
\begin{center}
\includegraphics[width=160pt]{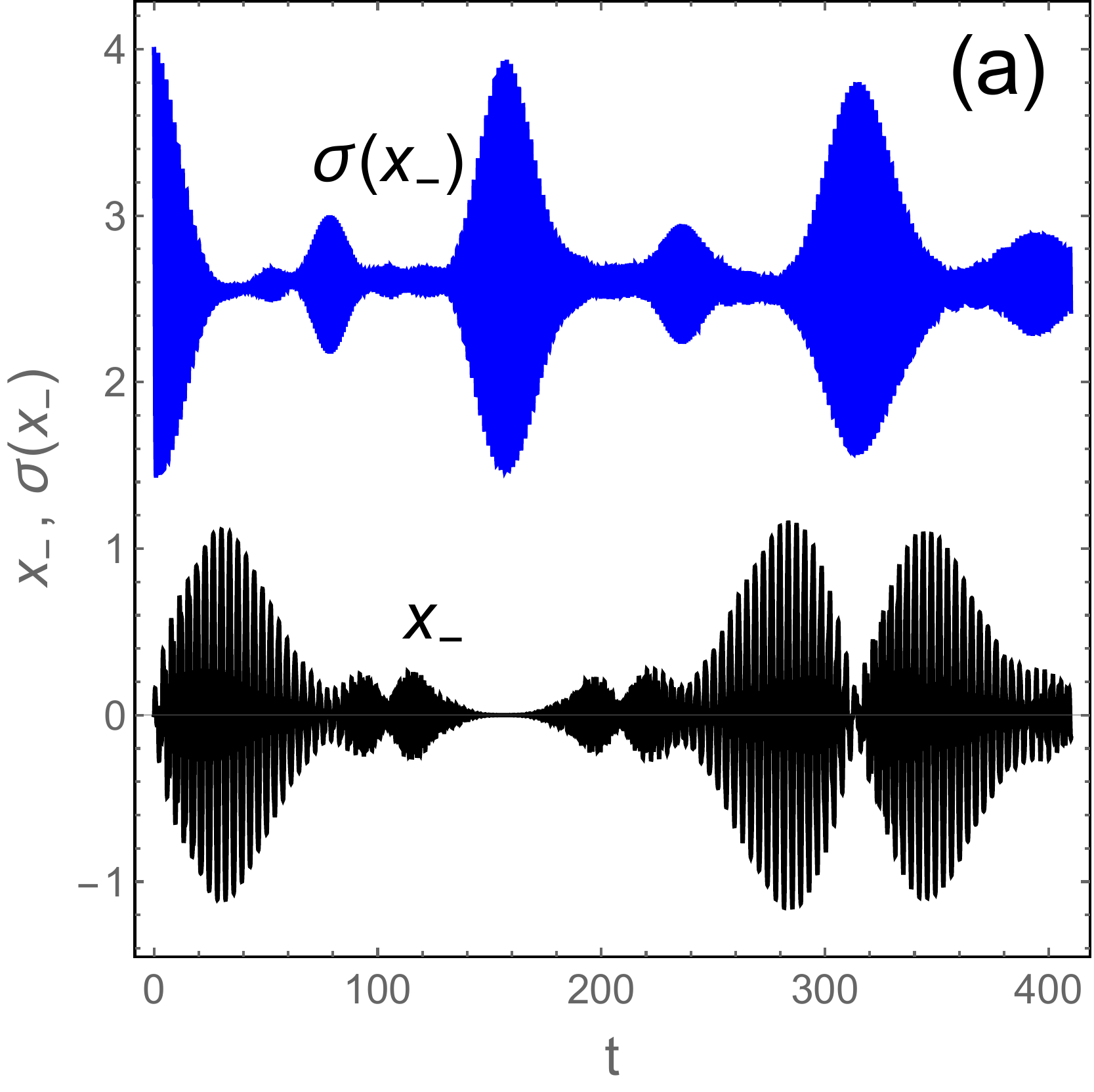}%
\includegraphics[width=160pt]{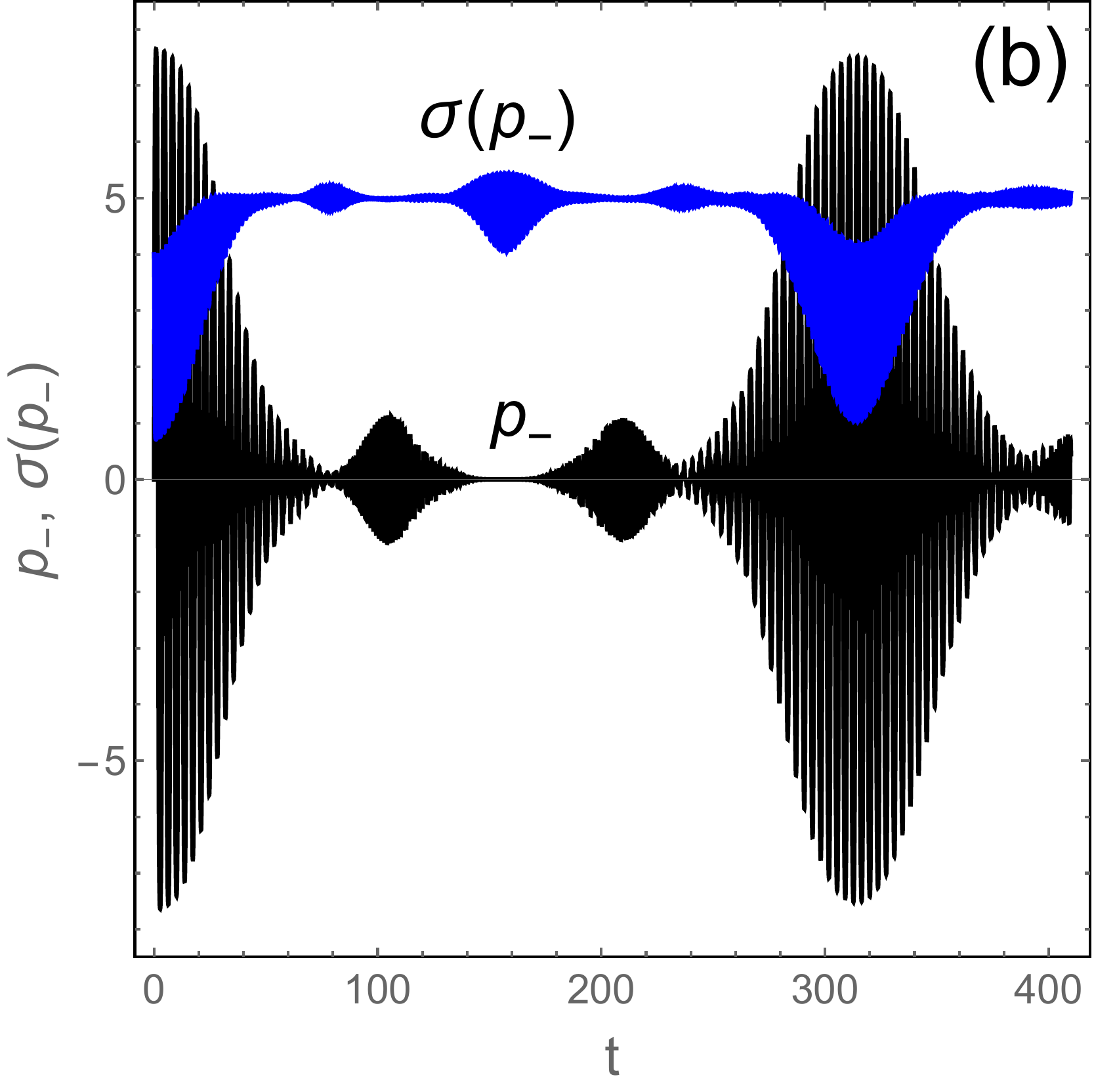}%
\includegraphics[width=160pt]{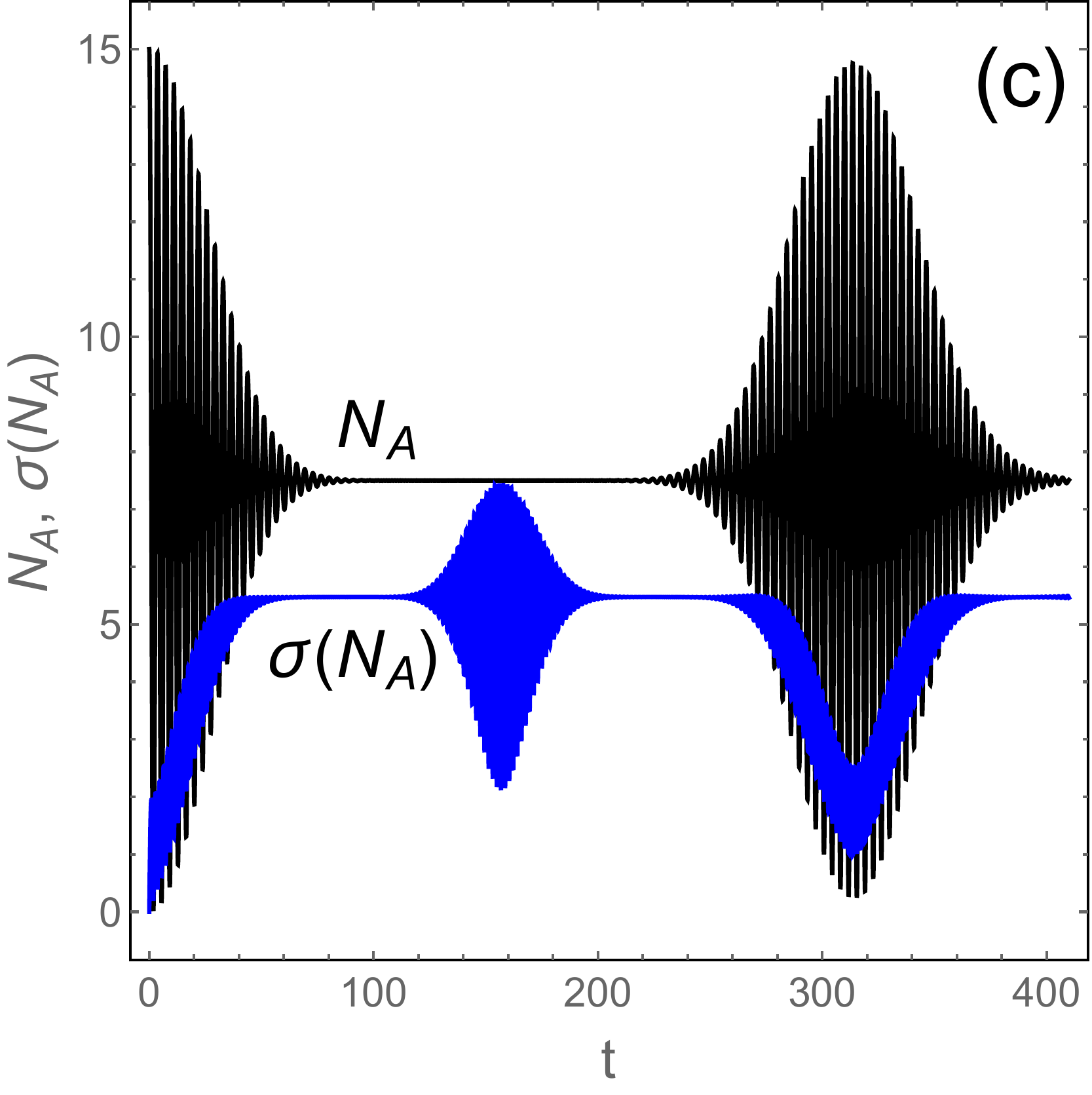}\\[0pt]
\includegraphics[width=160pt]{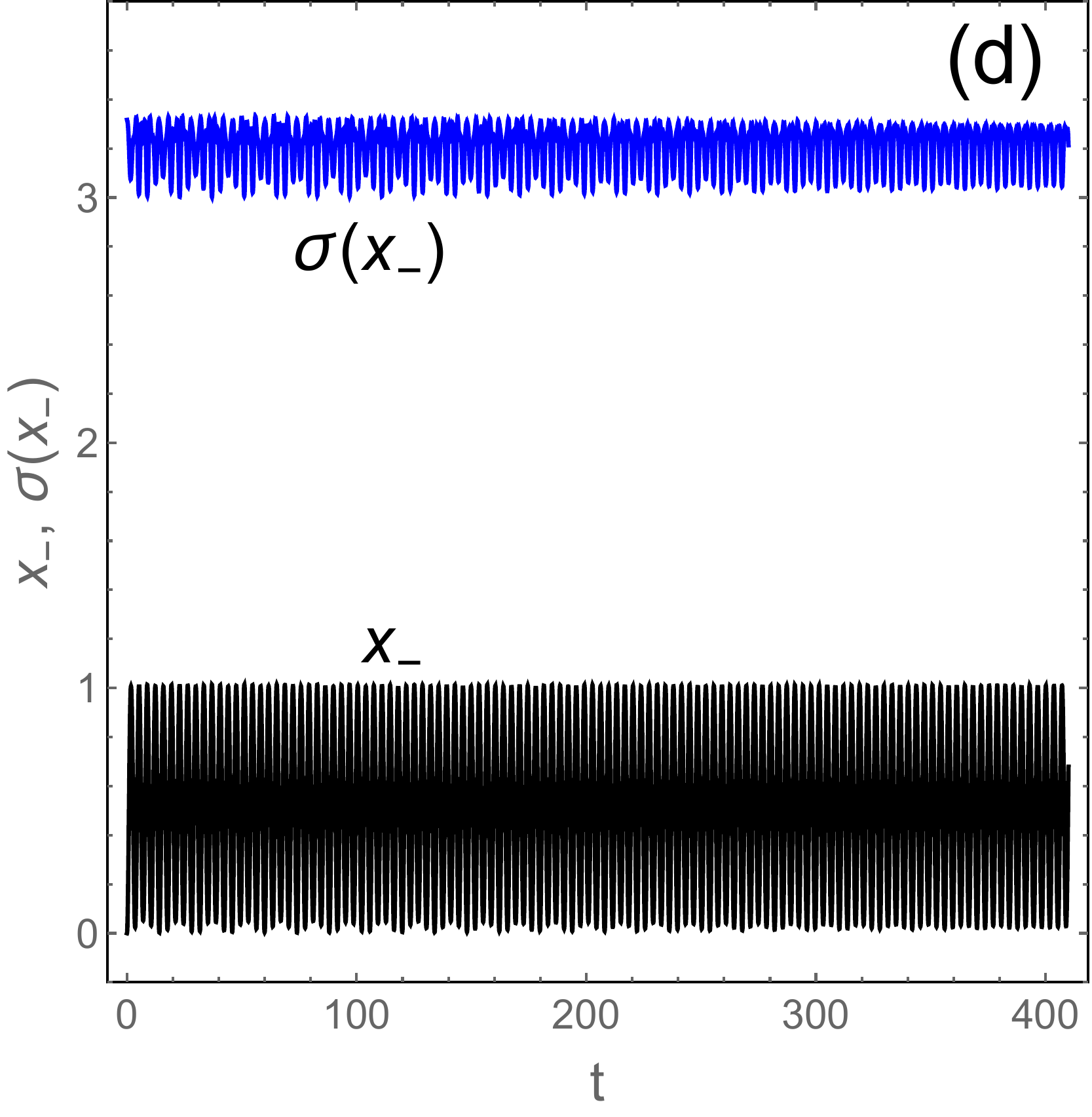}%
\includegraphics[width=160pt]{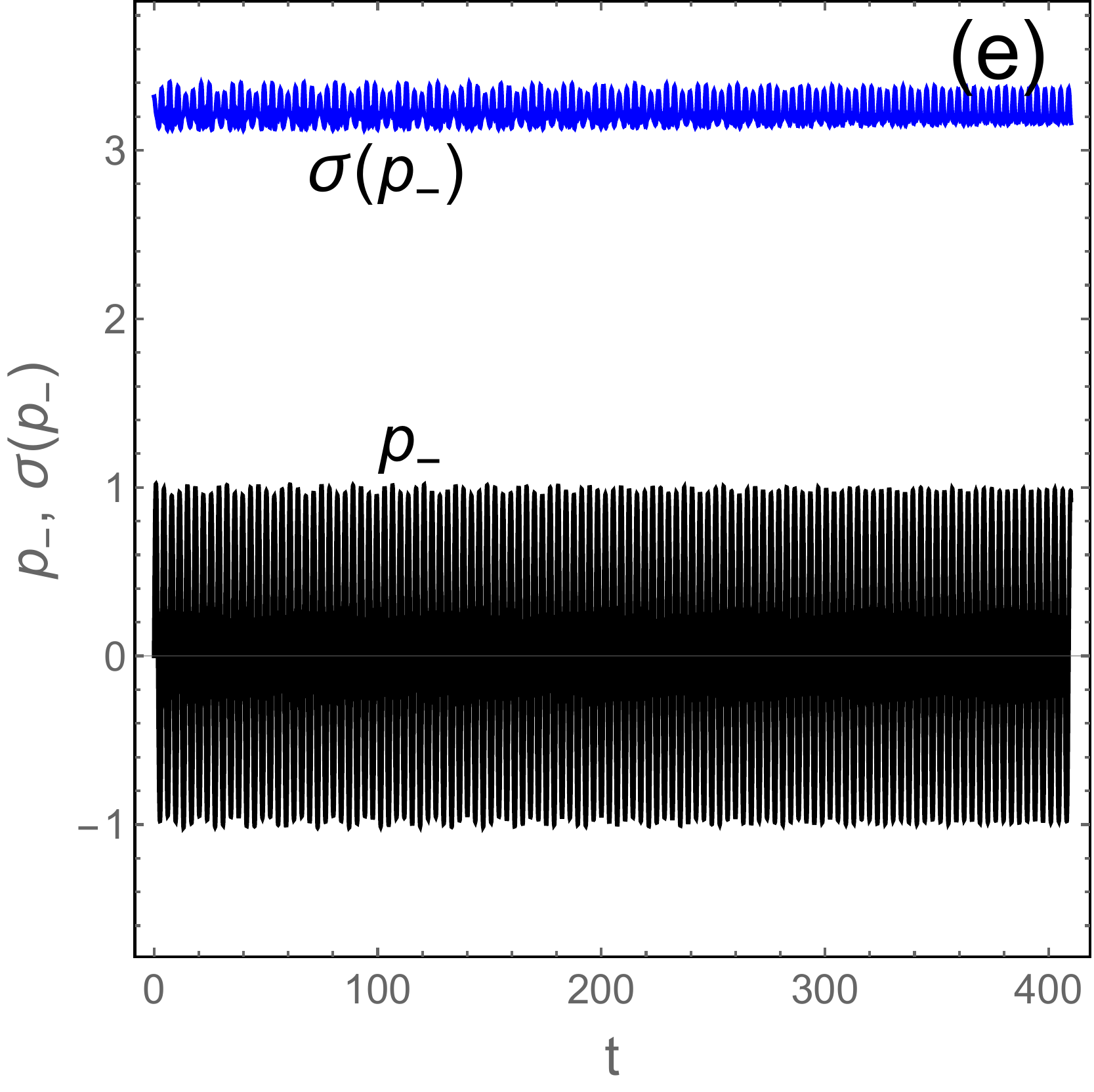}%
\includegraphics[width=160pt]{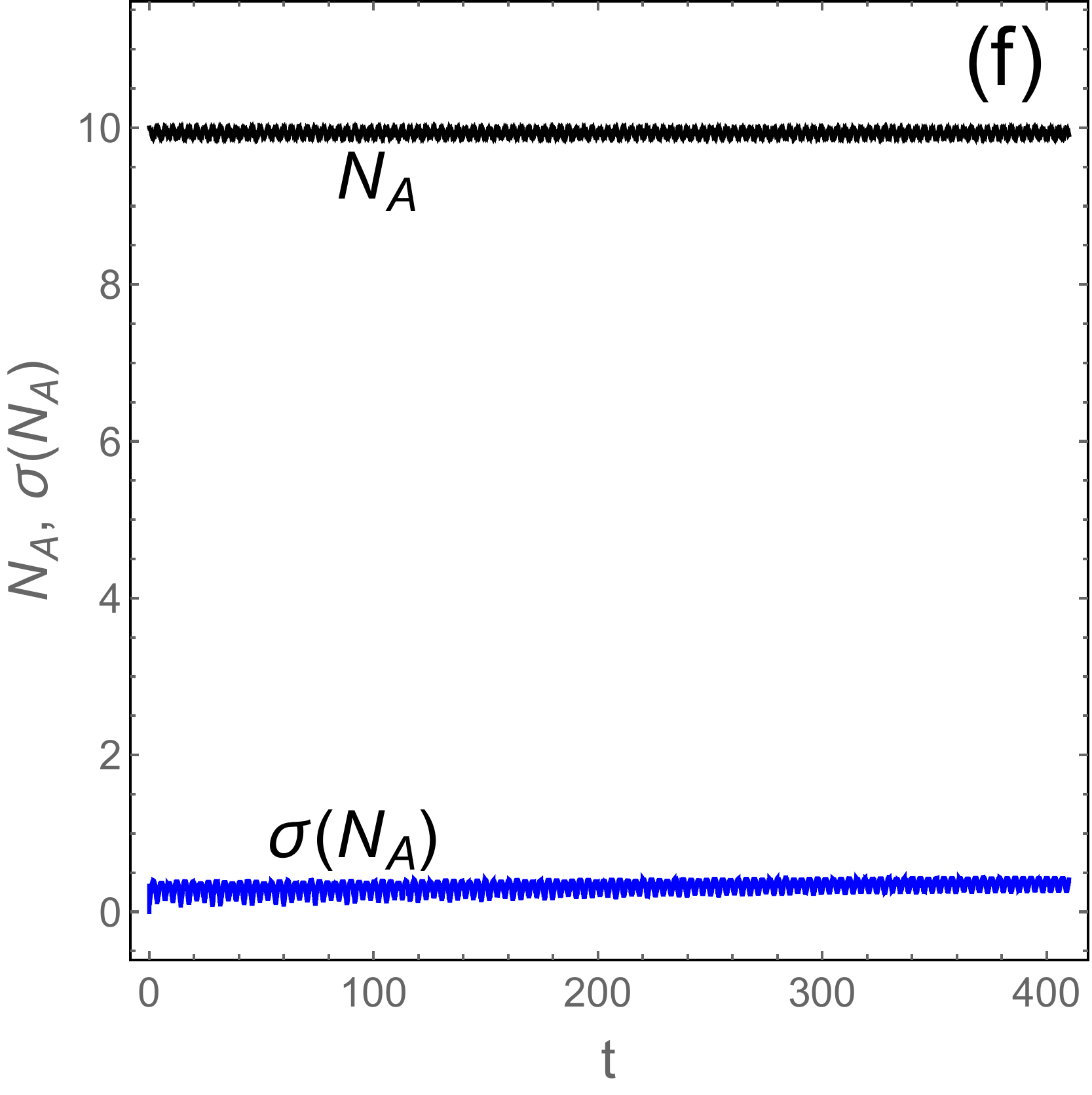}\\[0pt]
\end{center}
\caption{The dynamics of mean values and their relevant uncertainties for (a)
(b) (c) $N=15$ in the symmetric case of $\protect\delta =0$ and $\protect%
\chi/g =-0.01$, and (d) (e) (f) for $N=10$ with $\protect\delta =0$ and $%
\protect\chi/g=-0.1$. The initial condition is determined by Eq.(\protect\ref%
{Nini}).}
\label{figure7}
\end{figure}

However, for a quantum dynamics in a closed system, the high-order coherence of the error operators
should be included for a full synchronization judgement, which will lead non-negligible
modulations to the mean-value dynamics between two modes. The
fluctuations of the error operators beyond mean values can be estimated by
\begin{equation}
\sigma \left( x_{-}\right) =\frac{1}{\sqrt{2}}\sqrt{\sigma ^{2}\left(
x_{A}\right) +\sigma ^{2}\left( x_{B}\right)},  \label{dx}
\end{equation}%
where $\sigma^{2}(x_{A})$ and $\sigma^{2}(x_{B}) $ are position variances for modes $A$ and $B$,
%\begin{eqnarray*}
%\sigma ^{2}\left( x_{A}\right) &=&\sum_{j=0}^{N}\left[ \left( 2j+1\right)
%\left\vert C_{j}\right\vert ^{2}+\sqrt{\left( j+1\right) \left( j+2\right) }%
%\left( C_{j}C_{j+2}^{\ast }+C_{j}^{\ast }C_{j+2}\right) \right] \\
%&&-\left[ \sum_{j=0}^{N}\sqrt{j+1}\left( C_{j}C_{j+1}^{\ast }+C_{j}^{\ast
%}C_{j+1}\right) \right] ^{2},
%\end{eqnarray*}%
%\begin{eqnarray*}
%\sigma ^{2}\left( x_{B}\right) &=&\sum_{j=0}^{N}\left[ \left( 2N-2j+1\right)
%\left\vert C_{j}\right\vert ^{2}+\sqrt{\left( N-j\right) \left( N-j-1\right)
%}\left( C_{j}^{\ast }C_{j+2}+C_{j}C_{j+2}^{\ast }\right) \right] \\
%&&-\left[ \sum_{j=0}^{N}\sqrt{N-j}\left( C_{j}^{\ast
%}C_{j+1}+C_{j}C_{j+1}^{\ast }\right) \right] ^{2},
%\end{eqnarray*}%
and
\begin{equation}
\sigma \left( p_{-}\right) =\frac{1}{\sqrt{2}}\sqrt{\sigma ^{2}\left(
p_{A}\right) +\sigma ^{2}\left( p_{B}\right) },  \label{dp}
\end{equation}%
where $\sigma^{2}(p_{A})$ and $\sigma^{2}(p_{B}) $ are their momentum variances.
%\begin{eqnarray*}
%\sigma ^{2}\left( p_{A}\right) &=&\sum_{j=0}^{N}\left[ \left( 2j+1\right)
%\left\vert C_{j}\right\vert ^{2}-\sqrt{\left( j+2\right) \left( j+1\right) }%
%\left( C_{j}C_{j+2}^{\ast }+C_{j}^{\ast }C_{j+2}\right) \right] \\
%&&+\left[ \sum_{j=0}^{N}\sqrt{j+1}\left( C_{j}C_{j+1}^{\ast }-C_{j}^{\ast
%}C_{j+1}\right) \right] ^{2},
%\end{eqnarray*}%
%\begin{eqnarray*}
%\sigma ^{2}\left( p_{B}\right) &=&\sum_{j=0}^{N}\left[ \left( 2N-2j+1\right)
%\left\vert C_{j}\right\vert ^{2}-\sqrt{\left( N-j\right) \left( N-j-1\right)
%}\left( C_{j}C_{j+2}^{\ast }+C_{j}^{\ast }C_{j+2}\right) \right] \\
%&&+\left[ \sum_{j=0}^{N}\sqrt{N-j}\left( C_{j}C_{j+1}^{\ast }-C_{j}^{\ast
%}C_{j+1}\right) \right] ^{2}.
%\end{eqnarray*}
Fig.\ref{figure7} shows the mean-value dynamics of the error operators and
their corresponding fluctuations in a parametric region where the partial CS
in Fig.\ref{figure2} occurs. Fig.\ref{figure7}(a)(b) indicate that the varying
fluctuations over the mean-value dynamics sometimes plays a dominant role in the dynamics. We can
see that the error fluctuations become very large at the ``revival times" that
introduce large deviations of $x_{-}$ and $p_{-}$ to destroy the mean-value synchronization between
two modes. Both error quadratures conduct revival and
collapse dynamics due to coherent superpositions of different mode
frequencies of $C_{j}(t)$, $j=0,1,\dots,N$. The mode number of
$N_{A}\equiv \langle \hat{N}_{A}\rangle =\sum_{j=0}^{N}j\left\vert
C_{j}\right\vert ^{2}$ also exhibits a collapse and revival dynamics and the
number fluctuations increase during the collapse region and depresses
during the revival period of time (see Fig.\ref{figure7}(c)).

We can find a quantum trapping case of this model that, if the coupling parameter
satisfies $\chi =-g/(N-1)$, all the non-diagonal terms in Eq.(\ref{Cjs})
vanish and the quantum state will confine to the initial Fock state
$|N,0\rangle$ because it is the eigenstate of the total Hamiltonian. Surely,
this case is demanding and accidental because any fluctuations
on the parameters will destroy it and the transition from initial state to
other states will be activated. However, if $g+(N-1)\chi\rightarrow 0$, the
oscillating amplitudes and frequencies of $C_{j}(t)$ will be dramatically
decreased and the fluctuations of all the variables will be
suppressed. Fig.\ref{figure7}(d)-(f) demonstrate this case for $N=10$
giving $g+(N-1)\chi=0.1$ and we can see a clear suppression of varying
amplitudes of the mean values and their variances, both of them losing
the manifest ``collapse and revival" behaviors
compared with that in Fig.\ref{figure7}(a)-(c).

Because of the unceasing coherent revival of the quantum fluctuations, QS is
anyway different from its classical counterpart. In order to compare CS
with QS, Mari et. al. \cite{Mari}
introduced a generalized criterion which is based on the fluctuations of the error
operators as
\begin{equation*}
S_{c}\left( t\right) \equiv \frac{1}{\sigma ^{2}\left( x_{-}\right) +\sigma
^{2}\left( p_{-}\right) }=\frac{1}{\langle \left( \Delta \hat{x}_{-}\right)
^{2}+\left( \Delta \hat{p}_{-}\right) ^{2}\rangle }.
\end{equation*}%
This quantity gives the orbital deviation between two modes in the classical field and can measure
the level of QS with a quantum limit of
\begin{equation}
S_{c}\left( t\right) \leq \frac{1}{2\sqrt{\langle \left( \Delta \hat{x}%
_{-}\right) ^{2}\rangle \langle \left( \Delta \hat{p}_{-}\right) ^{2}\rangle
}}\leq \frac{1}{2}.  \label{bound}
\end{equation}%
In the classical dynamics, $%
S_{c}\left( t\right) $ is unbounded for a regular classical orbit in the
phase space (no deviation fluctuations) and will easily break the upper limit of 1/2 such as in an
irregular orbit of chaotic state.
\begin{figure}[t]
\begin{center}
\includegraphics[width=230pt]{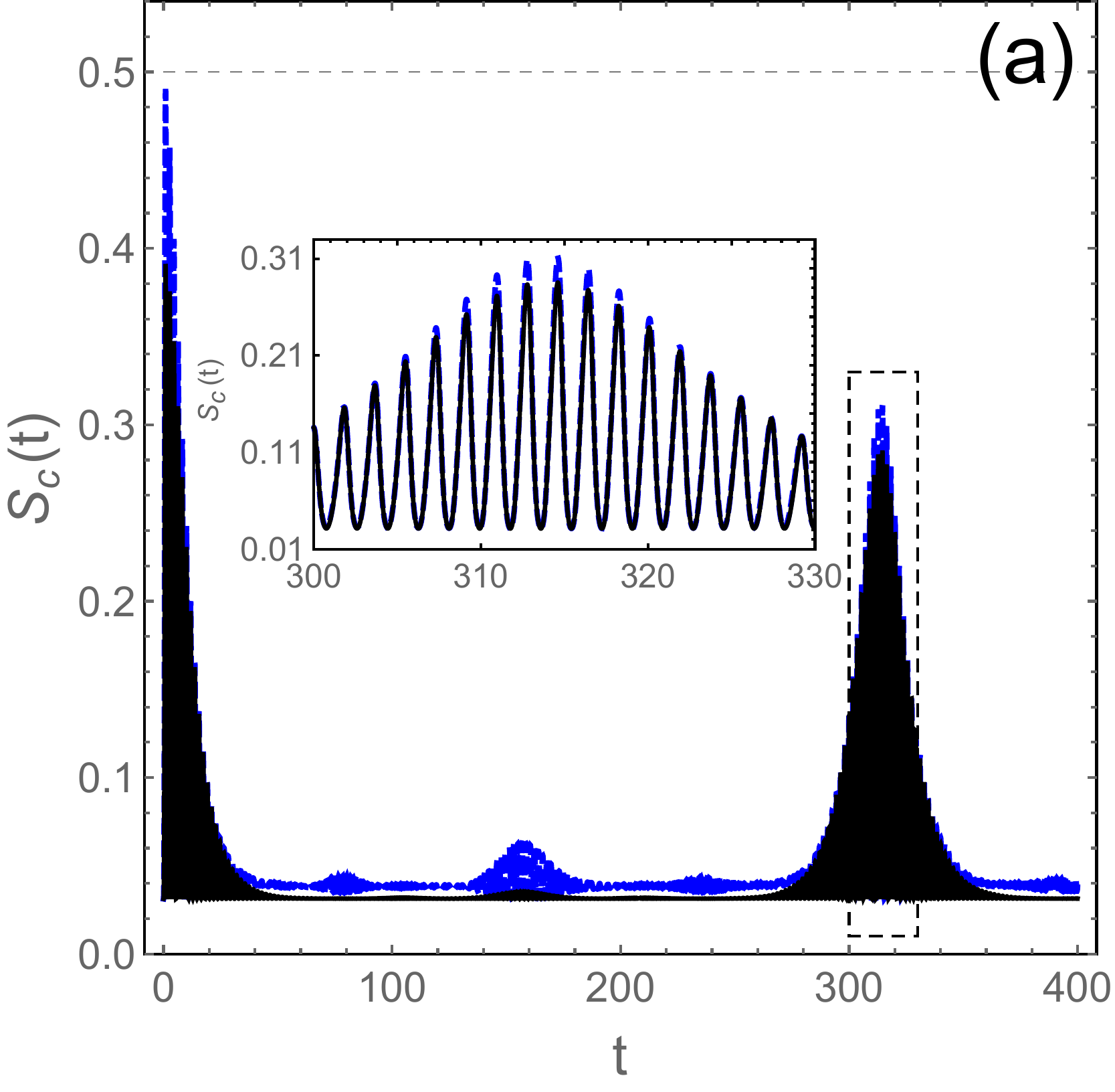}%
\includegraphics[width=230pt]{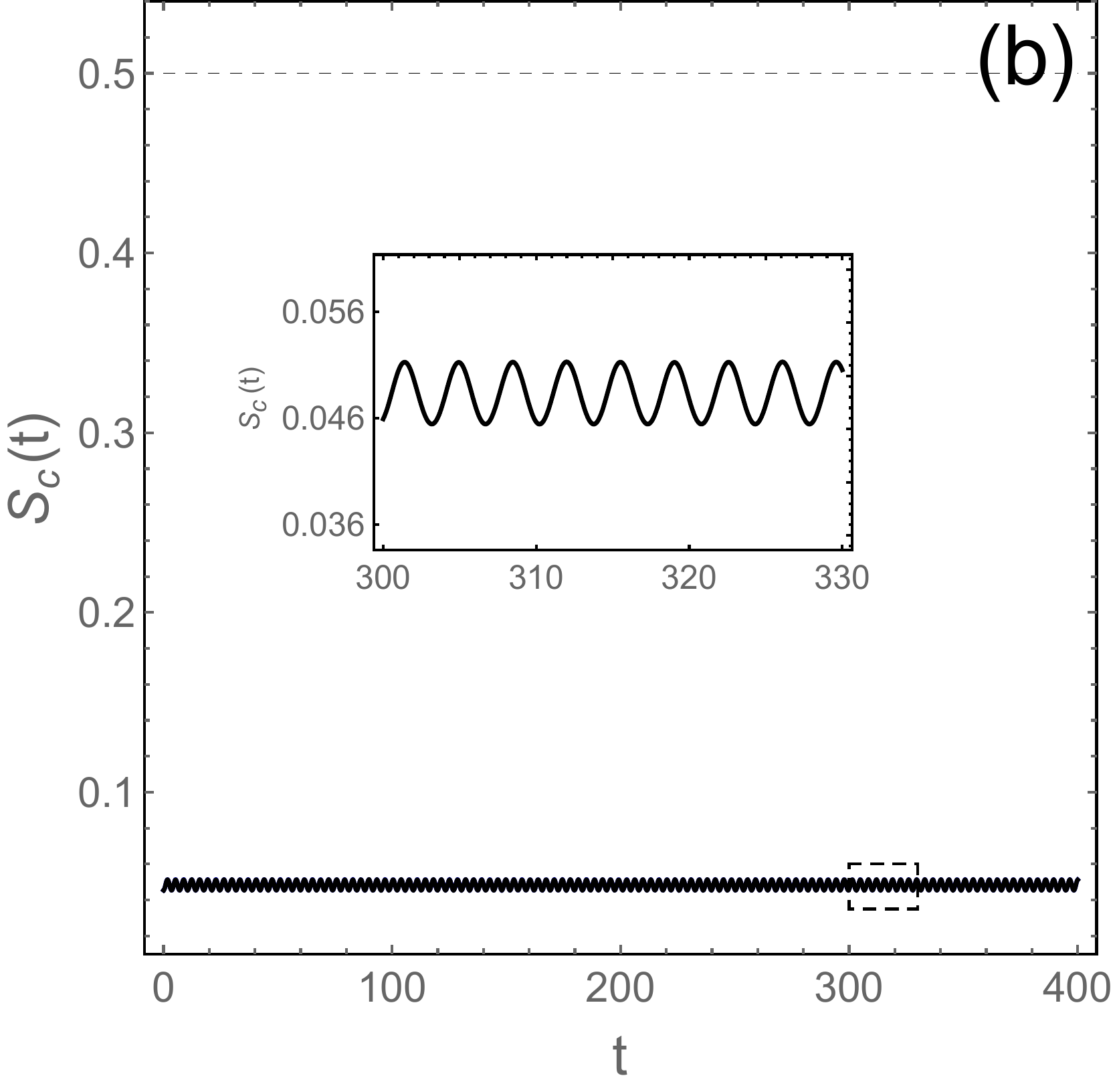}\\[0pt]
\end{center}
\caption{The dynamical measure of synchronization $S_{c}(t)$ for different
subspace of (a) $N=15$ and (b) $N=10$ with the same parameters as in
Fig.\protect\ref{figure7}. Insets: the zoomed-in images of $S_{c}(t)$ from the dashed rectangles.}
\label{figure8}
\end{figure}
Fig.\ref{figure8} demonstrates $S_c(t)$ with a limits of Eq.(\ref{bound}%
) in the model of two scattering modes of BEC
in different number states of $N$. Clearly, the upper limit of $1/2$
holds safely during the quantum dynamics and the measure $S_c(t)$ is closely
below the quantum uncertainty limitation: the middle term of Eq.(\ref{bound}) (the dashed blue curves in Fig.\ref{figure8}).
Fig.\ref{figure8}(b) is for the trapping case of $g+(N-1)\chi=0.1$. The calculation
verifies that the quantum fluctuation is a permanent property of quantum dynamics
and always exclude a full QS between the coupled quantum systems.

Recently, Ameri et al. \cite{Ameri} demonstrated that the fluctuation
measure of $S_{c}\left( t\right) $ has a similar behavior to the coherent
measure of the mutual information for modes $A$ and $B$, which, in a steady state
case, can be defined by%
\begin{equation*}
I=S\left( \hat{\rho}_{A}\right) +S\left( \hat{\rho}_{B}\right) -S\left( \hat{%
\rho}\right) ,
\end{equation*}%
where the Von Neumann entropy $S\left( \hat{\rho}\right) =-Tr\left( \hat{\rho%
}\ln \hat{\rho}\right) $ and $\hat{\rho}_{A,B}=Tr_{B,A}\left( \hat{\rho}%
\right) $. In our present model, the density operator is
time-dependent and can then be given by%
%\begin{equation*}
%\hat{\rho}\left( t\right) =\left[
%\begin{array}{cccc}
%C_{0}(t)C_{0}^{\ast }(t) & C_{0}(t)C_{1}^{\ast }(t) & C_{0}(t)C_{2}^{\ast
%}(t) & \cdots \\
%C_{1}(t)C_{0}^{\ast }(t) & C_{1}(t)C_{1}^{\ast }(t) & C_{1}(t)C_{2}^{\ast
%}(t) & \cdots \\
%C_{2}(t)C_{0}^{\ast }(t) & C_{2}(t)C_{1}^{\ast }(t) & C_{2}(t)C_{2}^{\ast
%}(t) & \cdots \\
%\vdots & \vdots & \vdots & \ddots%
%\end{array}%
%\right] .
%\end{equation*}%
%in a Fock representation. The Von Neumann entropy
%can then be given by
\begin{equation*}
S\left( \hat{\rho}\right) =-Tr\left( \hat{\rho}\ln \hat{\rho}\right)
=-\sum_{j}\lambda _{j}\left( t\right) \ln \lambda _{j}\left( t\right),
\end{equation*}%
where $\lambda _{j}(t)$ is the transient eigenvalue of the density operator. As the
present quantum system will keep a unitary evolution in the absence of
dissipation, the entropy reduces to $S(\hat{\rho})=0$, $S(\hat{\rho}%
_{A})=S(\hat{\rho}_{B})$, and the mutual information becomes $I(t)=2S(\hat{%
\rho}_{A})$.
\begin{figure}[htp]
\begin{center}
\includegraphics[width=230pt]{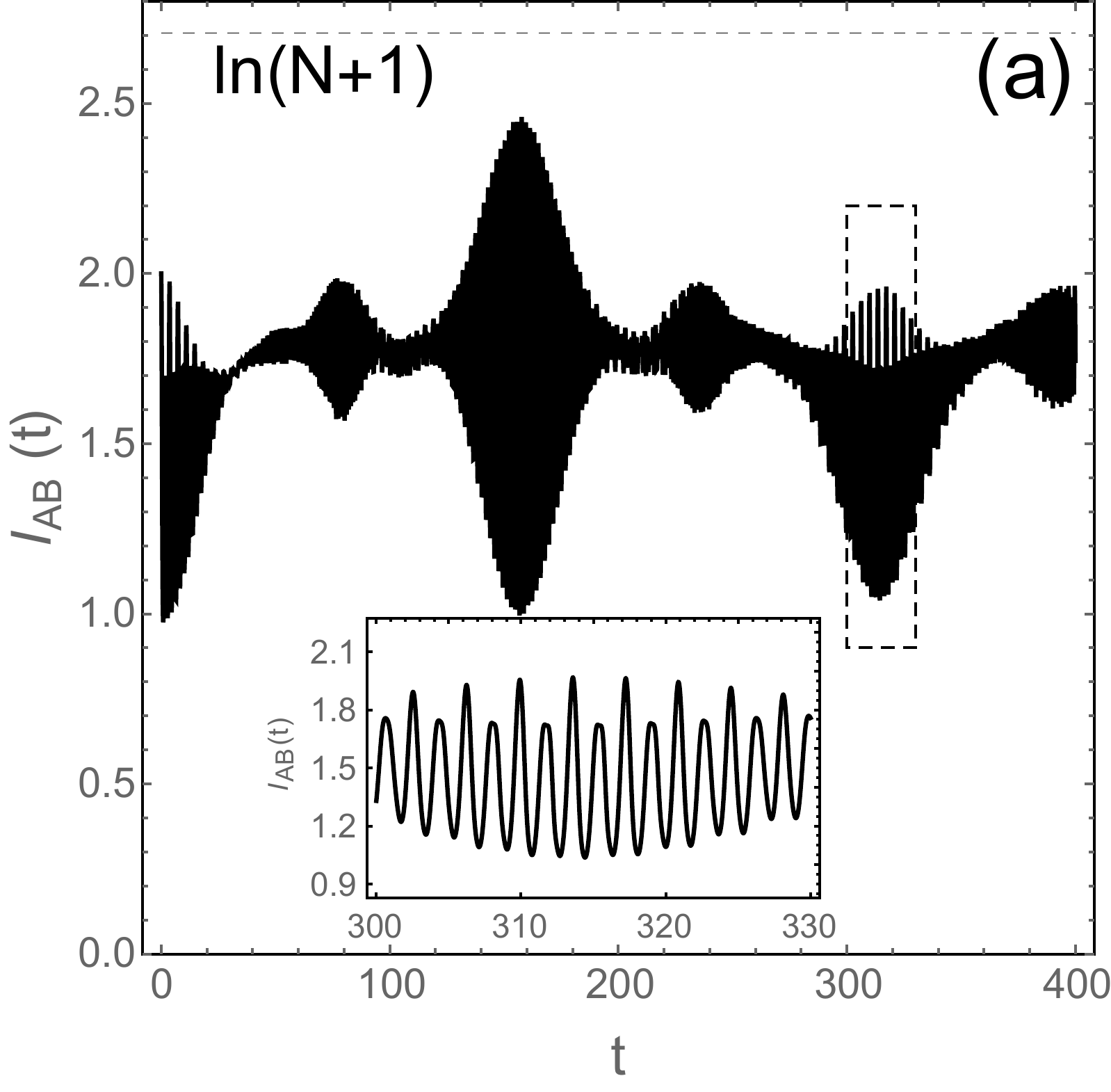}%
\includegraphics[width=230pt]{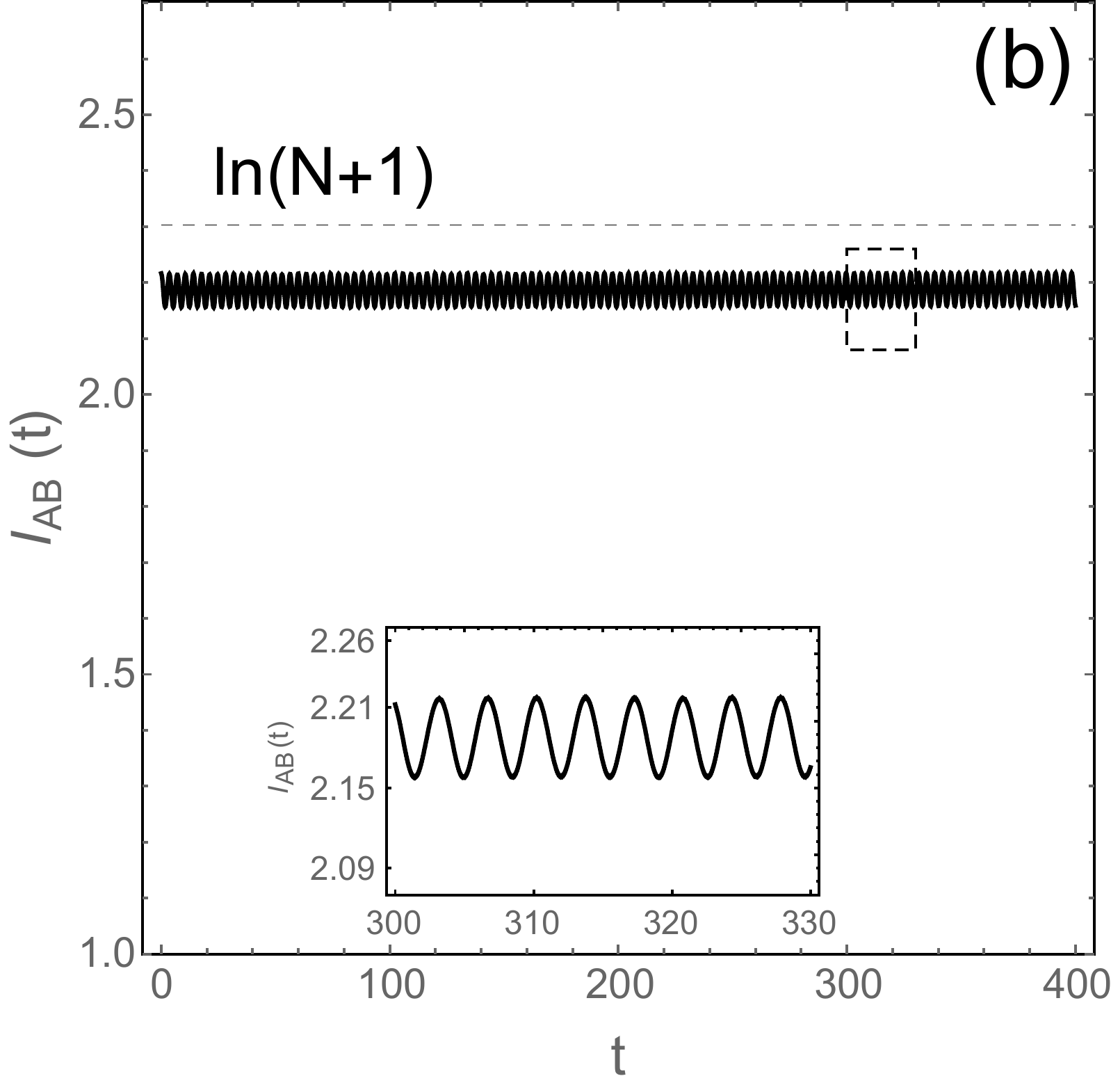}\\[0pt]
\end{center}
\caption{The dynamical measure of mutual information $I_{A,B}(t)$ for
different subspace of (a) $N=15$ and (b) $N=10$ with the same parameters as
in Fig.\protect\ref{figure7}. Insets: the zoomed-in images of $I_{A,B}(t)$ from the dashed rectangles.}
\label{figure9}
\end{figure}
Now we introduce the mutual information expressed by $Q$-function to investigate the
dynamical synchronization of two quantum modes as%
\begin{equation*}
I_{AB}\left( t\right) =\int \int Q\left( \alpha _{A},\alpha _{B},t\right)
\ln \left[ \frac{Q\left( \alpha _{A},\alpha _{B},t\right) }{Q\left( \alpha
_{A},t\right) Q\left( \alpha _{B},t\right) }\right] \frac{d^{2}\alpha _{A}}{%
\pi }\frac{d^{2}\alpha _{B}}{\pi },
\end{equation*}%
where the joint $Q$-function $Q(\alpha _{A},\alpha _{B},t)$ and the marginal
$Q$-functions $Q(\alpha _{A},t)$, $Q(\alpha _{B},t)$ have been given by Eq.(\ref%
{Q}) and Eq.(\ref{Qa})(\ref{Qb}), respectively. The mutual information
concerns about the information sharing between two modes with a
limit of%
\begin{equation*}
0\leq I_{AB}\left( t\right) \leq \ln (N+1),
\end{equation*}%
where $N+1$ is the dimension of the Hilbert space of $\mathcal{H}_{N}$. If the
$Q$-function can separate at a transient time of $t_{s}$
by satisfying $Q\left( \alpha _{A},\alpha _{B},t_{s}\right) =Q\left( \alpha
_{A},t_{s}\right) Q\left( \alpha _{B},t_{s}\right)$,
then $I_{AB}\left( t_{s}\right) =0$ and the two modes will be desynchronized
at time $t_s$ without sharing any information. This measure implies that
QS is somehow related to the quantum entanglement
between two modes. For a pure quantum state under unitary evolution, the
mutual information can be expressed by%
\begin{equation}
I_{AB}\left( t\right) =-2\int Q\left( \alpha _{A},t\right) \ln Q\left(
\alpha _{A},t\right) \frac{d^{2}\alpha _{A}}{\pi }.  \label{IAB}
\end{equation}%
By using the marginal $Q$-function defined by Eq.(\ref{Qa}), above mutual
information can be calculated by%
\begin{equation*}
I_{AB}\left( t\right) =-4\int_{0}^{\infty }Q\left( r,t\right) \ln Q\left(
r,t\right) dr,
\end{equation*}%
where%
\begin{equation*}
Q\left( r,t\right) =e^{-r^{2}}\sum_{j=0}^{N}\frac{\Gamma \left( N-j+\frac{1}{%
2}\right) }{j!\left( N-j\right) !}\left\vert C_{j}(t)\right\vert ^{2}r^{2j}.
\end{equation*}%
Fig.\ref{figure9} calculates the dynamics of the mutual information under
the same parameters as that in Fig.\ref{figure8} just for a dynamical comparison. We
can see, strictly, the mutual information $I_{AB}(t)$, exhibits a different
dynamics from that of measure $S_{c}(t)$. However,
both measures indicate correlated dynamics of synchronization
with a similar decreasing or increasing fluctuation at the same
time as that shown in Fig.\ref{figure7}. Although Fig.\ref{figure8} and Fig.\ref{figure9}
seem to have opposite mean-value tendencies,
their relevant ``collapse and revival" behaviors keep close connections. Fig.\ref{figure8}
and Fig.\ref{figure9} also show that the fluctuation amplitudes of two different measures
depends heavily on the initial states of the whole system (see the trapping case of $N=10$ is
very different from $N=15$).
Furthermore, based on $Q$-functions, we can see that QS
is somehow related to the quantum correlations (e.g., entanglement, discord, mutual
entropy) and connected to a mathematical problem of dynamical variable
separation for a joint probability function, such as $Q\left( \alpha _{A},\alpha
_{B},t_{s}\right)\rightarrow Q\left( \alpha _{A},t_{s}\right) Q\left( \alpha
_{B},t_{s}\right)$.

\subsection{QS of scattering modes from BEC in a coherent state}

As the importance of the initial state of BEC, we now consider QS again for two scattering modes generated from
a BEC in a coherent state. In this case, the whole Hilbert space of the system is a
direct sum of the subspaces of $\mathcal{H}_N$ with different atomic number of $N$, and the wave
function in the whole Hilbert space $\mathcal{H}=\bigoplus_{N}\mathcal{H}_N$ should be expanded by
\begin{equation}
\left\vert \Psi \left( t\right) \right\rangle
=\sum_{N_{A},N_{B}}C_{N_{A};N_{B}}\left\vert N_{A},N_{B}\right\rangle,
\label{psi2}%
\end{equation}%
where the number states of $\hat{N}_{A,B}\left\vert N_{A,B}\right\rangle =N_{A,B}\left\vert
N_{A,B}\right\rangle $.
Therefore the dynamical coefficients meet%
\begin{eqnarray}
i\dot{C}_{N_{A};N_{B}} &=&\left[ \delta \left( N_{A}-N_{B}\right) +\chi
\left( N_{A}^{2}+N_{B}^{2}+4N_{A}N_{B}-N_{A}-N_{B}\right) \right]
C_{N_{A};N_{B}}  \notag \\
&&+\left[ \left( g-\chi \right) +\chi \left( N_{A}+N_{B}\right) \right]
\sqrt{N_{B}\left( N_{A}+1\right) }C_{N_{A}+1;N_{B}-1}  \notag \\
&&+\left[ \left( g-\chi \right) +\chi \left( N_{A}+N_{B}\right) \right]
\sqrt{N_{A}\left( N_{B}+1\right) }C_{N_{A}-1;N_{B}+1},  \label{Cab}
\end{eqnarray}%
where the atomic number $N_{A},N_{B}$ of the two modes have no
limit in the whole Hilbert space. But within a certain subspace of specific $%
N=N_{A}+N_{B}$, the above Eq.(\ref{Cab}) returns back to Eq.(\ref{Cjs}).
%\begin{eqnarray*}
%i\dot{C}_{N_{A};N_{B}} &=&\sqrt{N_{A}\left( N_{B}+1\right) }\left[ g+\left(
%N-1\right) \chi \right] C_{N_{A}-1;N_{B}+1}+\left[ \delta \left(
%N_{A}-N_{B}\right) +\chi \left( N^{2}-N+2N_{A}N_{B}\right) \right]
%C_{N_{A};N_{B}} \\
%&&+\sqrt{\left( N_{A}+1\right) N_{B}}\left[ g+\left( N-1\right) \chi \right]
%C_{N_{A}+1;N_{B}-1}.
%\end{eqnarray*}
By using the coefficients $C_{N_{A};N_{B}},N_{A,B}=0,1,2\cdots $, the
phase-space quasi-probability distribution function of $Q(\alpha _{A},\alpha
_{B})$ can be written as%
\begin{equation*}
Q(\alpha _{A},\alpha _{B},t)=\left\vert \sum_{N_{A},N_{B}}C_{N_{A};N_{B}}(t)%
\frac{\left( \alpha _{A}^{\ast }\right) ^{N_{A}}\left( \alpha _{B}^{\ast
}\right) ^{N_{B}}}{\sqrt{N_{A}!N_{B}!}}\right\vert ^{2}e^{-\left( \left\vert
\alpha _{A}\right\vert ^{2}+\left\vert \alpha _{B}\right\vert ^{2}\right) },
\end{equation*}%
which satisfies $0\leq Q(\alpha _{A},\alpha _{B},t)\leq 1$ proved by
Cauchy's inequality.
As the $Q$-function $Q(\alpha _{A},\alpha _{B},t)$ describes the probability
distribution of a two-mode state in the phase space,
the single-mode quasi-probability distribution function for $%
A $ or $B$ is obtained by the integral of%
\begin{equation*}
Q(\alpha _{A,B},t)=\int Q(\alpha _{A},\alpha _{B},t)\frac{d^{2}\alpha _{B,A}%
}{\pi }.
\end{equation*}%
With polar coordinates of $\alpha _{A,B}=r_{A,B}e^{i\theta _{A,B}}$, the
marginal $Q$-functions for mode $A$ and $B$ can be obtained as
\begin{eqnarray*}
Q(\alpha _{A},t)&=&e^{-\left\vert \alpha _{A}\right\vert
^{2}}\sum_{N_{A},N_{A}^{\prime },N_{B}}^{\infty }\frac{\Gamma \left( \frac{%
2N_{B}+1}{2}\right) }{N_{B}!}\frac{C_{N_{A}N_{B}}C_{N_{A}^{\prime
}N_{B}}^{\ast }}{\sqrt{N_{A}!N_{A}^{\prime }!}}\left( \alpha _{A}^{\ast
}\right) ^{N_{A}}\left( \alpha _{A}\right) ^{N_{A}^{\prime }}, \\
Q(\alpha _{B},t)&=&e^{-\left\vert \alpha _{B}\right\vert
^{2}}\sum_{N_{A}}^{\infty }\frac{\Gamma \left( \frac{2N_{A}+1}{2}\right) }{%
N_{A}!}\sum_{N_{B}}^{\infty }\frac{\left( \alpha _{B}^{\ast }\right) ^{N_{B}}%
}{\sqrt{N_{B}!}}C_{N_{A}N_{B}}\sum_{N_{B}^{\prime }}^{\infty }\frac{\left(
\alpha _{B}\right) ^{N_{B}^{\prime }}}{\sqrt{N_{B}^{\prime }!}}%
C_{N_{A}N_{B}^{\prime }}^{\ast }.
\end{eqnarray*}%
The initial quantum state of the two-mode system in the whole Hilbert space is then
assumed to be
\begin{equation*}
\left\vert \Psi \left( 0\right) \right\rangle =\left\vert \alpha
_{0}\right\rangle_A \otimes \left\vert 0\right\rangle_B =e^{-\left\vert \alpha
_{0}\right\vert ^{2}/2}\sum_{j=0}^{\infty }\frac{\alpha _{0}^{j}}{\sqrt{j!}}%
\left\vert j,0\right\rangle ,
\end{equation*}%
which means that the coefficients are zero except for
$C_{N_{A};0}\left( 0\right) =e^{-\left\vert \alpha _{0}\right\vert ^{2}/2}%
\frac{\alpha _{0}^{N_{A}}}{\sqrt{N_{A}!}}$,
where $N_{A}=0,1,2,\cdots ,\infty$. In order for a numerical calculation, we
should introduce a specific number of $n$ for a truncated sum of the coefficients
\begin{equation*}
C_{N_{A};0}\left( 0\right) =e^{-\left\vert \alpha _{0}\right\vert ^{2}/2}%
\frac{\alpha _{0}^{N_{A}}}{\sqrt{N_{A}!}},\quad N_{A}=0,1,2,\cdots ,n,
\end{equation*}%
which can be simplified further by only choosing $N_{A}\in \lbrack
\vert \alpha _{0}\vert ^{2}-\sqrt{\vert \alpha
_{0}\vert ^{2}},\vert \alpha _{0}\vert ^{2}+\sqrt{%
\vert \alpha _{0}\vert ^{2}}]$ because of the initial Possion population
distribution. %Fig8%%
\begin{figure}[htp]
\begin{center}
\includegraphics[width=500pt]{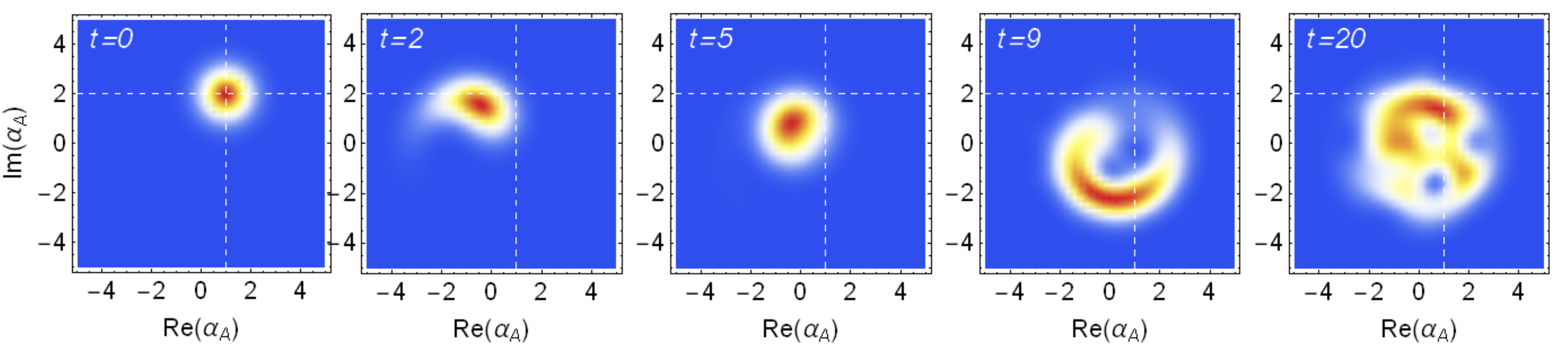}\\[0pt]
\includegraphics[width=500pt]{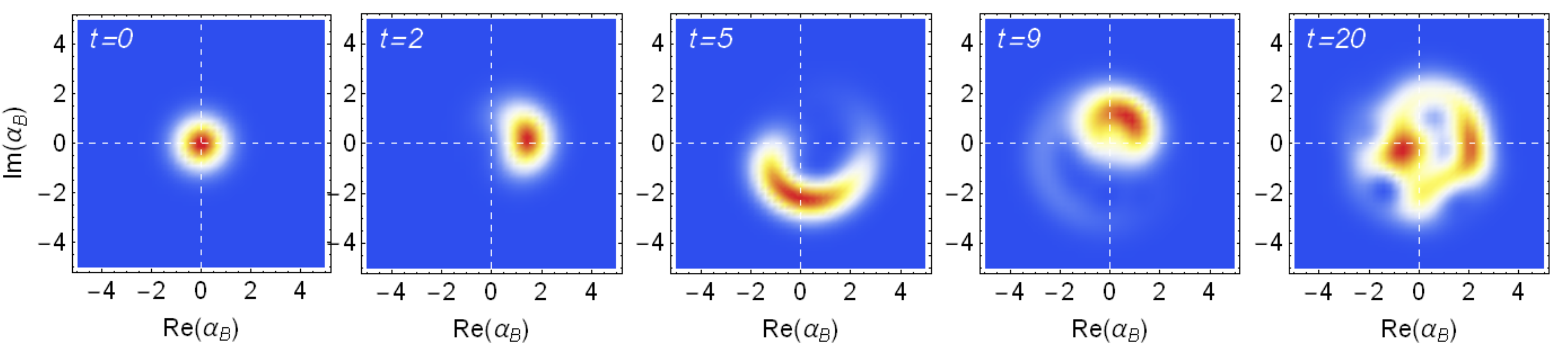}
\end{center}
\caption{The temporal distributions of $Q$-functions for two symmetric modes ($\delta=0$) of $Q(%
\alpha _{A},t)$ (upper panel) and $Q(\protect\alpha _{B},t)$ (lower
panel). The dashed lines indicate the central positions of the initial states for modes
$A$ and $B$. The initial state for the mode $A$
is in a coherent state of $|\alpha _{0}\rangle $ with $\protect\alpha %
_{0}=1+2i$ and the mode $B$ is in a vacuum state $|0\rangle $. The nonlinear coupling
rate $\chi/g=-0.01$ and the time unit of evolution is $\pi/[1+(N-1)\protect\chi/g]$.}
\label{figure10}
\end{figure}
For convenience, we change the subscripts of $C_{N_A;N_B}$ and use the equations of motion in a closed
truncated Hilbert space (because the total atomic number is still conserved) of%
\begin{eqnarray}
i\dot{C}_{kl} &=&\left[ \delta \left( k-l\right) +\chi \left(
k^{2}+l^{2}+4kl-k-l\right) \right] C_{kl}
+\sqrt{l\left( k+1\right) }\left[ g+\left( k+l-1\right) \chi \right]
C_{k+1;l-1}  \notag  \\
&&+\sqrt{k\left( l+1\right) }\left[ g+\left( k+l-1\right) \chi \right]
C_{k-1;l+1}, \label{initial}
\end{eqnarray}%
where $(k,l)$ are all the integer pairs satisfying
$k+l\leq N, N=0,1,2,\cdots ,n$,
and the total number of $C_{kl}$ or equations is $\left( n+1\right)
\left( n+2\right) /2$. Therefore, the marginal $Q$-functions will be%
\begin{equation}
Q(\alpha _{A},t)=e^{-\left\vert \alpha _{A}\right\vert
^{2}}\sum_{k=0}^{n}\sum_{l=0}^{n-k}\sum_{j=0}^{n-k}\frac{\Gamma \left( k+%
\frac{1}{2}\right) }{k!}\frac{C_{lk}C_{jk}^{\ast }}{\sqrt{l!j!}}\left(
\alpha _{A}^{\ast }\right) ^{l}\left( \alpha _{A}\right) ^{j},  \label{Qa2}
\end{equation}%
and%
\begin{equation}
Q(\alpha _{B},t)=e^{-\left\vert \alpha _{B}\right\vert
^{2}}\sum_{k=0}^{n}\sum_{l=0}^{n-k}\sum_{j=0}^{n-k}\frac{\Gamma \left( k+%
\frac{1}{2}\right) }{k!}\frac{C_{kl}C_{kj}^{\ast }}{\sqrt{l!j!}}\left(
\alpha _{B}^{\ast }\right) ^{l}\left( \alpha _{B}\right) ^{j}.  \label{Qb2}
\end{equation}

By using Eq.(\ref{initial}), the distribution patterns of $Q(\alpha _{A},t)$ and $%
Q(\alpha _{B},t)$ in the whole Hilbert space $\mathcal{H}$ are shown in Fig.\ref%
{figure10} with an initial $Q$-functions being $Q(\alpha _{A},0) =\sqrt{\pi }
e^{-\left(\vert \alpha _{0}\vert ^{2}+\vert \alpha _{A}\vert ^{2}\right)}
\sum_{i=0}^{n}\sum_{j=0}^{n}\left( \alpha _{A}^{\ast }\alpha _{0}\right)
^{i}\left( \alpha _{A}\alpha _{0}^{\ast }\right) ^{j}$ and $Q(\alpha _{B},0)
=\sqrt{\pi }e^{-\left( \left\vert \alpha _{0}\right\vert ^{2}+\left\vert
\alpha _{B}\right\vert ^{2}\right) }$. The dynamics of the $Q$-functions shows
that the probability distributions of modes $A$ and $B$ in the phase space can
never reach a same pattern at a same time if they starts from
different initial states due to the conservations of their respective coherences.
However their distribution patterns will come closer to share a common phase space
and perform a similar shape variation with
a time lag because of the nonlinear couplings between them. Therefore,
the synchronized motion of the mode states can be somehow indicated
by obtaining similar but complementary patterns in the phase space,
especially after a long time evolution (see the last column of Fig.\ref{figure10} for $t=20$).
The different quantum MS in a whole Hilbert space from that in Fig.\ref{figure6}
is only due to a broken of the number
conservation in one BEC beam within a subspace of $\mathcal{H}_N$.
For an atomic BEC in a coherent state, there are many BEC beams with
different atomic numbers of $N$ mixing together during the state
evolution in the whole Hilbert space of $\mathcal{H}=\bigoplus_{N}\mathcal{H}_N$. The nonlinear
mechanical modes $A$ and $B$ generated from different BEC beams can coherently mix
together via the linear coupling $g$ and the nonlinear coupling $\chi $ which give
the irregular patterns of the motion probability as shown in Fig.\ref{figure10}.
According to the symmetry of modes $A$ and $B$ in the scattering
process, the same $Q$ distributions of two modes in the phase space will own a lower total
energy for Boson modes and thus modes $A$ and $B$ intend to share a same phase area and evolve into a
similar distribution pattern indicated by Fig.\ref{figure10}.
\begin{figure}[t]
\begin{center}
\includegraphics[width=170pt]{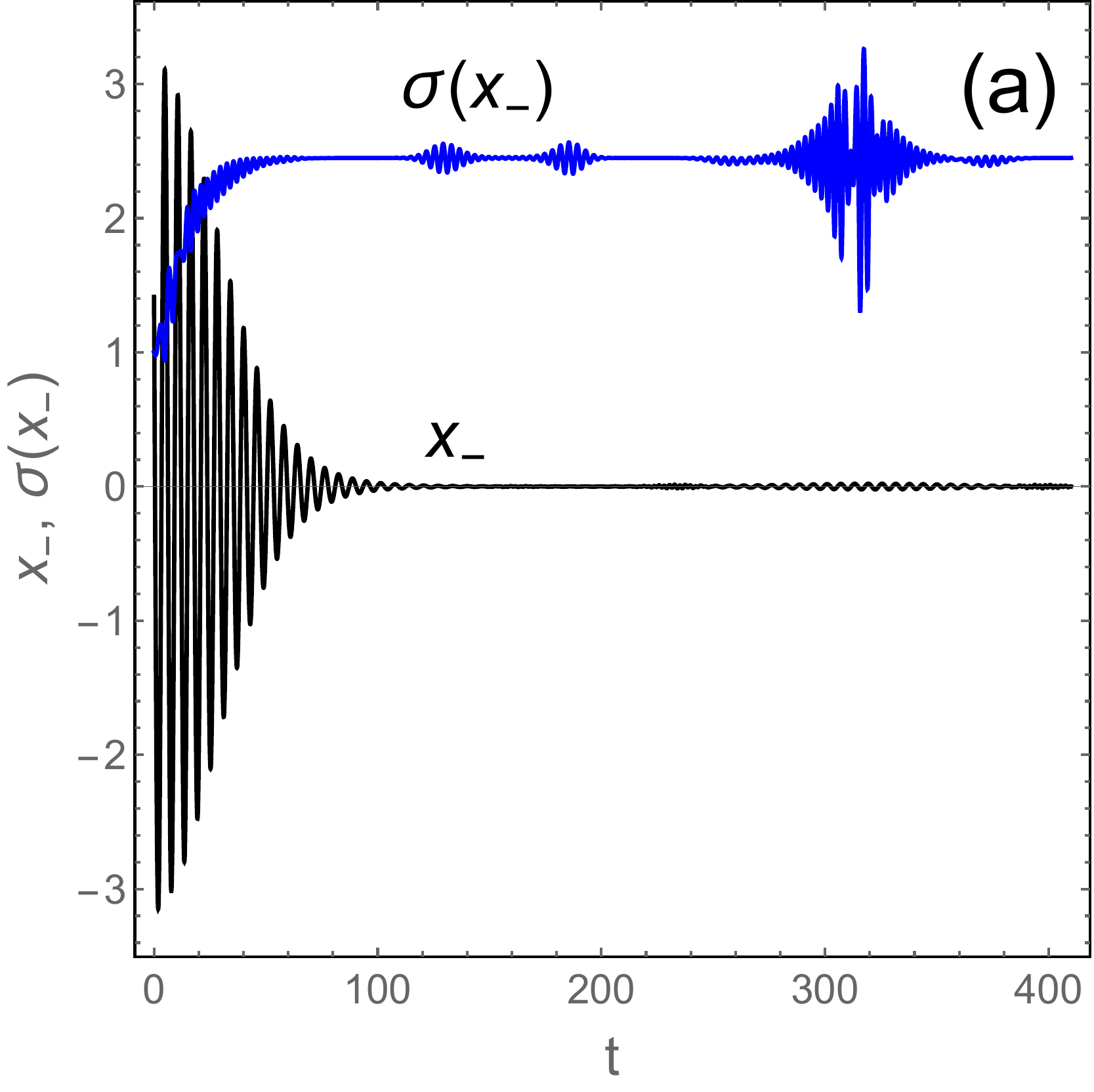}%
\includegraphics[width=170pt]{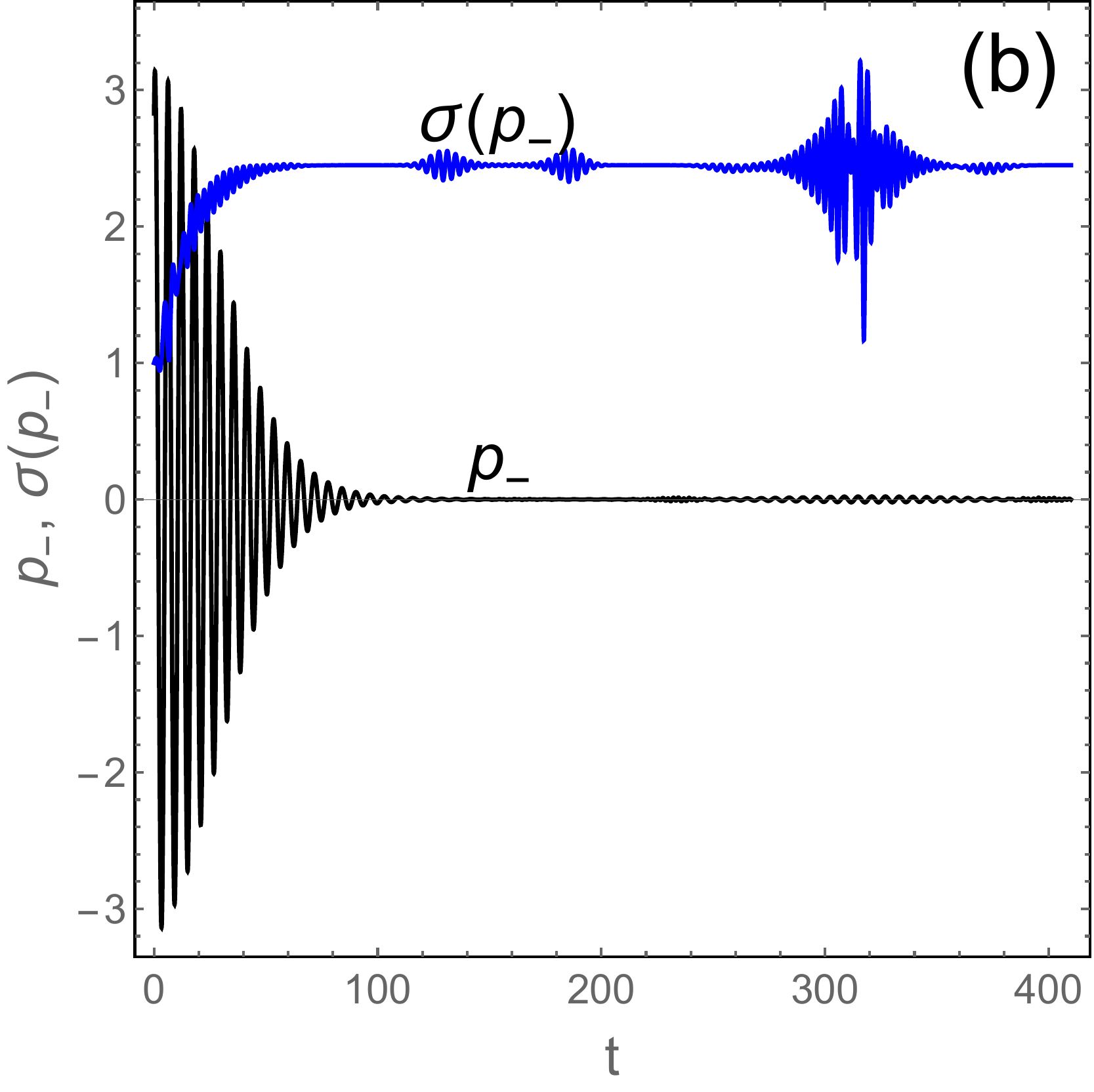}%
\includegraphics[width=170pt]{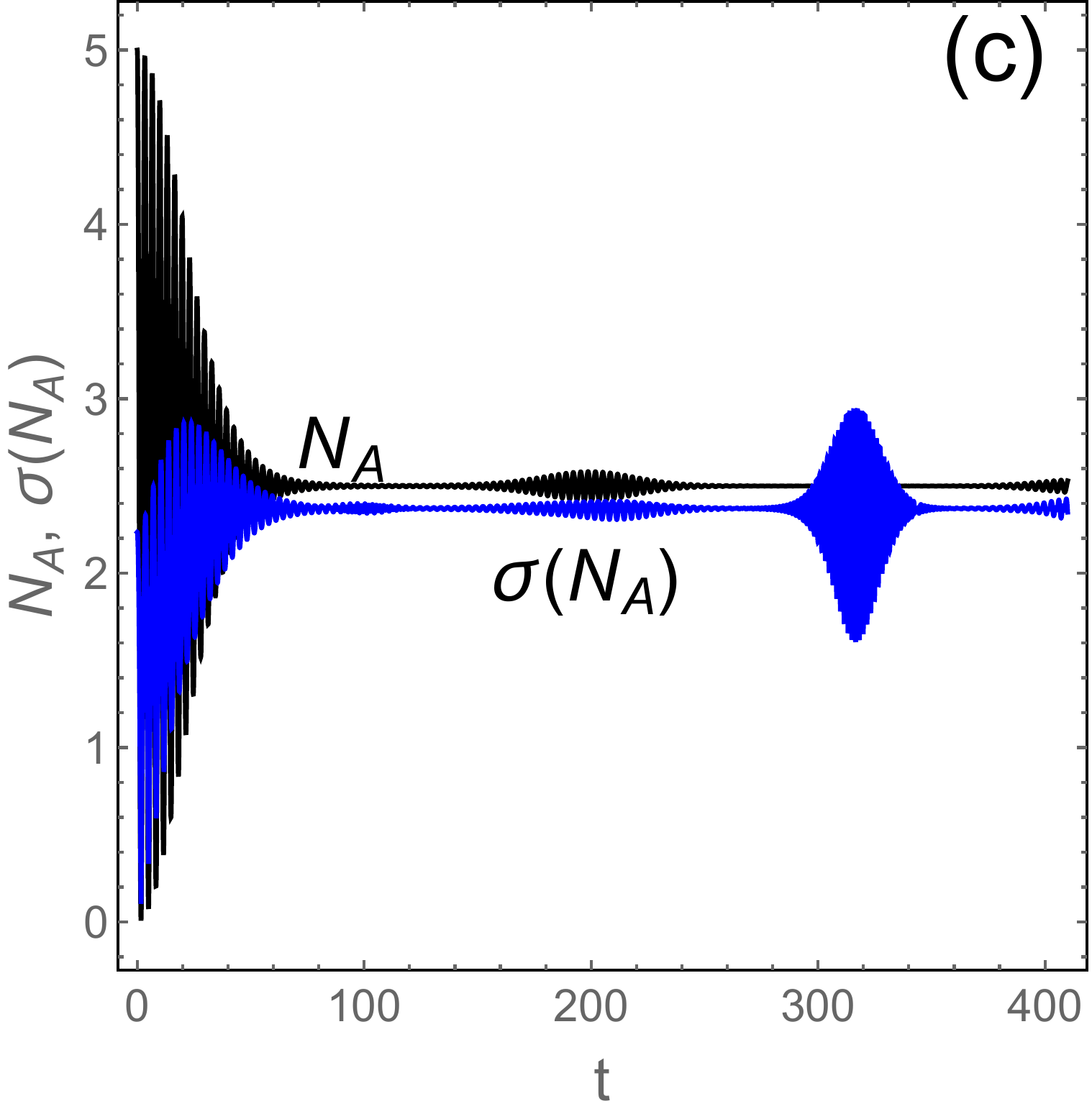}\\[0pt]
\includegraphics[width=170pt]{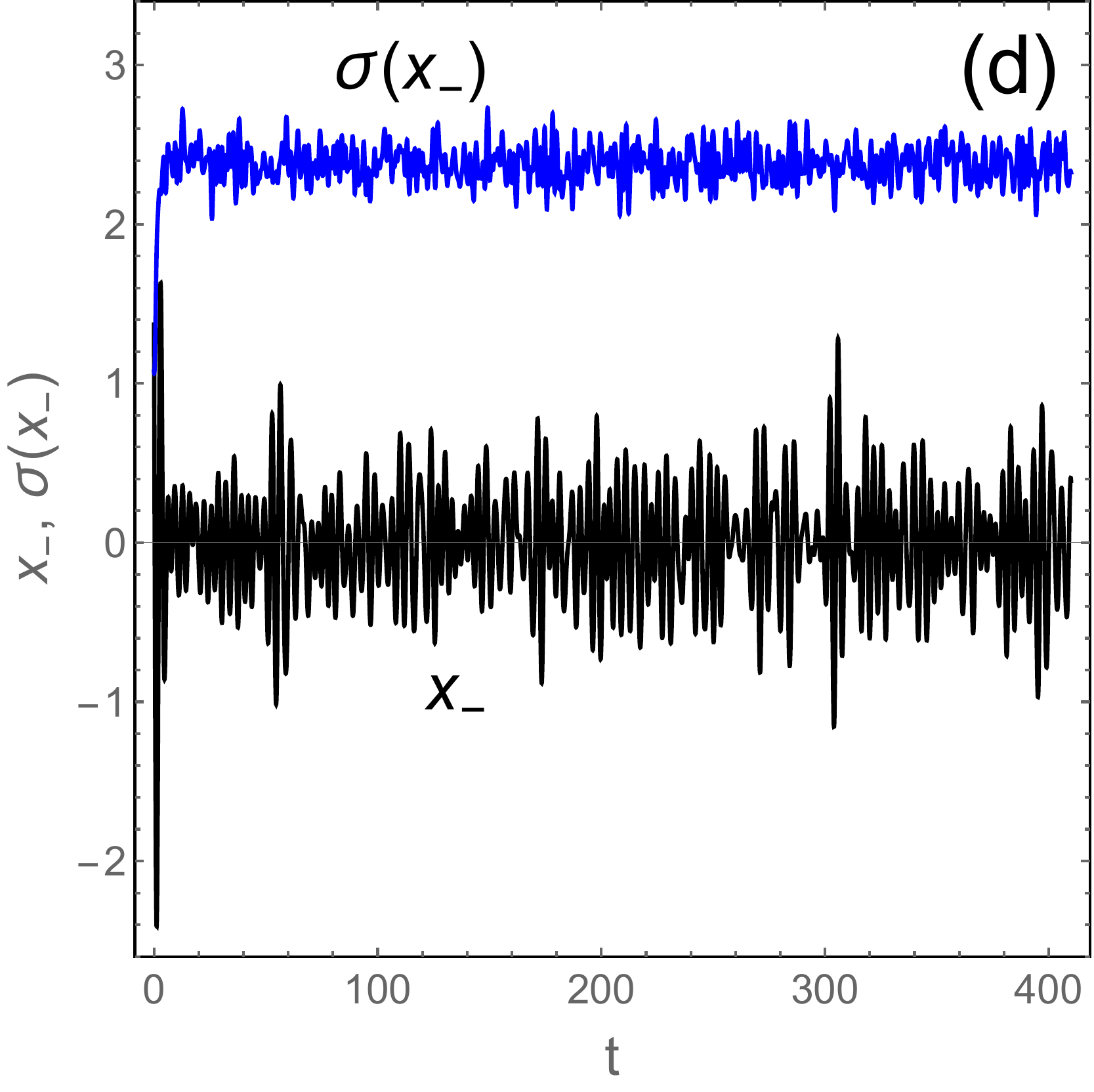}%
\includegraphics[width=170pt]{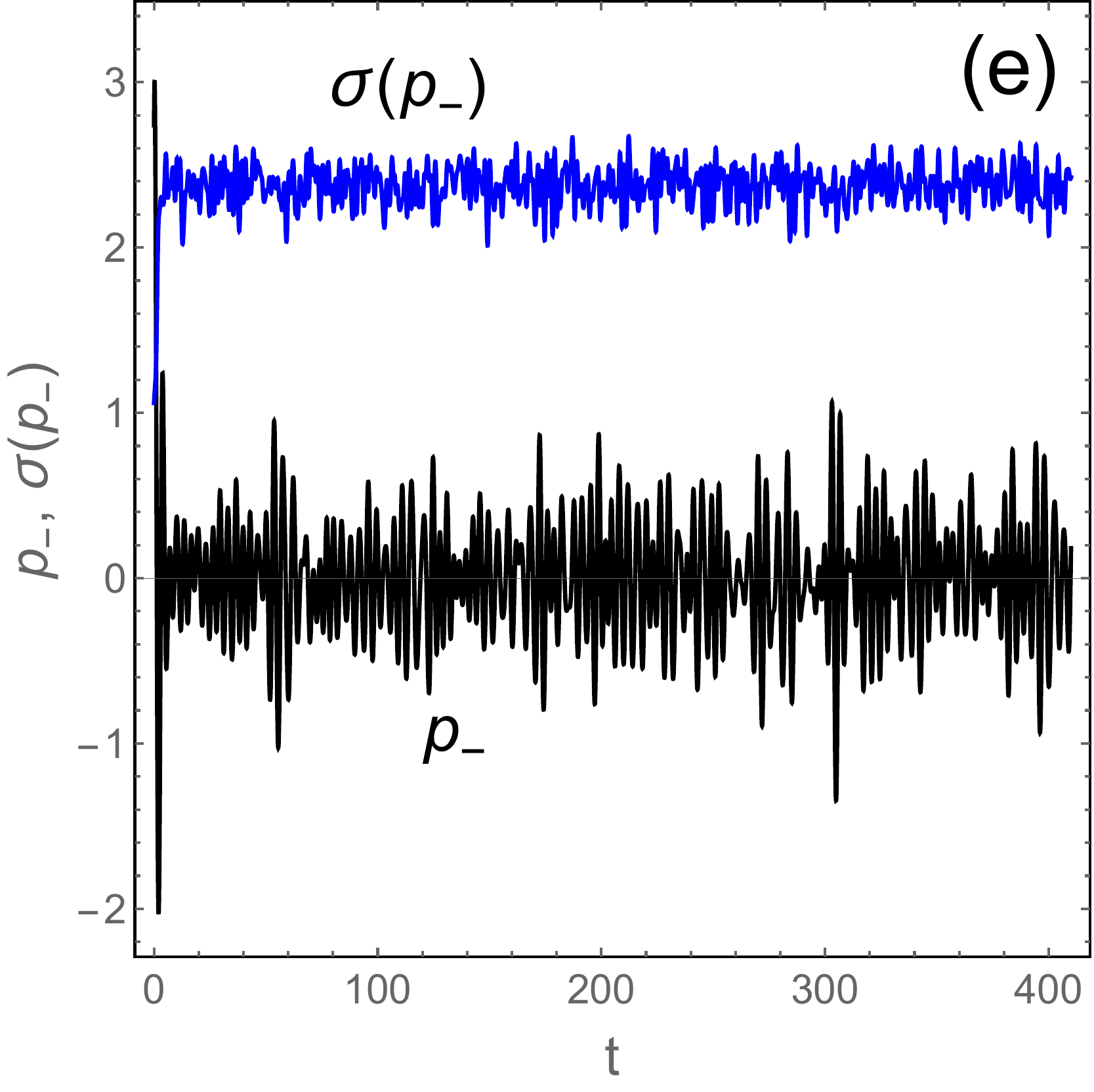}%
\includegraphics[width=170pt]{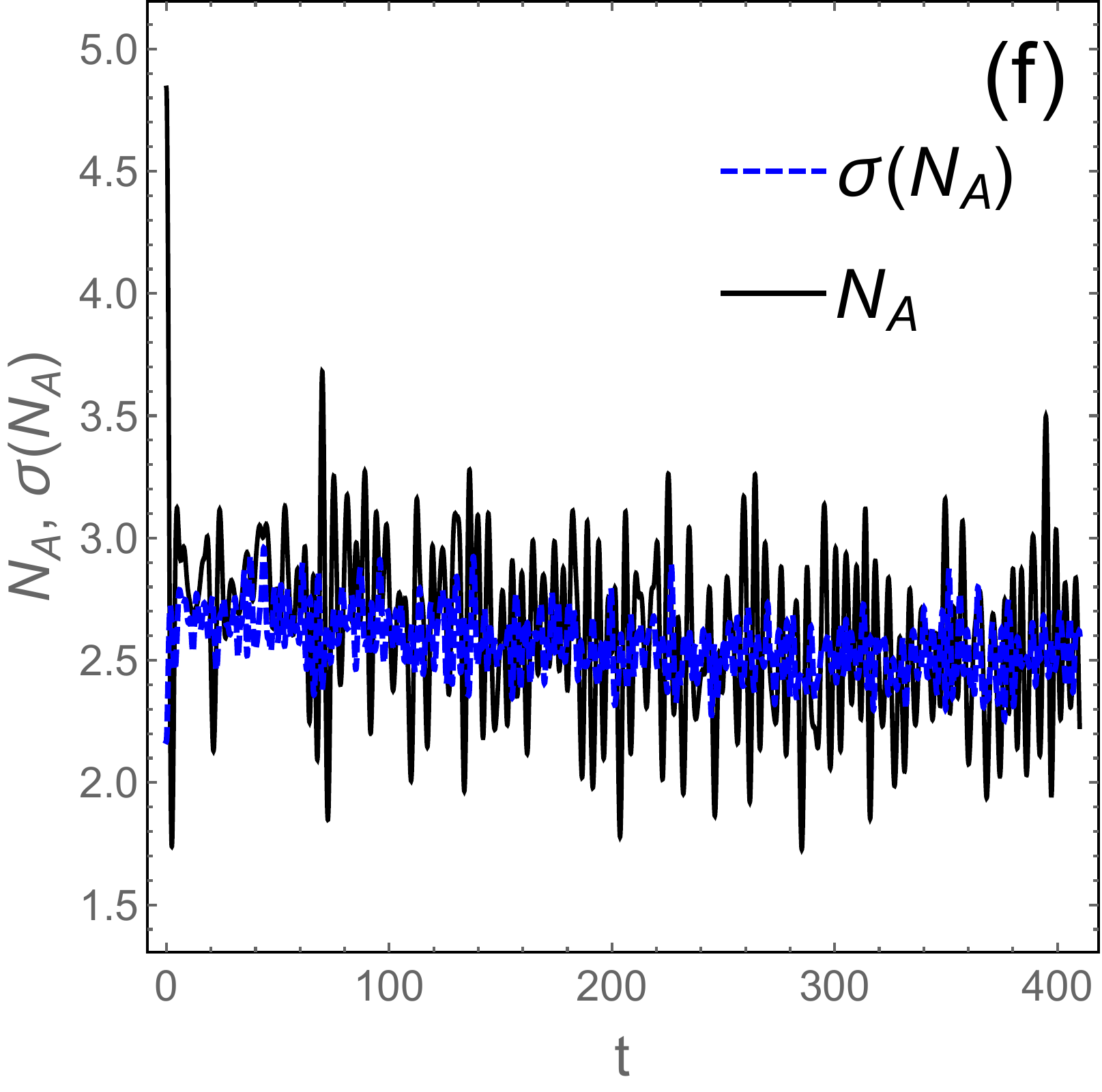}\\[0pt]
\end{center}
\caption{The dynamics of mean values and their relevant uncertainties in the
whole Hilbert space of $\bigoplus_{N}\mathcal{H}_{N}$ (a)-(c) for $N=0,1,2\cdots
,15 $ and (d)-(f) for $N=0,1,2\cdots ,10$. The other parameters are the same
as that in Fig.\protect\ref{figure7}.}
\label{figure11}
\end{figure}

In the Hilbert space of $\mathcal{H}$, the dynamics of CS and QS between scattering modes from
coherent state BEC will be different from that from number state BEC due to
the different dynamics of the mean values and the variances of the
error operators. In this case,
the mode number $\hat{N}_{A}$ and its variance can be
determined by%
\begin{equation*}
N_{A}\left( t\right) =\left\langle \hat{a}_{A}^{\dag }\hat{a}%
_{A}\right\rangle =\sum_{k,l}k\left\vert C_{kl}\left( t\right) \right\vert
^{2},\quad
\sigma \left( N_{A}\right) =\sqrt{%
\sum_{k,l}k^{2}\left\vert C_{kl}\left( t\right) \right\vert ^{2}-\left(
\sum_{k,l}k\left\vert C_{kl}\left( t\right) \right\vert ^{2}\right) ^{2}}.
\end{equation*}%
Fig.\ref{figure11} calculates the dynamics of the mean values and their
corresponding fluctuations, and it shows that
the quantum fluctuations (which are omitted in a classical dynamics)
still play an important roles in the synchronized behaviors in this case. For
a coherent-state BEC, two scattering modes can quickly reach an exact
CS in a sense of $x_{-}\rightarrow 0, p_{-}\rightarrow 0$ (see
Fig.\ref{figure11} (a)(b)) but their fluctuations present ``collapse and
revival" behaviors and will remain for a long time due to the weak dissipations
of the BEC scattering modes. But for a trapping truncation of $%
n=1-g/\chi\approx 10$, the two modes hardly reach to a complete CS but
more irregular fluctuations are produced than that in the number-state BECs (see Fig.\ref{figure7}(d)(e)(f))because
two scattering modes generated in the number-state BECs will no longer trap themselves
in their initial states within their respective Hilbert space of $\mathcal{H}_N$ and the
mode mixing between different number-state BECs will be involved.
\begin{figure}[htp]
\begin{center}
\includegraphics[width=240pt]{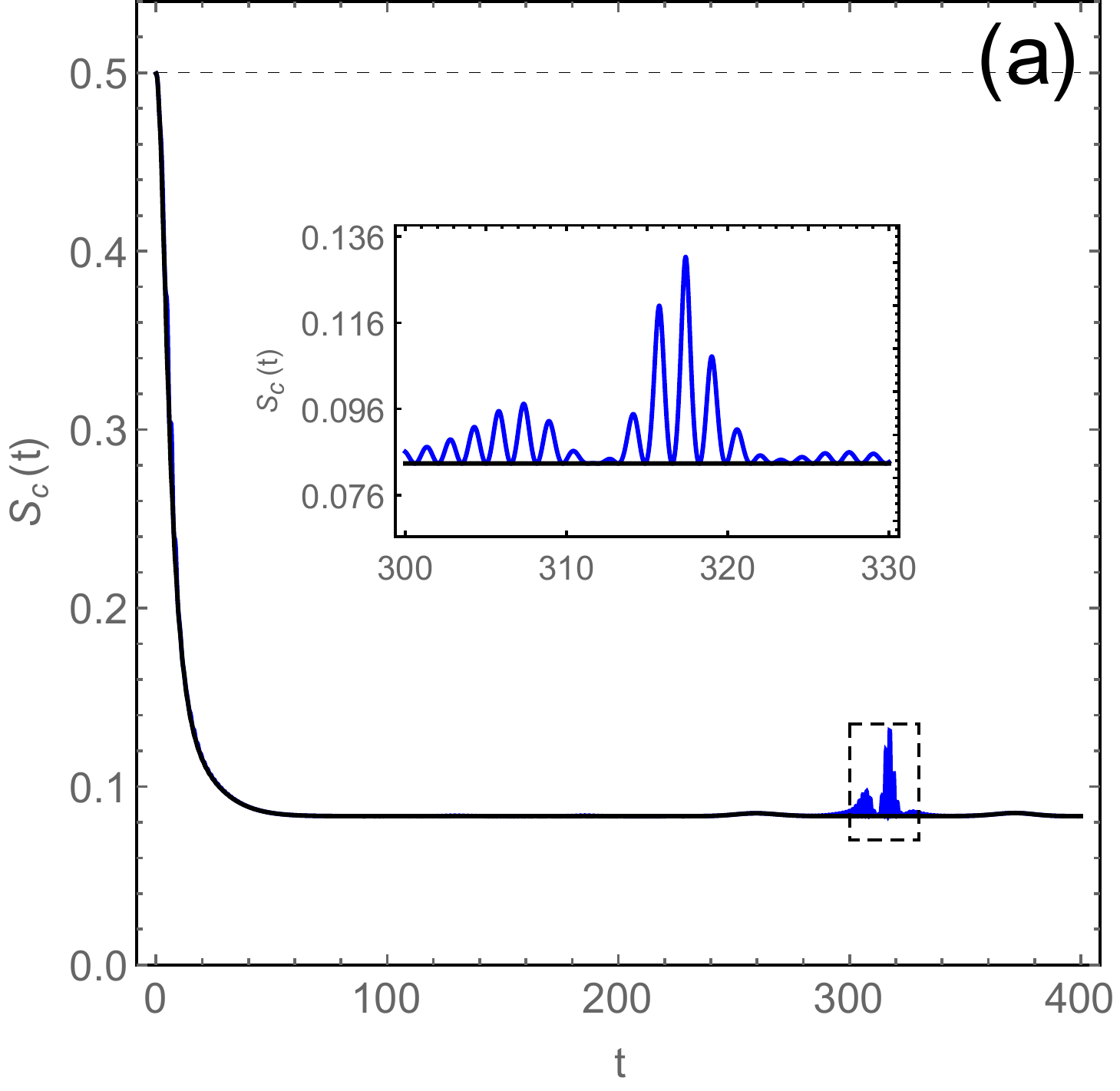}%
\includegraphics[width=240pt]{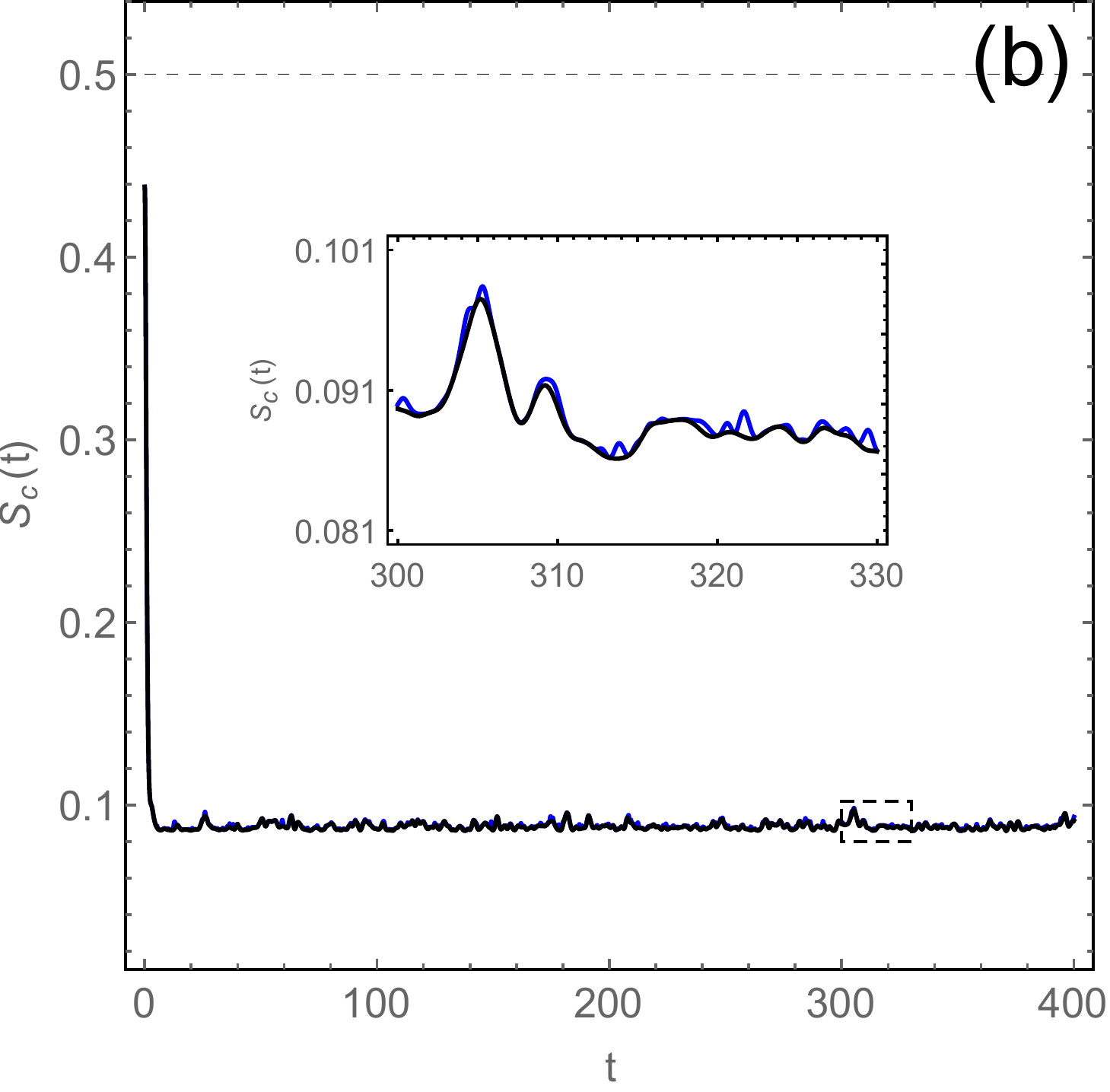}\\[0pt]
\includegraphics[width=240pt]{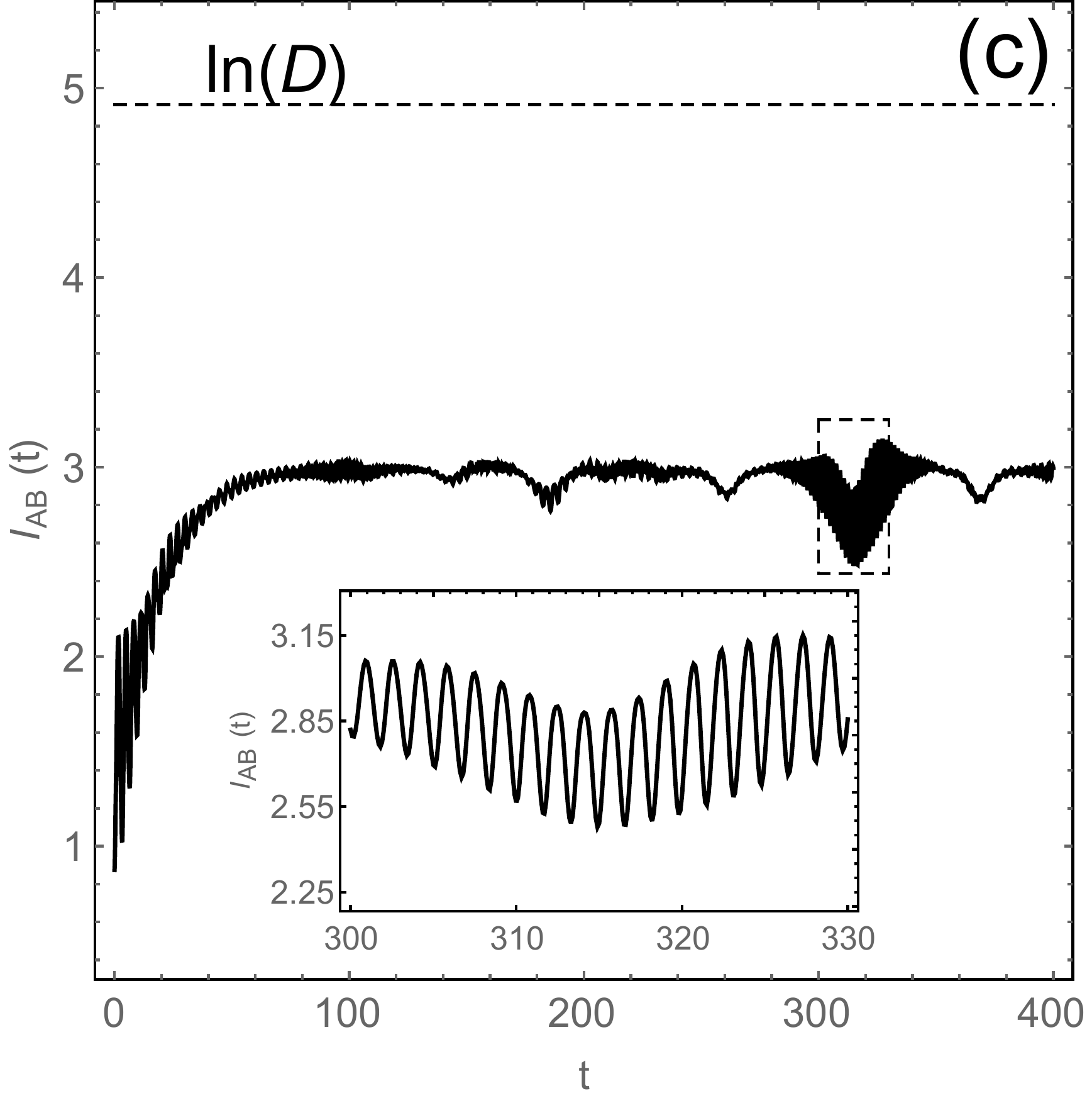}%
\includegraphics[width=240pt]{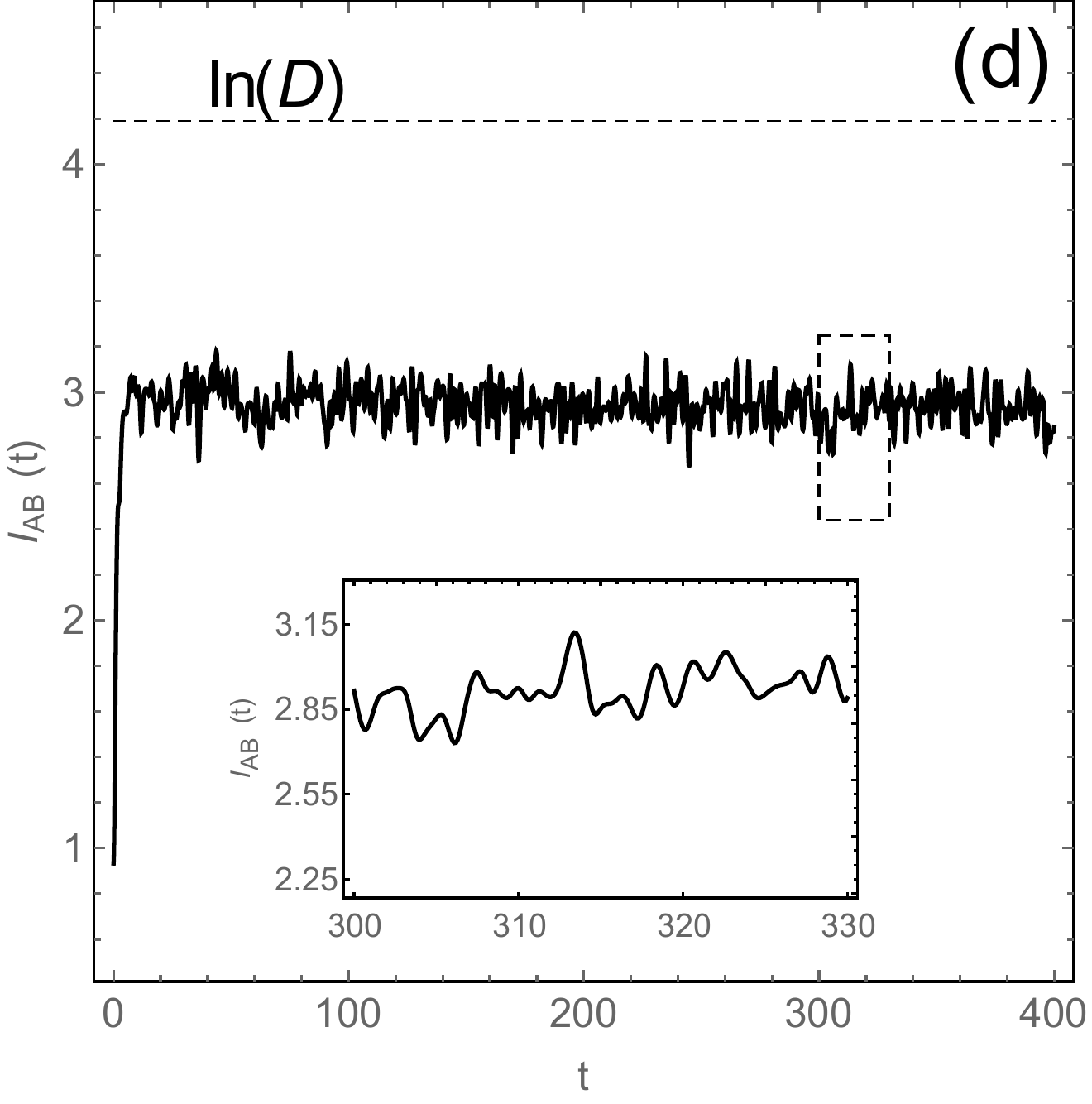}\\[0pt]
\end{center}
\caption{The dynamical measure synchronization of (a)(b) $S_{c}(t)$ and
(c)(d) the mutual information $I_{AB}(t)$ in the whole Hilbert space of $\bigoplus_{N}%
\mathcal{H}_{N}$ for $N=0,1,2,\cdots ,n$. The truncated number $n=15$ for (a) (c) and
$n=10$ for (b)(d). The other parameters are the same as that in Fig.%
\protect\ref{figure8} and Fig.\protect\ref{figure9}. Insets: the zoomed-in images from the dashed rectangles.}
\label{figure12}
\end{figure}

Similarly, we investigate the Mari measure of $%
S_{c}(t)$ in the mixed Hilbert space by calculating
\begin{equation*}
S_{c}\left( t\right) \leq \frac{1}{2\sigma \left( x_{-}\right) \sigma \left(
p_{-}\right) }\leq \frac{1}{2},
\end{equation*}%
and the $Q$-function mutual information (Eq.(\ref{IAB}))%
\begin{equation}
I_{AB}\left( t\right) =-\frac{2}{\pi }\int_{0}^{\infty }\int_{0}^{2\pi
}Q\left( r,\theta ,t\right) \ln \left[ Q\left( r,\theta ,t\right) \right]
drd\theta,
\label{Ir}
\end{equation}%
where
\begin{equation*}
Q\left( r,\theta ,t\right) =e^{-r^{2}}\sum_{k=0}^{n}\frac{\Gamma \left( k+%
\frac{1}{2}\right) }{k!}\left[ \sum_{l=0}^{n-k}\sum_{j=0}^{n-k}\frac{%
C_{lk}\left( t\right) C_{jk}^{\ast }\left( t\right) }{\sqrt{l!j!}}%
r^{l+j}e^{i\left( j-l\right) \theta }\right] .
\end{equation*}

In this case, the mutual information between the two modes has a limit of $%
0\leq I_{AB}\left( t\right) \leq \ln D$, where the full dimension of the
truncate Hilbert space is $D=(n+1)(n+2)/2$. Both the temporal evolutions of $%
S_c(t) $ and $I_{AB}(t)$ are comparatively demonstrated in Fig.\ref{figure12}%
(a)-(d). The calculation of the mutual information in this case is a little
more complicated because the integral dimension of $Q$-function in Eq.(\ref{Ir}) is
doubled with respect to both the amplitude $r$ and the phase $\theta$.
According to our calculation, we can see that both Mari measure $S_{c}(t)$ and mutual
information $I_{AB}(t)$ exhibit smoother characteristics of motion, but their opposite mean-value
tendencies and relevant details of ``revival and collapse" still indicate a close
connection with each other. Anyway, based on the quantum measures expressed by $Q$-function,
we show that the synchronized behaviors between two nonlinear modes in a closed quantum
system can exhibit microscopic dynamics of QS by establishing
similar irregular patterns in the phase space. The QS measures estimated either by $S_{c}(t)$ or
by $I_{AB}(t)$ are closely connected to the everlasting fluctuations maintained by the quantum correlations
during the quantum evolution beyond mean-value dynamics.

\section{Conclusion}
Usually, the synchronization refers to an emergent behavior that often occurs between coupled
self-sustained oscillators (the driven dissipative systems) \cite{Janson,Pikovsky} by means of
establishing collective correlated dynamics among different autonomous oscillators. However, in
a problem of quantum control of quantum units with nonlinear couplings, only the synchronized
dynamics beyond decoherence is concerned about, which enables the validity of treating an open system
as a closed non-dissipative system. In the closed quantum systems, the synchronized behaviors at
macroscopic and microscopic levels both play important roles during the control process and a
complete investigation of synchronization for a whole quantum system should include both CS and QS.
Therefore, in this paper, we choose BEC as an example to investigate the dynamics of CS and QS
between two nonlinear scattering modes both from the classical and quantum points of view. Our
study shows that the evolution of two scattering modes exhibits strong synchronized behaviors in
both classical and quantum aspects in spite of their nonlinear dynamics. The CS in a closed quantum
system is based on the mean-value behavior due to its macroscopic dynamical scale and exhibits a
typical phenomenon of MS which gives attractive interwinding trajectories in the phase space sharing
similar phase areas because of the conservation of total phase volume and the equilibrium energy exchange
between modes. We propose a comparable method to use $Q$-function as a tool to study the corresponding
QS behavior and investigate the dynamics of QS through the overlapping patterns of probability density
in the phase space as well as the mutual information sharing between two nonlinear modes. The
calculations reveal that the ``revival and collapse" of the fluctuations, derived from the coherent
superposition of eigenmodes (see Eq.(\ref{psi}) or Eq.(\ref{psi2})), discriminates the dynamics of QS
from CS, and a full synchronization for CS is basically impossible for QS in a closed system.
The calculation indicates that the overlaps of the correlated trajectories established in CS are
statistically different from that of the similar density patterns characterized in QS.

In order to highlight the roles of quantum fluctuations in the synchronization, the Mari measure of
$S_c(t)$, which is a generalized measure of CS based on the dynamical variances (e.g. $\sigma(x)$
and $\sigma(p)$)\cite{Mari}, is calculated in the quantum regime. The result indicates that the upper
limitation of $S_c(t)$, which excludes an exact synchronization between two non-dissipative quantum
modes, is due to the permanent fluctuations of the unitary evolutions of a closed quantum system, and
the dynamics of  fluctuations is hence sensitively dependent on the initial states of the system (e.g.
number state BEC or coherent state BEC). In the other hand, the QS measure estimated by the mutual information
$I_{AB}(t)$ gives similar dynamics of the fluctuations compared with that of $S_c(t)$ in spite of the
opposite mean time dynamical tendencies of decreasing $S_c(t)$ and increasing $I_{AB}(t)$ starting from
the same initial states. The evolution of the mutual information calculated by $Q$-function also implies
a close connection of QS with the quantum entanglement \cite{Manzano} in a sense of variable separation
of $Q$-functions. The synchronized behavior of a closed quantum system analyzed by $Q$-function reveals
that the distance measure of $S_c(t)$ and the correlation measure of $I_{AB}(t)$ are dynamically equivalent
in spite of providing different aspects of synchronization, and which further implies that the measure
to describe QS can not be uniquely defined as that for CS.

Anyway, based on the different measures calculated by $Q$-function, we can find that the synchronized
behaviors between two nonlinear modes in a closed quantum system exhibit not only the macroscopic behaviors
of CS by establishing interwinding trajectories in the classical phase space, but also the microscopic behaviors
of QS by conducting similar dynamics of irregular distribution patterns in the quantum phase space. The
QS measures estimated either by the distance variance $S_{c}(t)$ or by the mutual information $I_{AB}(t)$
are connected with each other by a relevant dynamics of fluctuation beyond the mean-value dynamics. As the intrinsic
fluctuations related to quantum correlations are perfectly protected in a closed quantum system, the uncertainty
principle inevitably excludes the complete synchronization for QS and basically discriminates the QS from CS.

\begin{acknowledgments}
This work is supported by the National Natural Science Foundation of China
(Grants No.11447025 and No.11234003) and the National Basic Research Program of China (973 Program)
under Grant No. 2011CB921604.
\end{acknowledgments}

\end{document}